%% file: axion_mass_so10.tex
\documentclass[a4paper,11pt]{article}
\pdfoutput=1 

\usepackage{jheppub} 
\usepackage{hyperref}
\usepackage[T1]{fontenc} 

\newcommand{\hts}{$\overline{126}_H$ }
\def\vev#1{\left\langle #1\right\rangle}

\newcommand{\fPQ}{f_{\rm PQ}}
\newcommand{\MPQ}{M_{\mathrm{PQ}}}
\newcommand{\MBL}{M_{\mathrm{BL}}}
\newcommand{\MU}{M_{\mathrm U}}
\newcommand{\GeV}{\,\mathrm{GeV}}
\newcommand{\eV}{\,\mathrm{eV}}
\newcommand{\vbl}{v_{\mathrm{BL}}}

\title{\boldmath Axion Predictions in $SO(10)\times U(1)_{\rm PQ}$ Models}

\author[a]{Anne Ernst,}
\author[a]{Andreas Ringwald,}
\author[b]{and Carlos Tamarit}

\affiliation[a]{Deutsches Elektronen-Synchrotron DESY,\\ Notkestra\ss e 85, D-22607 Hamburg, Germany}
\affiliation[b]{Physik Department T70, Technische Universit\"at M\"unchen,\\ James Franck Stra\ss e 1, 85748 Garching, Germany}

\emailAdd{anne.ernst@desy.de}
\emailAdd{andreas.ringwald@desy.de}
\emailAdd{carlos.tamarit@tum.de}

\abstract{
Non-supersymmetric Grand Unified $SO(10)\times U(1)_{\rm PQ}$ models have all the ingredients to solve several fundamental 
problems of particle physics and cosmology -- neutrino masses and mixing, baryogenesis, the non-observation of strong CP violation, dark matter, inflation -- in one stroke. The axion - the pseudo Nambu-Goldstone boson arising from the 
spontaneous breaking of the $U(1)_{\rm PQ}$ Peccei-Quinn symmetry - is the prime dark matter candidate in this setup. 
We determine the axion mass  and the low energy couplings of the axion to the Standard Model particles, 
in terms of the relevant gauge symmetry breaking scales. We 
work out the constraints imposed on the latter by gauge coupling unification. We discuss the cosmological and phenomenological 
implications. 
}

\begin{document} 

\begin{flushright}
{\large \tt DESY 17-213}\\
{\large \tt TUM-HEP-1127/18}
\end{flushright}

\maketitle
\flushbottom

\input{axion_mass_so10_intro}

\input{axion_mass_so10_case_for_pq}

\input{axion_generalities}

\input{axion_mass_so10_models}

\input{axion_mass_so10_constraints_unification}

\input{axion_mass_so10_conclusions}

\newpage

\appendix
\input{app_phases}
\input{app_roots}

\input{app_beta_functions}

\input{app_discrete_symmetry}

\acknowledgments

We would like to thank Karin Daum, Ketan Patel, Werner Rodejohann,  Ken'ichi Saikawa  and Javier Redondo
for stimulating discussions and useful hints. C.T. acknowledges partial support from the 
Collaborative Research Centre SFB1258 of the Deutsche 
Forschungsgemeinschaft (DFG).



%
%
%




\bibliographystyle{JHEP}
\bibliography{gutbib}{}

\end{document}

%% file: axion_mass_so10_intro.tex
\section{Introduction}

Observations in particle physics and cosmology have revealed five fundamental problems which can not be solved
by the field and particle content of the Standard Model (SM): 
\begin{itemize}
\item[{\em i)}] neutrino masses and mixing,
\item[{\em ii)}] the baryon asymmetry of the universe,
\item[{\em iii)}] the non-observation of strong CP violation,
\item[{\em iv)}] dark matter, and 
\item[{\em v)}] inflation.
\end{itemize}

Employing a bottom-up approach, it has been shown recently that a minimal extension model of the SM -- dubbed Standard Model - Axion - Seesaw - Higgs portal inflation model or SMASH --  may explain these problems in one stroke \cite{Ballesteros:2016euj,Ballesteros:2016xej}.  It consists in extending the SM by 
three right-handed extra neutrinos, by an extra quark, and a complex scalar field, which are charged under a new global $U(1)_{\rm PQ}$ Peccei-Quinn (PQ) symmetry. The latter is assumed to be broken spontaneously at an intermediate scale $v_\sigma\sim 10^{11}$\,GeV given by the vacuum expectation value of the scalar field. 
The neutrino flavour oscillation puzzle is solved by the well-known type-I 
seesaw mechanism ~\cite{Minkowski:1977sc,GellMann:1980vs,Yanagida:1979as,Mohapatra:1979ia}: the neutrino mass eigenstates split into a
heavy set comprising three states with masses proportional
to $v_\sigma$, composed of mixtures of the new right-handed
neutrinos, and a light set of three states with masses
inversely proportional to $v_\sigma$, composed of mixtures of the SM
left-handed neutrinos. 
The extra quark and the excitation of the modulus of the new
scalar field also get large masses proportional to $v_\sigma$, while the
excitation of the phase of the new scalar field stays very light,
its mass being inversely proportional to $v_\sigma$. Crucially, this
phase field acquires a linear coupling to the gluonic
topological charge density from a loop correction due to the
extra quark. Correspondingly, it replaces the $\theta$ angle of QCD by 
a dynamical field 
and thus solves the strong CP problem \cite{Peccei:1977hh}. Its particle excitation can
therefore be identified with the axion \cite{Weinberg:1977ma,Wilczek:1977pj}. 
Loop effects involving gravitons induce non-minimal gravitational 
couplings of the Higgs boson and of the new scalar field.
These couplings make the scalar potential energy in the Einstein frame convex  and
asymptotically flat at very large field values. Correspondingly, 
the modulus of the new scalar field or a mixture of it with the modulus of the
Higgs field can play the role of the inflaton. 
The baryon asymmetry is produced  via thermal leptogenesis~\cite{Fukugita:1986hr}. 
Soon after reheating of the universe and breaking of the 
$U(1)_{\rm PQ}$ symmetry, the decays of the right-handed neutrinos
produce a lepton asymmetry, which {is partly converted into a baryon asymmetry by sphaleron interactions before the breaking of the electroweak symmetry. } 
Finally, at temperatures around the QCD transition between a
quark-gluon and a hadron plasma phase, dark matter is
produced in the form of a condensate of extremely nonrelativistic
axions \cite{Preskill:1982cy,Abbott:1982af,Dine:1982ah}. To account for all of the cold dark matter in
the universe, the $U(1)_{\rm PQ}$ symmetry breaking scale is required to
be around $v_\sigma \sim 2\times 10^{11}$\,GeV,
corresponding an axion mass around
$30\,\mu$eV \cite{Ballesteros:2016euj,Ballesteros:2016xej,Borsanyi:2016ksw,Klaer:2017ond}. 
Adding a cosmological constant to account for the present
acceleration of the universe, SMASH offers a
self-contained description of particle physics, from the
electroweak scale to the Planck scale, and of cosmology, from
inflation until today.

It is an interesting question whether one can get a similar self-contained description by exploiting a grand unified theory (GUT)
extension of the SM: a GUT SMASH variant\footnote{For a recent first attempt in this direction exploiting a non-supersymmetric $SU(5)$ setup see ref.  \cite{Boucenna:2017fna}.}. 
In fact, it is well established that GUTs based on the gauge group $SO(10)$ 
\cite{Georgi:1974my,Fritzsch:1974nn} may solve the fundamental problems {\em i)-iv)}
discussed above exploiting the same mechanisms as our bottom-up variant of SMASH \cite{Reiss:1981nd,Mohapatra:1982tc,Holman:1982tb,Bajc:2005zf,Altarelli:2013aqa,Babu:2015bna}. In fact, the right-handed neutrinos and thus the seesaw mechanism and the possibility of baryogenesis via thermal leptogenesis 
\cite{Fong:2014gea}
occur automatically 
in these models. Moreover, an axion  suitable to solve the strong CP problem and to account for the observed 
amount of dark matter can arise from the rich Higgs sector of these models. An intermediate scale 
$M_{\rm I}\sim 10^{11}$\,GeV (as required in order to have axion dark matter) between the GUT scale and the electroweak scale may arise naturally from the necessity of one or more intermediate gauge 
groups between the SM gauge group and $SO(10)$ to get  gauge coupling unification without invoking TeV scale supersymmetry \cite{Deshpande:1992au,Deshpande:1992em,Bertolini:2009qj}. 
Finally, 
the Higgs sector of these models necessary for the breaking of $SO(10)$ and its intermediate scale subgroups also provides candidates for the inflaton if they are  non-minimally coupled to gravity \cite{Leontaris:2016jty}. 

As a first step towards an $SO(10)$ GUT SMASH model, in this paper we identify the physical axion field and determine its low energy effective Lagrangian in a set of well-motivated non-supersymmetric $SO(10)\times U(1)_{\rm PQ}$ models -- a missing piece in the existing literature. In our treatment we bridge the gap between GUT and low scales: we pay particular attention to low energy constraints, ensuring orthogonality of the physical axion with respect to the gauge bosons of all the broken gauge groups, and we are able to identify the global symmetry associated with the physical axion, given by a combination of the original PQ symmetry and transformations in the Cartan subalgebra of $SO(10)$. We  provide calculations of the domain-wall number for the physical axion --which match the expectations from the simple UV symmetries-- and we also explore how gauge coupling unification, proton decay, B-L, black hole superradiance and stellar cooling constraints affect the allowed window of axion masses. 

The paper is structured as follows. In Section \ref{thecaseforapqsymmetry} we revisit 
how a PQ symmetry can be motivated in non-supersymmetric $SO(10)$ models independently of the strong CP problem: it forbids  some terms in the Yukawa interactions
responsible for the fermion masses and their mixing, thereby crucially improving the economy and predictivity 
of the models. 
The general construction of the low energy effective Lagrangian of the axion in theories with multiple scalar 
fields is reviewed in Section \ref{axiongeneralities}. 
This formalism is then used in Section \ref{so10xu1pqmodels} to 
work out the axion predictions for a number of promising $SO(10)\times U(1)_{\rm PQ}$ models. 
The constraints imposed by gauge coupling unification are explored in Section \ref{gaugecouplingunification}. We
summarise and discuss the cosmological and phenomenological 
implications of our results  in  Section \ref{discussionandconclusions}.

%% file: axion_mass_so10_case_for_pq.tex
\section{The case for a Peccei-Quinn symmetry}
\label{thecaseforapqsymmetry}

The SM matter content  nicely fits in three generations of a 16-dimensional spinorial representation 
$16_F$ of $SO(10)$, cf. Table \ref{tab:fermions}.  On the other hand, there are many possible Higgs representations corresponding to various possible 
symmetry stages between $SO(10)$ and 
$SU(3)_C \otimes U(1)_{\rm em}$, cf.  Table \ref{tab:vevs}.
\begin{table}[h]
\begin{align*}
\begin{array}{|c|c|c|c|r|}
\hline
  SO(10)	& 4_C 2_L 2_R	&  4_C 2_L 1_R				&  3_C 2_L 1_R 1_{B-L}			&\ \hskip-2cm{3_C 2_L 1_Y}								\\
  \hline
  \hline
  16_F		&(4,2,1)	&(4,2,0)				&   \left(3,2,0,\frac{1}{3}\right)		&   \left(3,2,\frac{1}{6}\right)   :=q  			\\
                &		&					&   (1,2,0,-1)            		&   \left(1,2,-\frac{1}{2}\right)  :=l					\\
                \cline{2-5}
		&(\bar 4,1,2)	&\left(\bar4,1,\frac{1}{2}\right)	&   \left(\bar 3,1,\frac{1}{2},-\frac{1}{3}\right)		&   \left(\bar 3,1,\frac{1}{3}\right):=d         \\
		&		&					&   \left(1,1,\frac{1}{2},1\right)         	   	&   \left(1,1,1\right)  :=e      		        \\
                                \cline{3-5}
		&               &\left(\bar4,1,-\frac{1}{2}\right)	&   \left(\bar 3,1,-\frac{1}{2},-\frac{1}{3}\right)		&   \left(\bar 3,1,-\frac{2}{3}\right):=u       \\
		&		&					&   \left(1,1,-\frac{1}{2},1\right)         	   	&   \left(1,1,0\right):=n	                        \\
  \hline
 \end{array} 
 \end{align*}
 \caption{\label{tab:fermions}Decomposition of the fermion multiplets according to the various subgroups in our breaking chains.
 All SM fermions have masses set by the Higgs mechanism, the heavy right handed neutrinos acquire their mass at the BL breaking scale from the coupling to the $\overline{126}_H$. 
 }
\end{table}
Group theory requires 
at least the  following representations  in order to achieve a full breaking of the rank five group $SO(10)$ down to the rank 4 SM group $SU(3)\times SU(2)\times U(1)$:
\begin{itemize}
\item $16_H$ or $\overline{126}_H$: they reduce the rank by at least one unit,  either leaving a rank four $SU(5)$ little group unbroken, or else breaking the SM group.
\item $45_H$ or $54_H$ or $210_H$: they admit for rank five little groups, either $SU(5) \otimes U(1)$ or different ones, like  the Pati-Salam (PS) group $SU(4)\otimes SU(2)\otimes SU(2)$ \cite{Pati:1974yy}. In the latter case, the intersection
of the little group with the $SU(5)$ preserved by a $16_H$ or $126_H$
can give the SM gauge group.
\end{itemize}

{\renewcommand{\arraystretch}{1.2}
\begin{table}[t]
\begin{align*}
\begin{array}{|c|c|c|c|c||c|c|}
\hline
  SO(10)	& 4_C2_L2_R	&  4_C2_L1_R		&  3_C2_L1_R1_{B-L}	& 3_C2_L1_Y		& 3_C 1_{\rm em}		&\text{VEV}\\
  \hline
  \hline
  10_H		&(1,2,2)	&(1,2,\frac{1}{2})	& (1,2,\frac{1}{2},0)	& (1,2,\frac{1}{2})	&(1,0)=:H_u		& v^{10}_u\\
		& 		& (1,2,-\frac{1}{2})	&(1,2,-\frac{1}{2},0)	&(1,2,-\frac{1}{2})	&(1,0)=:H_d	& v^{10}_d\\
  \hline
\overline{126}_H &(10,1,3)	&(10,1,1)	&(1,1,1,-2)		&(1,1,0)		&(1,0)	:=\Delta_R		&v_R\\
		&(15,2,2)	&(15,2,\frac{1}{2})	&(1,2,\frac{1}{2},0)	&(1,2,\frac{1}{2})	&(1,0):=\Sigma_u 	 &v^{126}_u\\
		& 		& (15,2,-\frac{1}{2})	&(1,2,-\frac{1}{2},0)	& (1,2,-\frac{1}{2})   	&(1,0):=\Sigma_d	& v^{126}_d\\
  \hline 
  210_H		&(1,1,1)  &(1,1,0)		&(1,1,0,0)		&(1,1,0)		&(1,0):=\phi			&v^{210}\\
  \hline
 \end{array} 
 \end{align*}
 \caption{\label{tab:vevs}Decomposition of the scalar multiplets according to the various subgroups in our breaking chains. We only display the multiplets which get nonzero vacuum expectation values (VEVs) in the different models considered in the paper.}
\end{table}

We will exploit in our explicit models the $126_H$ and the $210_H$ representations. 
Since $16_F\times 16_F= 10_H+120_H+\overline{126}_H$, the most general Yukawa couplings involve at most  three possible Higgs representations, 
\begin{equation}
{\mathcal L}_Y = 16_F \left( Y_{10} 10_H + Y_{120} 120_H + Y_{126} \overline{126}_H \right) 16_F 
+ {\rm h.c.}, 
\end{equation}
where $Y_{10}$ and $Y_{126}$ are complex symmetric matrices, while $Y_{120}$ is complex antisymmetric. 
It is then natural to ask: what is the minimal Higgs sector to reproduce the observed fermion masses and mixings?
Clearly, in order to get fermion mixing at all, one needs at least two distinctive Higgs representations\footnote{A single Yukawa 
matrix can always be diagonalised by rotating the $16_F$ fields.}.  
Out of the six remaining combinations, however, only three turn out to give realistic fermion mass 
and mixing patterns: $10_H + \overline{126}_H$, $120_H + \overline{126}_H$, and $10_H + 120_H$ 
(see for example \cite{Senjanovic:2006nc,DiLuzio:2011my} and references therein). 
From these combinations, the first two are phenomenologically preferred since the $\overline{126}_H$ is required 
for neutrino mass generation via the seesaw mechanism. The first one is the most studied, in particular because it is the one occurring in the
minimal supersymmetric version of $SO(10)$. We will  also  exploit it in our PQ extensions of $SO(10)$, as elaborated next.

First of all, it is important to note that the components of $10_H$ can be chosen to be either real or complex. 
In the non-supersymmetric case it is natural to assume a real $10_H$ representation.  
However, as pointed out in \cite{Babu:1992ia,Bajc:2005zf}, this is phenomenologically unacceptable, 
because it predicts $m_t \sim m_b$. In the  
alternative case in which 
the complex conjugate fields differ from the original ones by some extra charge, 
$10_H \neq 10_H^\ast$, both components are allowed in the Yukawa Lagrangian,
\begin{equation}
\label{SO10Yukawacomplex}
\mathcal{L}_Y = 16_F \left( Y_{10} 10_H + \tilde{Y}_{10} 10_H^\ast  + Y_{126} \overline{126}_H \right) 16_F + {\rm h.c.} \, ,
\end{equation}
since they transform in the same way under $SO(10)$.
The representations in \eqref{SO10Yukawacomplex} decompose under the Pati-Salam group $SU(4)_C\otimes SU(2)_L\otimes SU(2)_R$ as 
\begin{equation}\begin{aligned}
& 16_F = (4,2,1) \oplus (\overline{4},1,2) \, , \\
& 10_H = (1,2,2) \oplus (6,1,1) \, , \\
& \overline{126}_H = (6,1,1) \oplus (10,1,3) \oplus (\overline{10},3,1) \oplus (15,2,2) \,.
\end{aligned}\end{equation}
(Throughout the paper we
will consider decompositions of representations under the PS gauge group by default).
From the above it follows that
the fields which can develop a VEV in which the SM subgroup $SU(3)_C \otimes SU(2)_L \otimes U(1)_Y$ is only broken by $SU(2)_L$ doublets, as in the standard Higgs mechanism, are $(1,2,2)$,
$(\overline{10},3,1)$, $(10,1,3)$,
and $(15,2,2)$: as seen in table \ref{tab:vevs}, the above PS representations include singlets under $SU(3)_C\otimes U(1)_{\rm em}$.
We denote the associated 
 VEVs as
\begin{equation}
\label{eq:VEVS_10_126}
\begin{aligned}
v_L \equiv \vev{(\overline{10},3,1)_{126}} \, , \qquad 
v_R \equiv \vev{({10},1,3)_{126}} \, , \\
v^{10}_{u,d} \equiv \vev{(1,2,2)^{10}_{u,d}} \, , \qquad
v^{126}_{u,d} \equiv \vev{(15,2,2)^{126}_{u,d}} \, .
\end{aligned}
\end{equation}
The $(1,2,2)$ bi-doublet can be further decomposed under the SM gauge group, 
yielding $(1,2,2)_{\rm PS} = [(1,2,+\tfrac{1}{2})_{\rm SM} \equiv H_u] \oplus [(1,2,-\tfrac{1}{2})_{\rm SM} \equiv H_d]$, where the suffixes PS and SM refer to decompositions of representations 
under 
the Pati-Salam and SM gauge 
groups, respectively. 
Now if $10_H = 10_H^\ast$ we have $H_u^\ast = H_d$ as in the SM, 
while if $10_H \neq 10_H^\ast$ then $H_u^\ast \neq H_d$ as in the MSSM or in the Two Higgs Doublet Model (2HDM).

As can be seen in table \ref{tab:fermions}, each  generation of SM fermions in the $16_F$ of $SO(10)$ transforms as 
$(4,2,1)$ and $(\overline{4},1,2)$ under $SU(4)_C \otimes SU(2)_L \otimes SU(2)_R$. The SM colour group $SU(3)_C$ is embedded within the $SU(4)$ of the PS group, 
$ SU(4)_C\supset SU(3)_C \otimes U(1)_{B-L}$, while  SM hypercharge is identified as
\begin{align}
 Y=U(1)_R+\frac{1}{2}U(1)_{B-L},
\end{align}
with $U(1)_R$ being the usual $T^3$ generator within the Lie algebra of $SU(2)_R$.
Given this embedding of the SM fermion families into PS representations, we can express
the fermion mass matrices arising from the interactions in \eqref{SO10Yukawacomplex} after electroweak symmetry breaking 
as
\begin{align}
\label{nonsusysumrule}
M_u    &= Y_{10} v_u^{10} + \tilde{Y}_{10} v_d^{10^*} +  Y_{126} v_u^{126} \,,\nonumber\\
M_d    &= Y_{10} v_d^{10} + \tilde{Y}_{10} v_u^{10^*} +  Y_{126} v_d^{126} \,,\nonumber\\
M_e &= Y_{10} v_d^{10} + \tilde{Y}_{10} v_u^{10^*} -3 Y_{126} v_d^{126} \,, \\
M_D    &= Y_{10} v_u^{10} + \tilde{Y}_{10} v_d^{10^*} -3 Y_{126} v_u^{126} \,,\nonumber\\
M_R    &= Y_{126} v_R \,, \nonumber \\
M_L    &= Y_{126} v_L \,. \nonumber
\end{align}
Here, $M_D$, $M_R$ and $M_L$ enter the neutrino mass matrix defined on the symmetric basis $(\nu, n)$\footnote{In the notation of table \ref{tab:fermions}, 
$\nu$ denotes the left-handed neutrinos included in the lepton doublets $l$, and $n$ designates the right-handed neutrinos.}, 
\begin{equation}
\label{seesawmatrix}
\left(
\begin{array}{cc}
M_L & M_D \\
M_D^T & M_R
\end{array}
\right) \, .
\end{equation} 

The three different Yukawa coupling matrices in 
\eqref{nonsusysumrule} weaken the predictive power of the model. 
This motivated the authors of Ref.~\cite{Bajc:2005zf} to 
impose a PQ symmetry~\cite{Peccei:1977hh}, under which the fields transform as 
\begin{align}
  16_F&\rightarrow16_F e^{i\alpha}, \nonumber\\
\label{pgsymmetry_yukawas}
  10_H&\rightarrow10_H e^{-2i\alpha},\\
  \overline{126}_H&\rightarrow\overline{126}_H e^{-2i\alpha}\,,
\nonumber
\end{align} 
which forbids the coupling 
$\tilde{Y}_{10}$ in \eqref{nonsusysumrule} (see also Ref.~\cite{Babu:1992ia}).

As mentioned above, the $126_H$ alone breaks $SO(10)$ to the experimentally disfavoured $SU(5)$ --or else it would also break the SM group-- 
so that we have to introduce a third Higgs representation to achieve 
a symmetry breaking pattern that arrives at the SM  gauge group at a scale above that of electroweak symmetry breaking. 
We exploit in this paper the $210_H$ representation, which has the following PS decomposition:
\begin{equation}
210_H = (1,1,1)\oplus (15,1,3)\oplus (15,1,1) \oplus (15,3,1) \oplus (\overline{10},2,2) \oplus (10,2,2) \oplus (6,2,2)
\,.
\end{equation}
The former allows for a VEV that preserves the SM gauge group,
\begin{equation}
v^{210} \equiv \vev{(1,1,1)_{210}}‚ \,  .
\end{equation}
We will further assume  $v_L=0$ (see equation \eqref{eq:VEVS_10_126}), which implies $M_L=0$ in the mass matrices in equations 
\eqref{nonsusysumrule} and \eqref{seesawmatrix}, thus giving a type-I seesaw, and yielding the following two-step breaking chain:
\begin{eqnarray} 
\label{chaintwostep}
SO(10)&\stackrel{v^{210}-210_H}{\longrightarrow}
&4_{C}\, 2_{L}\, 2_{R}\ \stackrel{v_R -\overline{126}_H}{\longrightarrow}3_{C}\, 2_{L}\, 1_{Y}\ \stackrel{v_{u,d}^{10,126}-10_H}{\longrightarrow} \ 3_{C}\,1_{\rm em}\,.
\end{eqnarray}
The symmetry breaking  VEVs are constrained by the requirement of gauge coupling unification 
and can be calculated from the renormalisation group running of the coupling constants, see Section \ref{gaugecouplingunification}. $v_R$ and $v^{210}$ 
are further constrained by proton decay and lepton-number violation bounds, but the former still allow for
excellent fits to the fermion masses and mixings, as was seen in \cite{Joshipura:2011nn,Altarelli:2013aqa,Dueck:2013gca} and references therein. For a recent
analysis of unification with intermediate left-right groups, see  \cite{Chakrabortty:2017mgi}.

%% file: axion_generalities.tex
\section{Axion generalities}
\label{axiongeneralities}
 
 As argued before, within the $SO(10)$ framework one can use predictivity to motivate a global PQ symmetry under which the fermions are charged, see \eqref{pgsymmetry_yukawas}.
 The chiral fermionic content of the theory ensures that the symmetry is anomalous under the GUT group, and by extension under the subgroups that survive at low energies, such as $SU(3)_C$.
 This allows to embed the axion solution of the strong CP problem in the GUT theory, as such solution
 requires a spontaneously broken global $U(1)$ symmetry with an $SU(3)_C$ anomaly. Moreover, the resulting axion excitation can play the role of dark matter. 
 
 In this section we will review generalities of axion fields in models with multiple scalar fields. We will first introduce the strong CP problem and its axionic solution, followed by a review on how the  axion excitation is identified in terms of the VEVs and PQ charges of the fields, and how its effective Lagrangian is determined. Then we will elaborate on the orthogonality conditions of the physical axion --which imply that the global $U(1)$ symmetry of the axion is not simply given by the PQ symmetry in \eqref{pgsymmetry_yukawas} -- and on the axion domain-wall number.

 \subsection{The guts of the strong CP problem}
 \label{strongCP}
 
 Gauge theories with field strength $F_{\mu\nu}=F_{\mu\nu}^a T^a=\partial_\mu A_\nu-\partial_\nu A_\mu-ig[A_\mu,A_\nu]$ admit renormalisable, CP-violating interactions of the form
 \begin{align}
 \label{eq:Ltheta}
  {\cal L}_\theta={\rm \bar{T}r}\,\frac{g^2\theta}{16\pi^2} \tilde F_{\mu\nu} F^{\mu\nu} ={\rm \bar{T}r}\,\frac{g^2\theta}{32\pi^2}
  \epsilon^{\mu\nu\rho\sigma} F_{\mu\nu} F_{\rho\sigma},
 \end{align}
 where $\epsilon^{\mu\nu\rho\sigma}$ is the Levi-Civita antisymmetric tensor with $\epsilon^{1234}=1$, and ${\rm \bar{T}r}$ denotes a normalised trace over an arbitrary representation $\rho$ of the Lie 
 algebra. Denoting  the Dynkin index of $\rho$  as $S(\rho)$ --defined from the identity
$\text{Tr}_\rho \rho(T^m)\rho( T^n)=S(\rho)\delta^{mn}$-- one has
 \begin{align}
 \label{eq:barTr}
  {\rm \bar{T}r}\equiv \frac{1}{2S(\rho)} {\rm Tr}_\rho,
 \end{align}
 which implies ${\rm \bar{T}r}\, T^a T^b=1/2\,\delta^{ab}$.

 The theta term ${\cal L}_\theta$ can be seen to be total derivative, so that its contribution to the action only picks a surface term and becomes topological. In fact this contribution
 is proportional to the integer topological charge $n_{\rm top}$ of a given gauge field configuration\footnote{In Euclidean space, $n_{\rm top}$ can be interpreted
 as a Pontryagin index, while in Minkowski space in the so called ``topological gauge'', 
 with $A_0=0$ and $A_i\rightarrow0$ for
 $|\vec{x}|\rightarrow\infty$, $n_{\rm top}$ becomes equal to the difference of the Chern-Simons numbers at $t=\infty$ and $t=-\infty$.}:
 \begin{align}
  {S_\theta}=\int d^4 x{\cal L}_\theta= \theta\, n_{\rm top}.
 \end{align}
 The coupling $\theta$ is known as the ``$\theta$ angle'', because physics is invariant under shifts $\theta\rightarrow\theta+2n\pi,n\in\mathbb{Z}$. This follows 
 simply from the fact that the partition function of the theory involves the functional integral  (after rotation to Euclidean time):
 \begin{align}
 \label{eq:Z}
  Z=\sum_{n_{\rm top}=-\infty}^\infty\int[d\varphi]\exp\left[-S_E[\varphi]+i n_{\rm top}\theta\right],
 \end{align}
 (where $\varphi$ is a shorthand for all the fields in the theory), which is invariant under the above shifts of $\theta$.

 Crucially, ${\cal L}_\theta$ is of the same form as the chiral anomaly. Indeed, let's assume that the gauge theory has Weyl fermions $\psi_a$ in 
 representations $T^a$ of a non-Abelian group.
Let's further assume that the fermions have mass-terms (which could be field-dependent) ${\cal L}\supset -1/2M^{ab}\psi_a\psi_b+c.c.\,$ . 
 Then under chiral transformations
\begin{align}
  \psi_a\rightarrow e^{i\alpha} \psi_a,
\end{align}
the associated chiral current ${J}^\mu$ is anomalous \cite{Adler:1969gk,Bell:1969ts,Bardeen:1969md},
\begin{align}
\label{eq:chirala}
 \partial_\mu J^\mu=i M^{ab}\psi_a \psi_b+c.c.+\frac{2g^2\sum_a  S(\rho_a)}{16\pi^2} \,{\rm \bar{T}r}\,\tilde F_{\mu\nu} F^{\mu\nu},
\end{align}
where $S(\rho_a)$ is the Dynkin index of the representation $\rho_a$ associated with the Weyl fermion $\psi_a$.

Under a chiral transformation the quantum effective action $\Gamma$ transforms as\footnote{This can be derived using path integral methods when accounting for the non-invariance of
the fermionic measure \cite{Fujikawa:1979ay}, or, within dimensional regularisation, when accounting for the non-anticommuting character of the regularised chirality
operator \cite{Breitenlohner:1977hr}.}
\begin{align}
\label{eq:anomaly0}
 \delta_\alpha \Gamma=-\alpha\int d^4x \partial_\mu J^\mu = -\int d^4x\,i\alpha 2M^{ab}\langle\psi_a \psi_b\rangle+c.c.-{\cal N}\alpha n_{\rm top},
\end{align}
with 
\begin{align}
\label{eq:calN}
{\cal N}\equiv2\sum_a  S(\rho_a).
\end{align}
\eqref{eq:chirala} is equivalent to a simultaneous transformation of $M^{ab}$ and $\theta$, $\theta\rightarrow \theta-{\cal N}\alpha$, $M_{ab}\rightarrow e^{2i\alpha} M_{ab}$.
This means that the CP-violating phase $\theta$ is unphysical, as it is not invariant under field redefinitions such as chiral rotations. In fact, if there is at least one massless fermion charged under the gauge group, one can always 
rotate  $\theta$ away without affecting the rest of the parameters by just rephasing the massless field. Similarly, if there is at least one mass term pairing a singlet fermion with a fermion charged under the gauge group (which requires a Yukawa interaction in order to preserve gauge invariance) $\theta$ is again unphysical: one may change $\theta$ without altering $M$ by compensating rephasings of the charged and singlet fermion,
the charged fermion's rephasing inducing a change of $\theta$ and $M$, and the transformation of the singlet leaving $\theta$ unaffected but driving back $M$ to its original form. 
However, if there are no massless charged fermions and the mass terms only couple charged fermions among themselves (so that a change
in ${\rm Arg}\,\det M$ necessarily implies a change in $\theta$), then the combination
\begin{align}
\label{eq:thetaphys}
 \theta_{\rm phys}=\theta+\frac{{\cal N}}{2N}{\rm Arg}\,\det M,
\end{align}
where $N$ is the number of Weyl fermions in nontrivial representations, and $\cal N$ is given in \eqref{eq:calN},
is a chiral invariant that will show-up in CP-violating observables.

In the SM one could in principle have $\theta$ angles for all gauge groups. The $U(1)_Y$ angle $\theta_1$ cannot have any effect,
as there are no finite-action hypercharge field configurations with nonzero topological charge. Since the Weyl fermions charged under $SU(2)_L$ only get masses 
by coupling to the right-handed $SU(2)_L$ singlets, the arguments given before equation \eqref{eq:thetaphys} imply that the corresponding $\theta_2$ angle is unobservable.\footnote{The fact that $\theta_2$ can be driven to zero
without affecting fermion masses --but changing the unobservable $\theta_1$-- can be also understood in terms of B and L symmetries \cite{Perez:2014fja}, as they are also associated with 
opposite rephasings of left and right-handed Weyl spinors.} However, for the strong interactions there is a physical $\theta$ angle, 
which would contribute to  flavor conserving CP-violating observables such as the neutron dipole moment. Current experimental bounds, however, imply $|\theta_{\rm phys}|<10^{-10}$ \cite{Baker:2006ts}, from which the strong CP problem
follows: why is $\theta_{\rm phys}$ so small?

In a GUT completion of the SM, the $SU(3)_C$ $\theta$ term can only come from a GUT $\theta$ term as in equation \eqref{eq:Ltheta}, modulo rotations of fermion fields.
The SM $\theta^{\rm SM}$ term can be related to the GUT $\theta^{\rm GUT}$ by matching physical invariants in the GUT and its low-energy effective description (SM), so that
\begin{align}
\label{eq:thetasSMGUT}
 \theta_{\rm phys}=\theta^{\rm SM}_{3}+\frac{{\cal N}^{\rm SM}_3}{2N_3}{\rm Arg}\,\det M^{\rm SM}_3 =  \theta^{\rm GUT}+\frac{{\cal N}^{\rm GUT}}{2N^{\rm GUT}}{\rm Arg}\,\det M^{\rm GUT},
\end{align}
where ${\cal N}^{\rm SM}_3,N_3,M^{\rm SM}_3$ refer to the SM Weyl fermions charged under $SU(3)_C$.
Since ${\rm Arg}\,\det M$ is a sum of phases over the different eigenvalues, it is clear that $\theta^{\rm SM}$ is equal to $\theta^{\rm GUT}$ plus a combination of phases associated with the
heavy fermion states. In any case, the strong CP problem has again a reflection in the GUT theory, as again the physical invariant combination to the right of \eqref{eq:thetasSMGUT} 
appears to be tuned to a small value. Solving the CP problem at the level of the GUT by explaining the smallness of $\theta_{\rm phys}$ 
 guarantees a solution of the strong CP problem, given that $\theta_{\rm phys}$ has to match in both the UV and IR descriptions.

\subsection{The axion solution}

The axion solution relies on replacing the chiral invariant $\theta_{\rm phys}$ with a combination involving a dynamical field. Once this is achieved, the CP-problem is solved if
it can be shown that the now dynamical $\theta_{\rm phys}(x)$ gets a potential with a minimum at $\theta_{\rm phys}(x)=0$. Given \eqref{eq:thetaphys}, having a dynamical $\theta_{\rm phys}$ is  suggestive
of a field-dependent ${\rm Arg}\,{\det M}$, which can be achieved if the mass-matrices depend on some complex scalar fields with dynamical phases. 
With part of the mass-matrices becoming field-dependent, one can combine chiral rotations of the fermions with a rephasing of the complex scalars to define 
a classical $U(1)$ symmetry of the theory, which, being chiral, becomes anomalous at the quantum level. In order to avoid massless fermions and have a well-defined ${\rm Arg}\,{\det M}$, the 
complex scalars must get VEVs. Thus, as anticipated, the axion solution involves an anomalous chiral $U(1)$ symmetry
acting on fermions coupled to scalars --the PQ symmetry-- which must be spontaneously
broken. Then $\theta_{\rm phys}$ will involve a combination of phases of the scalars which contains the pseudo-Goldstone excitation
$A(x)$ corresponding to 
the spontaneously broken $U(1)$ symmetry, i.e. the axion.  Indeed, at low energies, when the massive excitations of the phases are decoupled, one can recover --as will be shown next-- 
an effective Lagrangian involving the axion 
 in which $\theta$ always enters in a combination  
 \begin{align}
 \label{eq:bartheta}
 \bar \theta\equiv
 \theta+\hat {N} A/f_{\rm PQ},
 \end{align}
 for some dimensionful scale $f_{\rm PQ}$ and constant $\hat {N}$, so that 
 $\theta_{\rm phys}=\theta+{\cal N}/(2N){\rm Arg}\,{\det M'}+\hat{N} A/f_{\rm PQ}$, where $\det M'$ designates the determinant restricted to the fermions which are not 
 charged under the global $U(1)$ symmetry. 
 
The last ingredient of the solution to the strong CP problem is the fact that nonperturbative QCD effects generate a nonzero 
 mass for $\theta_{\rm phys}$. As said before, at low energies one predicts that the QCD $\theta$
 term will involve the effective combination $\bar\theta=\theta+\hat{N} A/f_{\rm PQ}$, which coincides with $\theta_{\rm phys}$ if the quark masses are chosen to be real.
 Since the partition function of a theory is related to the vacuum energy density, one can obtain an effective potential for $\bar\theta$ 
 from the Euclidean partition function of QCD with real masses, supplemented with the topological term ${\cal L}_{\bar\theta}$ (see \eqref{eq:Ltheta}): 
 \begin{align}
  Z_{\rm QCD}[\bar\theta]=\exp\left[-V V_{\rm eff}(\bar\theta)\right],
 \end{align}
 where $V$ designates the four-dimensional Euclidean volume. The effective potential can be shown to have a minimum at $\bar\theta=0$, because for vector-like fermions
 $Z$ can be written as a path integral involving positive functions  times a $\bar\theta$-dependent phase factor \cite{Vafa:1984xg}. Then the axion  solves indeed the 
 CP problem, and the axion mass is given by
 \begin{align}
 \label{eq:m2A}
  m^2_A=\frac{\hat{N}^2}{f_{\rm PQ}^2}\frac{d^2V_{\rm eff}(\bar\theta)}{d\bar\theta^2}\Big|_{\bar\theta=0}\equiv \frac{1}{f_{A}^2}\frac{d^2V_{\rm eff}(\bar\theta)}{d\bar\theta^2}\Big|_{\bar\theta=0}\equiv \frac{\chi}{f^2_A}.
 \end{align}
 Above, we defined the axion decay constant $f_A$ as
 \begin{align}
 \label{eq:fPQ}
  f_A\equiv \frac{f_{\rm PQ}}{\hat{N}},
 \end{align}
 with $\hat{N}$ appearing in the combination $\overline{\theta}$ in equation \eqref{eq:bartheta} that enters the axion effective Lagrangian. In \eqref{eq:m2A},
 $\chi=d^2V_{\rm eff}(\bar\theta)/d\bar\theta^2|_{\bar\theta=0}$ is the topological susceptibility -- the variance of the topological charge distribution divided by the four-dimensional Euclidean volume, $\chi = \langle n_{\rm top}^2\rangle/V$. 
 It can be calculated from chiral perturbation theory or with lattice techniques, with recent agreement \cite{diCortona:2015ldu,Borsanyi:2016ksw}. Within the error of the NLO calculation 
 of \cite{diCortona:2015ldu}, one has
\begin{equation}
\label{zeroTma}
m_A= 
{57.0(7)\,   \left(\frac{10^{11}\rm GeV}{f_A}\right)\mu \text{eV}. }
\end{equation}

\subsection{Constructing the axion and its effective Lagrangian}
\label{subsec:effL}
Next we may explicitly construct the axion field and its interactions in a theory with Weyl fermions $\psi_a$ and complex scalars $\phi_i$, and $N_g$ gauge groups. We assume a 
PQ symmetry under which the fermions and scalars have charges $q_a$ and $q_i$, respectively, and which is broken spontaneously by VEVs $\langle\psi_i\rangle= v_i/\sqrt{2}$. 
In the broken phase, we may
parameterise scalar excitations as
\begin{align}
\label{eq:scalar_param}
 \phi_j=\frac{1}{\sqrt{2}}(v_j+\rho_j)e^{iA_j/v_j}.
\end{align}
The spontaneous breaking of the global PQ symmetry implies the existence of a Goldstone state $A$ \cite{Weinberg:1996kr}, the axion, which corresponds to the following excitation of the phases:
\begin{align}
\label{eq:AiA}
 A_i=\frac{q_i v_i}{f_{\rm PQ}}A+{\text{ orthogonal excitations}},
\end{align}
where $f_{\rm PQ}$ is a dimensionful scale.
Canonical normalisation of $A$ --whose kinetic term follows from applying \eqref{eq:AiA} to the sum of kinetic terms of the complex scalars-- implies
\begin{align}
\label{eq:fA}
  f_{\rm PQ}=\sqrt{\sum_j q^2_j v^2_j}.
\end{align}
From \eqref{eq:AiA} one may then derive 
\begin{align}
\label{eq:A0}
 A=\frac{1}{f_{\rm PQ}}\sum_i q_i v_i {A_i}.
\end{align}
The effective Lagrangian for the axion can be obtained from the anomalous conservation of the PQ current \cite{Srednicki:1985xd}. The latter is given by
\begin{align}
\label{eq:J}
J^\mu=\sum_a q_a \psi_a^\dagger \bar\sigma^\mu \psi_a+i\sum_j q_j (\partial_\mu\phi^\dagger_j \phi_j-\phi^\dagger_j \partial_\mu\phi_j),
\end{align}
and satisfies the anomaly equation
\begin{align}
\label{eq:anomaly}
 \partial_\mu J^\mu = \sum_{k=1}^{N_g}\frac{g_k^2\hat{N}_k}{16\pi^2} \,{\rm \bar{T}r}\,\tilde F^k_{\mu\nu} F^{k,\mu\nu},\quad \hat{N}_k=2\sum_a q_a T_k(\rho_a).
\end{align}
Using \eqref{eq:AiA} in \eqref{eq:J}, one has that at low-energies --when the heavier excitations of the $A_i$ are decoupled and can be ignored on the r.h.s of \eqref{eq:AiA}-- the
anomaly equation \eqref{eq:anomaly} becomes
\begin{align}
\label{eq:current}f_{\rm PQ}\square A+\sum_a q_a \partial_\mu(\psi_a^\dagger \bar\sigma^\mu \psi_a)=\sum_{k=1}^{N_g}\frac{g^2\hat{N}_k}{16\pi^2}\,{\rm \bar{T}r}\, \tilde F^k_{\mu\nu} F^{k,\mu\nu}.
\end{align}
The latter is equivalent to the Euler-Lagrange equations of the following effective interaction Lagrangian ${\cal L}[A]_{\rm eff}$ \cite{Srednicki:1985xd},
\begin{align}
\label{eq:Lint}
 {\cal L}[A]_{\rm eff}=\frac{1}{2}\partial_\mu A\partial^\mu A+\partial_\mu A\sum_a \frac{q_a}{f_{\rm PQ}} (\psi_a^\dagger \bar\sigma^\mu \psi_a)+A\sum_{k=1}^{N_g}\frac{1}{f_{A,k}}\frac{g_k^2}{16\pi^2} \,{\rm \bar{T}r}\,\tilde F^k_{\mu\nu} F^{k,\mu\nu},
\end{align}
where 
\begin{align}
\label{eq:fAk}
 f_{A,k}=\frac{\fPQ}{\hat{N}_k},
\end{align}
with $f_{\rm PQ}$ and $\hat{N}_k$ determined from the VEVs and PQ charges
as in equations \eqref{eq:fA} and \eqref{eq:anomaly}. As anticipated earlier, the $\theta_k$ parameter for a given group (see  \eqref{eq:Ltheta}) and the axion enter the low-energy Lagrangian in the combination $\theta_k+1/f_{A,k} A$.

The above effective interactions can also be recovered by rotating the scalar phases away from the Yukawa couplings \cite{Dias:2014osa}. The PQ-invariant Yukawa couplings induce contributions to the Lagrangian of the form
\begin{align}
\label{eq:fermionsA}
 {\cal L}\supset y^i_{ab}\phi_i \psi_a\psi_b+c.c.\supset\frac{y^i_{ab}v_i}{\sqrt{2}}e^{iq_i A/f_{\rm PQ}} \psi_a\psi_b+c.c.=\frac{y^i_{ab}v_i}{\sqrt{2}}e^{-i(q_a+q_b) A/f_{\rm PQ}} \psi_a\psi_b+c.c.,
\end{align}
where we used \eqref{eq:AiA} and the fact that the PQ invariance of the above Yukawa coupling demands $q_i+q_a+q_b=0$; for simplicity, we suppressed the internal indices for the different representations
of the gauge groups, and assumed the appropriate gauge-invariant contractions. The phase factors in \eqref{eq:fermionsA} can be removed by field-dependent 
chiral rotations
of the fermions,
\begin{align}
\label{eq:rotation}
 \psi_a\rightarrow e^{iq_a A/f_{\rm PQ}}\psi_a.
\end{align}
After the previous rotations, the  kinetic terms of the fermions pick extra contributions, given --up to a minus sign-- by the axion-fermion interactions in \eqref{eq:Lint}. Accounting for the fact that 
the  effective action picks up an anomalous term after chiral rotations, one gets the axion-gauge boson interactions in \eqref{eq:Lint}, again up to a minus sign. The difference in sign in the terms
linear in the axion can be removed with a physically irrelevant redefinition $A\rightarrow -A$, so that the results are equivalent. 

It should be noted that the fermion rotations in \eqref{eq:rotation} are  not the only possibility to eliminate the axion dependence in the Yukawas as in \eqref{eq:fermionsA}. One may make different choices of fermionic rephasings that will give rise to different effective actions. However, as these just differ by redefinitions of the phases of the fermion fields, they will be physically equivalent. In this respect, one may wonder whether these alternative Lagrangians give different bosonic interactions than the ones following from equations \eqref{eq:Lint}, \eqref{eq:fAk}, \eqref{eq:fA} and \eqref{eq:anomaly}; in particular, if the resulting value of $f_{A,3_C}$ for the $SU(3)_C$ gauge group were to be sensitive to the chosen rephasings, one would predict different axion masses (see \eqref{zeroTma}), in contradiction with the expectation that physical quantities should  remain invariant under field redefinitions. Fortunately this is not the case, as is explicitly shown in Appendix \ref{app:ferphases}. An important consequence from this is that $f_{A,3_C}$ is fixed by the scalar PQ charges, as is clear from the fact that one can always remove the phases in \eqref{eq:fermionsA} by rotating a single fermion per Yukawa interaction, with a phase fixed by the phase of the scalar field entering the Yukawa coupling. An explicit formula for $f_{A,3_C}$  in terms of the scalar PQ charges is given in equation \eqref{eq:fA3} of Appendix \ref{app:ferphases}.

All the $SO(10)$ models considered here have the same couplings to neutral gauge bosons, aside from variations in the scale $f_A\equiv f_{A,3_C}$. This is not surprising, as the GUT symmetry relates the 
different SM gauge groups \cite{Srednicki:1985xd}. Starting from the anomalous Ward identity of  ${\rm PQ}_{\rm phys}$ in the full GUT theory, it follows that the effective Lagrangian of the axion must 
contain axion-gauge boson interactions as in \eqref{eq:Lint}, but with a single gauge group $SO(10)$:
\begin{align}
\label{eq:axionLagGUT}
 {\cal L}\supset \frac{ A}{f_{A,
 \rm GUT}}\frac{g_{\rm GUT}^2}{16\pi^2} \,{\rm \bar{T}r}^{\rm GUT}\,\tilde F^{\rm GUT}_{\mu\nu} F^{{\rm GUT}\,\mu\nu}.
\end{align}
Then the effective Lagrangian in terms of the SM gauge bosons can be recovered from \eqref{eq:axionLagGUT} by selecting the contributions of the SM fields. Consider the definition of ${\rm \overline{T}r}$ in equation \eqref{eq:barTr} in terms of an arbitrary representation $R$ (the division by $S(R)$ ensures that the trace gives a representation-independent result). Without loss of generality, we may pick the $10$ of $SO(10)$, which decomposes under $SU(3)_\times SU(2)_L\times U(1)_Y$ as in table \ref{tab:fermions}:
\begin{align}
\label{eq:10dec}
 10=\left(3,1,-\frac{1}{3}\right)_{\rm SM}+ \left(\bar 3,1,\frac{1}{3}\right)_{\rm SM}+\left(1,2,\frac{1}{2}\right)_{\rm SM}+ \left(1,2,-\frac{1}{2}\right)_{\rm SM}.
\end{align}

One has $S(10)=1$, and using the orthogonality of the GUT generators belonging to different subgroups, we may write
\begin{equation}\begin{aligned}
{\rm \bar{T}r}^{\rm GUT}\,\tilde F^{\rm GUT}_{\mu\nu} F^{{\rm GUT},\mu\nu}=&\,\frac{1}{2}F^{{\rm GUT},a}_{\mu\nu} F^{{\rm GUT}\,b\mu\nu}\,{\rm Tr}^{{\rm GUT}}_{10 }\,T^a T^b\\
&\supset \frac{1}{2}\tilde F^{ 3_C,a}_{\mu\nu} F^{3_C\,b\mu\nu}\,{\rm Tr}^{{\rm GUT}}_{10} \,T_{3_C}^a T_{3_C}^b+\frac{1}{2} \tilde F^{ 2_L,a}_{\mu\nu} F^{2_L\,b\mu\nu}\,{\rm Tr}^{{\rm GUT}}_{10}\,T_{2_L}^a T_{2_L}^b\\
&+\frac{1}{2} \tilde F^{ Y}_{\mu\nu} F^{Y\,\mu\nu}\,{\rm Tr}^{{\rm GUT}}_{10}\, T_{Y} T_{Y}.
\end{aligned}\end{equation}
Using the decomposition in \eqref{eq:10dec}, and using  ${\rm Tr}\, T^a T^b=1/2 \delta^{ab}$ in the (anti)fundamental representations of $SU(N)$, one gets
\begin{equation}\label{eq:LintGUT}\begin{aligned}
 &\frac{ A}{f_{A,
 \rm GUT}}\frac{g_{\rm GUT}^2}{16\pi^2} \,{\rm \bar{T}r}^{{\rm GUT}}\,\tilde F^{{\rm GUT}}_{\mu\nu} F^{{\rm GUT},\mu\nu}=\frac{A}{f_{A,\rm GUT}}\frac{\alpha_{\rm GUT} }{8\pi }\Big(\tilde F^{{3_C}\,a}_{\mu\nu} F^{{3_C} a\,\mu\nu}+\tilde F^{{2_L}\,a}_{\mu\nu} F^{{2_L} a\,\mu\nu}\\
 &\left.+\frac{5}{3}\tilde F^{{Y}}_{\mu\nu} F^{{Y}\,\mu\nu}\right)\supset\frac{A}{f_{A,\rm GUT}}\frac{\alpha_{\rm GUT} }{8\pi } \tilde G^{a\,\mu\nu}G^a_{\mu\nu}+\frac{A}{f_{A,\rm GUT}}\frac{\alpha_{\rm GUT}\sin^2\theta_W }{8\pi }\frac{8}{3}\tilde F^{\mu\nu}F_{\mu\nu}.
\end{aligned}\end{equation}
At low energies $\alpha_{\rm GUT}$ renormalises differently in each term, and identifying $\alpha_{\rm GUT}\rightarrow\alpha_s$ in the QCD term, and $\alpha_{\rm GUT}\sin^2\theta_W\rightarrow\alpha$ in the
electromagnetic term, one recovers the result in \eqref{eq:axionLaggen} --derived with the method of fermion rephasings-- but without any assumption on the matter representations. The ratio of the couplings of the axion to gauge bosons is thus predicted to be the same for any $SO(10)$ theory regardless of the matter content \cite{Srednicki:1985xd}. Moreover, the $f_A$ scale of the QCD interactions corresponds to that in the grand unified theory, $f_{A,{\rm GUT}}$.

\subsubsection{Axial basis}
\label{subsec:axial_basis}

It is customary to write the axion-SM fermion couplings in terms of chiral currents of the massive SM fermions:
\begin{align}
\label{eq:axial}
 {\cal L}[A]_{\rm eff}\supset \sum_f c_f\frac{\partial_\mu A}{f_A}\,\overline{\Psi}_f\gamma^\mu\gamma_5\Psi_f,
\end{align}
where $\Psi=\{\psi_\alpha,\tilde\psi^{\dagger,\dot{\alpha}}\}$ are Dirac fermions constructed from the Weyl spinors paired by mass terms.\footnote{We use a notation in which conjugates
of Lorentz spinors are denoted with dotted indices, $(\psi_\alpha)^\dagger=\psi^{\dagger}_{\dot{\alpha}}$,  and indices are lowered and raised with antisymmetric matrices $\epsilon^{\alpha\beta},\epsilon^{\dot{\alpha}\dot{\beta}}$, e.g. $\tilde\psi^{\dagger,\dot{\alpha}}=\epsilon^{\dot{\alpha}\dot{\beta}}\tilde\psi^\dagger_\beta$, with $\epsilon^{12}=\epsilon_{12}=1$. The chirality operator $\gamma_5$ is defined in such a way that a Dirac spinor $\Psi=\{\psi_\alpha,0\}$ has a negative eigenvalue.} The axial basis is particularly useful when accounting for nonperturbative QCD effects in the axion-nucleon interactions, either because
one may use current algebra techniques \cite{Weinberg:1977ma,Srednicki:1985xd}, or because
the matching between the UV and nucleon theory is simplified when using axial currents \cite{diCortona:2015ldu}. Moreover, as will be seen, the coefficients of the fermion-axion
interactions in the axial basis depend only on the scalar PQ charges.
One has $\overline{\Psi}\gamma^\mu\gamma_5\Psi=-\psi^\dagger\overline{\sigma}^\mu\psi-\tilde\psi^\dagger\overline{\sigma}^\mu\tilde\psi$, from
which it follows that the axion-fermion interactions in the general formula \eqref{eq:Lint} can be recasted in terms of chiral currents as in \eqref{eq:axial} if the Weyl fermions connected by mass terms have 
equal PQ charges. This won't be the case in the GUT models considered here, for which the global symmetry associated with the physical axion enforces different charges for the  fermions interacting through  Yukawas. However, one can always redefine the fermion fields with axion-dependent phases
in such a way that one recovers interactions of the form in \eqref{eq:axial}, without affecting the axion coupling to neutral gauge bosons or the Yukawa couplings.
Consider for example two SM Weyl spinors $\psi$, $\tilde\psi$ with PQ charges $q_1,q_2$, and which can be grouped into a massive Dirac fermion after electroweak symmetry breaking --e.g. $\{u_L, u\}$, where $u_L$ is the upper component of a $q$ doublet, in the notation of table \ref{tab:fermions}. One can always redefine
\begin{align}
\psi\rightarrow e^{i\Delta q A/(2f_{PQ})}\psi,\quad \tilde\psi\rightarrow e^{-i\Delta q A/(2f_{PQ})}\tilde\psi, \quad\Delta q=q_1-q_2.
\end{align}
Under such redefinition with opposite phases, the axion couplings to neutral gauge bosons remain invariant, as the redefinition is a non-anomalous vector transformation, rather than a chiral one. On the other hand, the axion couplings to $\psi,\tilde\psi$ change to
\begin{align}
\label{eq:axial_rephasings}
 \partial_\mu A\left[ \frac{q_1}{f_{\rm PQ}} \psi^\dagger \bar\sigma^\mu \psi+ \frac{q_2}{f_{\rm PQ}} \tilde\psi^\dagger \bar\sigma^\mu \tilde\psi\right]\rightarrow
  -\partial_\mu A\,\frac{q_1+q_2}{2f_{\rm PQ}}\,\overline{\Psi}\gamma^\mu\gamma_5\Psi. 
\end{align}
The combination of charges $q_1+q_2$ above can be related to the PQ charge of the Higgs that gives a mass to the SM fermion in question, because the Yukawas have to be invariant under the PQ symmetry. From \eqref{nonsusysumrule} it is clear that in the presence of a PQ symmetry (enforcing $\tilde Y=0$ in \eqref{nonsusysumrule}) the up quarks receive their masses from the scalars $H_u,\Sigma_u$, and the down quarks and charged leptons from $H_d,\Sigma_d$. Then in the axial basis the axion interaction with  $u,d$ quarks and the electron can be expressed in terms of axial currents involving the corresponding Dirac fields $U,D,E$ as
\begin{align}
\label{eq:axial2}
  {\cal L}[A]_{\rm eff}\supset  \partial_\mu A\,\frac{q_{H_u}}{2f_{\rm PQ}}\,\overline{U}\gamma^\mu\gamma_5 U+ \partial_\mu A\,\frac{q_{H_d}}{2f_{\rm PQ}}\,\overline{D}\gamma^\mu\gamma_5 D+ \partial_\mu A\,\frac{q_{H_d}}{2f_{\rm PQ}}\,\overline{E}\gamma^\mu\gamma_5 E,
\end{align}
where $q_{H_u}$ and $q_{H_d}$ are the PQ charges of the Higgses. In regards to the neutrinos, in the models considered here the Weyl spinors $\nu\supset l$ and and the SM singlet $n$ (see table \ref{tab:fermions}) have nontrivial charges under the physical PQ symmetry. For a high seesaw scale $v_R$ (see equations \eqref{nonsusysumrule} and \eqref{seesawmatrix}), the light physical states in the neutrino sector will be mostly aligned with the $\nu_i$. One may always
do an axion-dependent phase rotation such that the $\nu_i$ end up carrying no PQ charge, and which again does not affect the axion couplings to neutral bosons because the neutrinos
are singlets under the strong and electromagnetic groups. The physical light neutrinos can then be described with  Majorana spinors constructed from the $\nu_i$ and which do not couple to the axion, in contrast to the other fermion fields in \eqref{eq:axial2}. 

The previous arguments seem to imply that the PQ charges of the fermions might be unobservable, since one can choose a basis of fields in which the axion coupling to charged fermions  only depends on the scalar PQ charges, and in which the light neutrinos do not couple to the axion. However, an important subtlety is that the axion-dependent rephasings needed to go to this special basis of fermion fields, although they do not affect the 
couplings of the axion to the neutral gauge bosons, remain anomalous under $SU(2)_L$ and thus alter the couplings of the axion to the weak gauge bosons. Thus the latter retain the lost information of the fermionic PQ charges, and if the corresponding couplings were measurable one could potentially distinguish theories differing in the fermionic PQ charges, as for example a GUT model in which the fermionic PQ charges are related to the global symmetries of the GUT theory, versus a DFSZ model in which the fermion charges are not constrained by the GUT symmetry.

\subsubsection{Mixing with mesons and the axion-photon coupling}

Although the effective Lagrangian in \eqref{eq:Lint} includes couplings of the axion to the photon, such interaction is further modified by QCD effects. The reason is essentially that
QCD induces a mixing between the axion and the neutral mesons, which in turn couple to photons through the chiral anomaly, involving the same $\tilde F F$ interaction that appears in the axion-photon coupling. 
The QCD-induced shift of the axion to photons can be computed with current algebra techniques \cite{Weinberg:1977ma,Srednicki:1985xd} or in chiral perturbation theory, with next-to-leading order results provided in \cite{diCortona:2015ldu}. At lowest order, the modification of the coupling can be recovered by noting that the mixing between the axion and neutral mesons can be removed with an appropriate axion-dependent rotation of the meson fields, which however induces an anomalous shift of the action which
is precisely the QCD-induced axion-to-photon coupling. This shift is universal and does not depend on the 
PQ charges of the quarks, and is given by
\begin{align}
\label{eq:photon}
 \delta{\cal L}_{\rm eff}=\frac{\alpha}{8\pi f_A}\,\delta C_{A\gamma}A\tilde F_{\mu\nu}F^{\mu\nu},\,\delta C_{A\gamma}=-\frac{2}{3}\left(\frac{4m_u+m_d}{m_u+m_d}\right)+\text{higher order}=-1.92(4).
\end{align}

\subsubsection{Axion-nucleon interactions}

In regards to axion-nucleon interactions, they can also be obtained by current algebra methods \cite{Donnelly:1978ty,Srednicki:1985xd}, or alternatively using a nonrelativistic effective theory for nucleons, 
with couplings determined from lattice data \cite{diCortona:2015ldu}.  The axion-nucleon interactions are not universal, and are given in \cite{diCortona:2015ldu} in terms of the coefficients of the 
UV axion-fermion effective Lagrangian in the axial basis, i.e. with fermion interactions as in \eqref{eq:axial}, \eqref{eq:axial2}, with coefficients fixed by the scalar PQ charges. Equation  \eqref{eq:axial2} shows that the UV coefficients are simply determined by the scalar charges of the Higgses $H_u,H_d$. Then the results in \cite{diCortona:2015ldu} imply the following axion-nucleon interactions in the chiral basis:
\begin{align}
 \label{eq:nucleon}
 \delta{\cal L}_{\rm eff}= -\partial_\mu A\,\frac{C_{AN}}{2f_{A}}\,\overline{N}\gamma^\mu\gamma_5 N- \partial_\mu A\,\frac{C_{AP}}{2f_{A}}\,\overline{P}\gamma^\mu\gamma_5 P,
\end{align}
with 
\begin{equation} \label{eq:nucleon2}\begin{aligned}
  C_{AN}=&\,-0.02(3)+0.41(2)\frac{q_{H_u}f_A}{f_{\rm PQ}}-0.83(3)\frac{q_{H_d}f_A}{f_{\rm PQ}},\\
  C_{AP}=&\,-0.47(3)-0.86(3)\frac{q_{H_u}f_A}{f_{\rm PQ}}+0.44(2)\frac{q_{H_d}f_A}{f_{\rm PQ}}.
\end{aligned}\end{equation}

\subsection{The physical axion: orthogonality conditions}
\label{subsec:axion_physical}

In the presence of both gauge and global symmetries, identifying the axion becomes a bit subtle, as the PQ symmetry is not uniquely defined. This is due to the fact that
the gauge symmetries themselves are associated with global symmetries, so that any combination of the PQ symmetry plus a global $U(1)$ symmetry associated with one of the gauge 
groups defines a new global $U(1)$ symmetry. 
This arbitrariness implies that one cannot readily identify the PQ charges $q_i$ that define the axion as in equation \eqref{eq:A0},
as well as determine the ensuing axion interactions and domain wall number, all of which depend on the $q_i$. Nevertheless, there is an important physical constraint that allows to uniquely single 
out a global PQ symmetry PQ$_{\rm phys}$: its associated axion must correspond to a physical, massless excitation, and thus it must remain orthogonal to the Goldstone bosons of the broken gauge symmetries. 
This allows to  identify the combination of phases that defines the axion, from which one can reconstruct the scalar charges of  PQ$_{\rm phys}$. 
This will be the procedure used in Section \ref{so10xu1pqmodels} when 
studying the properties of the axion in concrete $SO(10)$ models. 

First, the combination of phases should be massless. Suppose the Lagrangian generates a quadratic interaction for a combination of the phase fields,
\begin{align}
\label{eq:mAs}
 {\cal L}\supset m\left(\sum_m d_m A_m\right)^2 \,,
\end{align}
for some coefficients $d_m$. Then one can simply use \eqref{eq:AiA} and demand that the term becomes zero, which gives
\begin{align}
\label{eq:ds}
\sum_m d_m q_m v_m=0.
\end{align}
Writing the axion as 
\begin{equation}
\label{eq:axionsimple}
 A=\sum_i c_i A_i,
\end{equation}
then equation \eqref{eq:ds} is equivalent to 
\begin{align}
 \label{eq:orthomass}
\sum_m d_m c_m=0,
\end{align}
which can be interpreted as an orthogonality condition between the mass eigenstate $\sum_m d_m A_m$ (see \eqref{eq:mAs}) and the axion $\sum_m c_m A_m$.

Another physical constraint on the axion is the fact that it should not mix with the massive gauge bosons. When the complex scalars are charged under gauge groups, which are then broken by the VEVs, there are additional Goldstone bosons associated with the broken gauge generators. These Goldstones are related to the longitudinal polarisations of massive gauge bosons, and should be orthogonal to the axion. 
The interaction of the scalars
with a gauge field $A_\mu$ contains the following terms:
\begin{align}
 \int d^4x {\cal L}=&\int d^4x \sum_m D_\mu \phi_m^\dagger D^\mu \phi_m+\dots = \int d^4x \sum_{mn}(i A^a_\mu \phi^\dagger_m T^a_{mn}\partial^\mu\phi_n+c.c.)+\dots\\
 \nonumber &=\int d^4 x \,\partial_\mu A^{\mu,a}\left(\sum_{mn} v_m T^a_{mn}  A_n\right)+(\text{$A_m$ and $B_\mu$-independent}).
\end{align}
Using equation \eqref{eq:AiA},
the cancellation of the axion-gauge boson interaction requires the following constraint on the PQ charges:
\begin{align}
\label{eq:orthogauge0}
\sum_{mn} v_m T^a_{mn} q_n v_n=0.
\end{align}
For a $U(1)$ generator under which the scalars $\phi_i$ transform by simple rephasings with gauge charges $\tilde q_i$, this reduces to 
\begin{align}
\label{eq:orthogauge}
\sum_m\tilde q_m q_m  v_m^2=0.
\end{align}
(note that $\tilde q_m$ and $q_m$ represent the gauge and PQ charges, respectively). Again, the avoidance of axion-gauge boson mixing can be also interpreted as an orthogonality condition between the axion combination $A= \sum_i c_i A_i$ and the Goldstone $\tilde G\equiv \sum_i d_i A_i$ of the $U(1)$ gauge group. Indeed, the same reasoning behind equation \eqref{eq:A0} implies that $d_i=1/f_G\, \tilde q_i v_i$, so that \eqref{eq:orthogauge} is equivalent to the orthogonality relation $\sum_m c_m d_m=0$, or
\begin{align}
\label{eq:orthogauge2}
\sum_m c_m \tilde q_m  v_m=0.
\end{align}

Under some conditions satisfied in the theories studied in this paper, it will suffice to consider orthogonality of the axion with respect to the Goldstones associated with diagonal $U(1)$ generators in the Cartan subalgebra of the gauge group. This algebra, of dimension equal to the rank of the Lie group (5 for $SO(10)$) is spanned by the mutually commuting
generators $H_i,i=1,\dots5$ of the Lie algebra. The commutation property allows to choose representations of matter fields with well defined quantum numbers, or weights, under the elements of the Cartan algebra. In other words, the Cartan generators will be diagonal. Their associated $U(1)$ symmetries correspond to rephasings of the fields, and thus the orthogonality requirements for the axion 
along the Cartan generators are of the form of equation \eqref{eq:orthogauge}. In regards to the orthogonality conditions for the non-diagonal generators outside the Cartan subalgebra, they can be simplified in terms of the weights of the fields and the roots of the algebra (that is, the weights in the adjoint representation). As shown in Appendix \ref{app:roots}, orthogonality conditions for non-diagonal generators are satisfied if, within each $SO(10)$ multiplet, the field components that get nonzero VEVs have $SO(10)$ weights whose difference is not a root of the Lie algebra.
This is the case in the models considered in this paper, with the fields getting nonzero VEVs indicated in  table \ref{tab:vevs}; more details are given in Appendix \ref{app:roots}.

An important consequence following from the constraint of equation \eqref{eq:orthogauge} is that if a field $\phi_n$ is charged under a given diagonal generator, then if the axion decay constant
$f_A$ is to involve the VEV $\langle \phi_n\rangle=v_n/\sqrt{2} $, then there has to be at least another scalar charged under the same generator as $\phi_n$ \cite{Kim:1981cr}. This simply follows from the fact that, for a VEV $v_n$ to contribute to $f_A$, the associated PQ charge $q_n$ has to be nonzero (see equations \eqref{eq:fA} and \eqref{eq:fPQ}). Then in order to have a solution to \eqref{eq:orthogauge} with nonzero charge $\tilde q_n$ one needs at least another scalar charged under both PQ and the diagonal generator. This is a consequence of the fact that if
only one scalar charged under PQ and the diagonal gauge generator develops a VEV, a physical global unbroken symmetry survives the breaking. Other fields are needed in order to break this surviving symmetry and give rise to an axion.
Even when there are several scalars with nonzero VEVs and charged under both PQ and a diagonal generator, then if one expectation value is much larger than the rest, it follows that $f_A$ is bound
to be of the order of the smaller VEVs. The orthogonality condition \eqref{eq:orthogauge} implies that the PQ charge $q^{\rm h}$ of the heavy field goes as
\begin{align}
q^{\rm h}=-\sum_{m} \frac{\tilde q^{\rm l}_mq^{\rm l}_m {v^{\rm l}_m}^2}{\tilde q^{\rm h}{ v^{\rm h}}^2},
\end{align}
where the superscripts $h,l$ denote the heavy field and the light fields, respectively.
Plugging this into \eqref{eq:fPQ} and \eqref{eq:fA}, one gets
\begin{align}
 f^2_A=\frac{1}{\hat N^2}\left[\sum_{m} (q^l_m v^l_m)^2 +\left(\sum_{m}\frac{\tilde q^{\rm l}_mq^{\rm l}_m {v^{\rm l}_m}^2}{\tilde q^{\rm h} v^{\rm h}}\right)^2\right],
\end{align}
which shows explicitly that $f_A$ is determined by the light VEVs $v^l_m$. This can be interpreted in an effective theory framework as follows: as discussed before, the single large VEV $v^h$ leaves 
a global symmetry unbroken, so that the theory with the heavy field integrated out has a new PQ symmetry that can only be broken by the  VEVs of the light fields, which will determine the scale $f_A$.

Finally, once the axion has been identified by starting from a general linear combination as in \eqref{eq:axionsimple} and imposing the orthogonality and masslessness constraints, the effective Lagrangian
can be determined in terms of the coefficients $c_i$, which encode the charges of the physical PQ symmetry. Indeed, comparing \eqref{eq:axionsimple} with \eqref{eq:A0}, one has that
\begin{align}
\label{eq:csqs}
 \frac{q_i}{f_{\rm PQ}}=\frac{c_i}{v_i}.
\end{align}
The charges $q_i$ correspond to the physical global symmetry PQ$_{\rm phys}$ connected to the axion. This symmetry must be a combination of the original global symmetries in the Lagrangian, and
one may find the corresponding coefficients by solving a system of linear equations. Since we expect ${\rm PQ}_{\rm phys}$ to act as a rephasing of fields, and since all fields have well-defined quantum numbers (weights) under the generators of the Cartan subalgebra of $SO(10)$, it is natural to expect ${\rm PQ}_{\rm phys}$ to be a combination of the original PQ symmetry and the transformations in the Cartan subalgebra. The latter includes in particular the charges $R$ and $B-L$ of tables \ref{tab:fermions} and \ref{tab:vevs}. Once the relevant combination of global symmetries has been identified, then one can immediately obtain the
ratios  $q_a/f_{\rm PQ}$ for the fermion fields. This provides all the necessary information to construct the interaction Lagrangian \eqref{eq:Lint}, together with the QCD induced photon corrections \eqref{eq:photon} and the axion to nucleon interactions in \eqref{eq:nucleon},\eqref{eq:nucleon2}, which only depend on the former ratios. This is clear for the axion-fermion interactions, while for the axion-gauge boson interactions it follows 
from the fact that  \eqref{eq:fPQ}, \eqref{eq:fA} and \eqref{eq:anomaly} imply
\begin{align}
\label{eq:csfA}
 f_{A,k}^{-1}={2\sum_a\left(\frac{q_a}{f_{\rm PQ}}\right)T_k(\rho_a)}.
\end{align}
Note how the $q_i/f_{\rm PQ}, q_a/f_{\rm PQ}$, and  $f_A$ are invariant under rescalings of the PQ charges, because under $q_i\rightarrow c\, q_i, q_a\rightarrow cq_a$,  one also has $f_{\rm PQ}\rightarrow c f_{\rm PQ}$, as follows from \eqref{eq:fA}. 
Thus, as expected, the axion effective Lagrangian \eqref{eq:Lint} does not depend on the overall normalisation of the PQ symmetry. The same applies to the axion mass \eqref{zeroTma}.

We note that, since the ${\rm PQ}_{\rm phys}$ symmetry of which the axion is a pseudo-Goldstone boson is a combination of global symmetries of the GUT theory, the arguments leading to \eqref{eq:LintGUT} are 
still valid when one considers the axion of ${\rm PQ}_{\rm phys}$, rather than the original PQ symmetry.

\subsection{Remnant symmetry and domain-wall number}
\label{subsec:ND}
Under a PQ symmetry, which we may assume to be orthogonal to gauge transformations as discussed in the previous section,
the scalar phases transform as $\delta_\alpha A_i= q_i v_i$. Together with \eqref{eq:fA}, this implies that the axion \eqref{eq:A0} transforms as
\begin{align}
\delta_\alpha A= \alpha f_{\rm PQ}.
\end{align}
The effective Lagrangian accounting for the PQ anomaly, given in \eqref{eq:Lint}, breaks the continuous PQ symmetry to a discrete subset
\begin{align}
\label{eq:discrete}
 S(n): A\rightarrow A+\frac{2\pi n}{ \hat N}f_{\rm PQ},\quad n\in \mathbb{Z}.
\end{align}
Like the periodicity of $\theta$ discussed around \eqref{eq:Z}, this follows from the invariance of the partition function, once the contribution
$\int d^4x {\cal L}_{\rm eff}$ in \eqref{eq:Lint} is included. 

Within the previous transformations,
not all of them are necessarily nontrivial, as some may correspond to rephasings of the original scalar phases $A_i$ by $2\pi v_i$, which leaves all complex scalar fields unchanged. According to \eqref{eq:AiA}, these trivial symmetries generate the following group of transformations for the axion:
\begin{align}
\label{eq:Ptrans}
 P(n_i):  A\rightarrow A+\sum_i \frac{2\pi n_i q_i v^2_i}{f_{\rm PQ}}, \quad n_i\in \mathbb{Z}.
\end{align}
Thus the physical symmetry group left after the anomaly is the quotient $S_{\rm phys}=S/P$.
If $S_{\rm phys}$ is a finite group, then any potential generated for the axion will have a finite number of minima, and there will be domain walls. This is because the potential has to be invariant under $S_{\rm phys}$,
so  that its transformations relate minima with other degenerate minima. The number of vacua must be then an integer times the dimension of the finite group. The only truly protected degeneracy is that  induced by the finite group, and so we expect as many minima as the dimension of the finite group (as happens for the potential of the axion
generated by QCD effects). The domain wall number $N_{\rm DW}$ corresponds then to the dimension
of the finite group, ${\rm dim}(S/P)$:
\begin{align}
\label{eq:NDgen}
 \text{Domain wall number: } N_{\rm DW}={\rm dim}\left[\frac{S}{P}\right].
\end{align}

Next we elaborate on a procedure to determine $N_{\rm DW}$ in terms of $f_{\rm PQ}$, the PQ charges $q_i$ and the VEVs $v_i$. Suppose that $n_{\rm min}$ is the minimum 
number $n$ for which one transformation in $S$ (eq. \eqref{eq:discrete}) can be undone with a transformation in $P$ (eq. \eqref{eq:Ptrans}). This implies
\begin{align}
\label{eq:nin}
 \frac{2\pi n_{\rm min}}{\hat N}=\sum_i \frac{2\pi n_i q_i v^2_i}{f_{\rm PQ}^2}
\end{align}
for some values $n_i$. Then any transformation $S(k n_{\rm min})$,
$k\in \mathbb{Z}$, can also be undone with an element of $P$, as is clear by doing $n_{\rm min}\rightarrow k n_{\rm min}$, $n_i\rightarrow k n_i$ in \eqref{eq:nin}.
This means that any element in $S(n)$ with 
$k n_{\rm min}\leq n \leq (k+1) n_{\rm min} $ is equivalent, up to a $P$ transformation, to an element in $\{S(n), 0\leq n \leq n_{\rm min}\}$. For the extrema of the interval, this follows from our previous arguments showing that all
the $S(k n_{\rm min}), k\in \mathbb{Z}$ are equivalent to the trivial transformation $S(0)$. For the transformations inside the interval $(kn_{\rm min},(k+1)n_{\rm min})$ we have
\begin{align}
 n=k n_{\rm min}+\delta, \,\,0< \delta < n_{\rm min} \Rightarrow \delta A_{S(n)}=\frac{2\pi nf_{\rm PQ}}{\hat N}=\frac{2\pi k n_{\rm min}f_{\rm PQ}}{\hat N}+\frac{2\pi  \delta f_{\rm PQ}}{\hat N}.
\end{align}
The part involving $2\pi k n_{\rm min}f_{\rm PQ}/\hat N$ is by hypothesis equivalent to a transformation in $P$, and the part involving  $2\pi \delta f_{\rm PQ}/\hat N$ is a transformation in $\{S(n), 0< n < n_{\rm min}\}$. This proves that all $S(n)$ are equivalent under $P$ to $S(n), n\leq n_{\rm min}$.
Thus
\begin{align}
\label{eq:domainnumber}
 {\rm dim} \frac{S}{P}=N_{\rm DW}, \quad  \quad  {N_{\rm DW}}={\text{ minimum integer}}\left\{{\hat N}\sum_i \frac{ n_i q_i v^2_i}{f_{\rm PQ}^2},\,n_i\in\mathbb{Z}\right\}.
\end{align}
If there is a finite solution for  $N_{\rm DW}$, since $S(N_{\rm DW})\sim S(0)$ (equivalence up to a $P$ transformation), then one has in fact
\begin{align}
 \frac{S}{P}=Z_{{N_{\rm DW}}}, 
\end{align} 
which is the usual finite symmetry associated with domain walls. 

We may write $N_{\rm DW}$ in terms of the coefficients $c_i$ of the axion combination \eqref{eq:axionsimple}. Using  \eqref{eq:csqs} and \eqref{eq:fA} it follows that
\begin{align}
\label{eq:NDW}
 N_{\rm DW}={\text{ minimum integer}}\left\{\frac{1}{f_A}\sum_i  n_i c_i v_i,\,n_i\in\mathbb{Z}\right\}.
\end{align}
Again, $N_{\rm DW}$ is invariant under common rescalings of the PQ charges, as these leave $c_i$ and $f_A$ invariant. A simple case is that in which $\hat{N}$ is an integer and the scalar $q_i$ charges have at least one 
common divisor, which could be unity. Let $k$ denote the maximal common divisor. In this case 
the domain-wall number is simply $\hat{N}/k$. Indeed, writing
\begin{align}
 q_i\equiv k \tilde q_i,
\end{align}
then the term in brackets in equation \eqref{eq:domainnumber} reaches a minimum integer value when taking $n_i=q_i/k$:
\begin{align}
\label{eq:ND_rational}
  n_{\rm min}={\hat N}\sum_i \frac{ n_i q_i v^2_i}{\fPQ^2}=\hat{N}\sum_i \frac{  q_i^2 f^2_i}{k f^2}=\frac{\hat{N}}{k}.
\end{align}
$\hat{N}/k$ is an integer because  $\hat{N}$ is a sum of terms proportional to the charges (see \eqref{eq:anomaly}).  The latter have $k$ as their maximal common divisor, and so $k$ is a maximal divisor of $\hat{N}$.

It should be stressed that the domain-wall number for the  axion corresponding to ${\rm PQ}_{\rm phys}$, computed after imposing orthogonality with respect to the gauge bosons, is not the same as the domain-wall number calculated using the above formulae but with the charges of the original PQ symmetry. The reason is as follows. Starting from the original PQ symmetry, without imposing orthogonality conditions, one has a group of discrete transformations $S$ as in \eqref{eq:discrete}, but defined in terms of the original PQ charges. Similarly, one can define $P$ transformations as in \eqref{eq:Ptrans}. When 
identifying the physically relevant transformations within $S$, then one has to remove not only the trivial rephasings in $P$, but also the discrete transformations in the center $Z$ of the gauge group. Thus we may rewrite equation \eqref{eq:NDgen} more precisely, emphasising the fact that it has assumed orthogonality with respect to gauge transformations, as follows:
\begin{align}
\label{eq:NDphys}
 \text{Domain wall number: } N_{\rm DW}={\rm dim}\left[\frac{S_{\rm phys}}{P_{\rm phys}}\right]={\rm dim}\left[\frac{S}{Z P}\right].
\end{align}
For $SO(10)$ the center of the group is $Z=Z_2$, so that the naive domain wall number computed from the original PQ symmetry (e.g.  \eqref{pgsymmetry_yukawas}) using equations \eqref{eq:domainnumber} or \eqref{eq:NDW} will be two times larger than the actual
physical domain-wall number.

The domain-wall number $N_{\rm DW}$ is relevant because the existence of $N_{\rm DW}$ inequivalent, degenerate vacua implies that, once the PQ symmetry is broken and QCD effects
generate a nonzero axion mass, the universe becomes populated with patches in which the axion falls into one of the $N_{\rm DW}$ vacua. These patches are separated by domain walls 
that meet at axion strings, with each string attached to $N_{\rm DW}$ domain walls. Within a domain wall, the axion field has nonzero gradients, so that the walls store a large amount of energy which may in fact overclose the universe, unless the system of domain walls and strings is diluted by inflation --as is the case if the PQ symmetry is broken before the end of inflation and not restored afterwards-- or is unstable \cite{Zeldovich:1974uw}. The latter can happen if string-wall systems can reconnect in finite size configurations which may shrink to zero 
size by the emission of relativistic axions or gravitational waves. This is allowed for $N_{\rm DW}=1$ \cite{Vilenkin:1982ks}, when for example a string loop becomes the boundary of a single membrane; however, for $N_{\rm DW}>1$ the loops become boundaries of multiple membranes in between which the axion field takes different values, and such configurations cannot be shrunk continuously to a point, which prevents their decay.

%% file: axion_mass_so10_models.tex
\section{\boldmath Axion properties in various $SO(10)\times U(1)_{\rm PQ}$ models}
\label{so10xu1pqmodels}

After motivating the PQ symmetry in predictive $SO(10)$ constructions and reviewing its connection to the axion solution to the CP problem,
next we study the properties of the axion in $SO(10)$ models, using the results of the previous section. First, we will show that
the Peccei-Quinn symmetry \eqref{pgsymmetry_yukawas} --postulated to get a predictive scenario for fermion masses and mixing-- is phenomenologically unacceptable unless other scalar fields
with nonzero PQ charges are introduced. This is because the model with the $10_H$ and $\overline{126}_H$ scalars predicts an axion decay constant at the electroweak 
scale, which has been ruled out experimentally (for a review, see  \cite{Patrignani:2016xqp}). Then we will move on to consider models in which the axion decay constant lies at either the unification scale 
or in between the latter and the electroweak scale.

\subsection{Models with an axion decay constant at the electroweak scale\label{subsec:fAEW}}

Here we consider the minimal scalar content motivated in Section \ref{thecaseforapqsymmetry}, i.e. a $210_H$, a $10_H$ and a $\overline{126}_H$, with the latter two charged under the PQ symmetry
in accordance to equation \eqref{pgsymmetry_yukawas}. The scale $f_{A}$ will be a combination of the VEVs of the fields charged under PQ, i.e. $10_H, \overline{126}_H$. These VEVs determine the fermion masses,
which include the SM fermions --whose masses and associated VEVs must lie below the electroweak scale--  and the right-handed neutrinos, which are allowed to be heavy. The mass of the latter is set only by 
the VEV $v_R=\langle (10,1,3)_{126}\rangle$ within $\overline{126}_H$, as follows from equations \eqref{eq:VEVS_10_126}, \eqref{nonsusysumrule}, \eqref{seesawmatrix}. 
The $U(1)_{B-L}\supset SU(4)_C$ symmetry is broken only by the VEVs $v_R=\langle (10,1,3)_{126}\rangle$ and 
$v_L=\langle(10,3,1)_{126}\rangle$. The latter breaks the electroweak symmetry and contributes to light neutrino masses and low-energy lepton number violation, so that  $v_R\gg v_L$.
Then we are at the situation commented at the end of the previous section, in which a gauge symmetry is broken by several VEVs, with a single dominant one. It follows
that $f_A$ is of the order of the light VEVs, i.e. $v_L, v^{10}_{u,d},v^{126}_{u,d}$, which are at the electroweak scale or below. For an overview of the various VEVs and mass scales, see table \ref{tab:allvevs}. The corresponding mass scales of the fermions are presented in table \ref{tab:allfermions}.


\begin{table}[t]
\begin{align*}
\begin{array}{|c|c|c|c|c||c|c|c|}
\hline
  SO(10)	& 4_C2_L2_R	&  4_C2_L1_R		&  3_C2_L1_R1_{B-L}	& 3_C2_L1_Y		& 3_C 1_{\rm em}		&\text{scale}&\text{VEV}\\
  \hline
  \hline
  10_H		&(1,2,2)	&(1,2,\frac{1}{2})	& (1,2,\frac{1}{2},0)	& (1,2,\frac{1}{2})	&(1,0)=:H_u		&M_Z& v^{10}_u\\
		& 		& (1,2,-\frac{1}{2})	&(1,2,-\frac{1}{2},0)	&(1,2,-\frac{1}{2})	&(1,0)=:H_d	&M_Z & v^{10}_d\\
  \hline
  45_H 		&(1,1,3) 	&(1,1,0)	& (1,1,0,0)		&(1,1,0)		&(1,0):=\sigma			&M_{\rm PQ}&v_{\rm PQ}\\
    \hline
\overline{126}_H &(10,1,3)	&(10,1,1)	&(1,1,1,-2)		&(1,1,0)		&(1,0):=\Delta_R			&M_{\rm BL}&v_{\rm BL}\\
		&(15,2,2)	&(15,2,\frac{1}{2})	&(1,2,\frac{1}{2},0)	&(1,2,\frac{1}{2})	&(1,0):=\Sigma_u 	& M_Z &v^{126}_u\\
		& 		& (15,2,-\frac{1}{2})	&(1,2,-\frac{1}{2},0)	& (1,2,-\frac{1}{2})   	&(1,0):=\Sigma_d	&M_Z& v^{126}_d\\
  \hline 
  210_H		&(1,1,1)  &(1,1,0)		&(1,1,0,0)		&(1,1,0)		&(1,0):=\phi			&M_{\rm U}&v_{\rm U}\\
  \hline
 \end{array} 
 \end{align*}
 \caption{\label{tab:allvevs}Decomposition of the scalar multiplets according to the various subgroups in our breaking chains. We only display the multiplets which get nonzero vacuum expectation values (VEVs) in the different models considered in the paper. ``Scale'' refers to the contribution to gauge boson masses induced by the VEV of a multiplet, rather than to the mass of the multiplet itself.
 According to the extended survival hypothesis, we only keep  the  multiplets which acquire a VEV at lower scales, (with the exception of $\Sigma_u,\Sigma_d$, which decouple at $\MBL$ in order to 
give rise to a  low-energy 2HDM limit). All submultiplets \emph{not} in the list are assumed to be at the unification scale $M_{\rm U}‚$. In all cases, we have $M_Z<\{M_{\rm BL},M_{\rm PQ}\}<M_{\rm U}$. The different relations between $M_{\rm BL}$ and $M_{\rm PQ}$ are considered in the cases A and B. Depending on the model, not all listed multiplets are included. The various models are described in the text.}
\end{table}
\begin{table}[h]
\begin{align*}
\begin{array}{|c|c|c|c||r|c|}
\hline
  SO(10)	& 4_C 2_L 2_R	&  4_C 2_L 1_R				&  3_C 2_L 1_R 1_{B-L}			&\ \hskip-2cm{3_C 2_L 1_Y}								& {\rm scale}\\
  \hline
  \hline
  16_F		&(4,2,1)	&(4,2,0)				&   \left(3,2,0,\frac{1}{3}\right)		&   \left(3,2,\frac{1}{6}\right)   :=q  			&  M_Z\\
                &		&					&   (1,2,0,-1)            		&   \left(1,2,-\frac{1}{2}\right)  :=l					&  M_Z\\
                \cline{2-6}
		&(\bar 4,1,2)	&\left(\bar4,1,\frac{1}{2}\right)	&   \left(\bar 3,1,\frac{1}{2},-\frac{1}{3}\right)		&   \left(\bar 3,1,\frac{1}{3}\right):=d        &  M_Z	 \\
		&		&					&   \left(1,1,\frac{1}{2},1\right)         	   	&   \left(1,1,1\right)  :=e      		        & M_Z\\
                                \cline{3-6}
		&               &\left(\bar4,1,-\frac{1}{2}\right)	&   \left(\bar 3,1,-\frac{1}{2},-\frac{1}{3}\right)		&   \left(\bar 3,1,-\frac{2}{3}\right):=u       & M_Z\\
		&		&					&   \left(1,1,-\frac{1}{2},1\right)         	   	&   \left(1,1,0\right):=n	                        & M_{\rm BL}\\
  \hline
  10_F		& (6,1,1) 	&(6,1,0)				&  \left(3,1,0,-\frac{2}{3}\right)	&  \left(3,1,-\frac{1}{3}\right):={\tiny{\tilde D}}                     & M_{\rm PQ}\\
  &		&					&  \left(\bar 3,1,0,\frac{2}{3}\right)	&  \left(\bar 3,1,\frac{1}{3}\right):=D							& M_{\rm PQ}\\
		 \cline{2-6}
		& (1,2,2)	&\left(1,2,\frac{1}{2}\right)		&  \left(1,2,\frac{1}{2},0\right)	&  \left(1,2,\frac{1}{2}\right):=\tilde L				& M_{\rm PQ}\\
		\cline{3-6}
		&		&\left(1,2,-\frac{1}{2}\right)		&  \left(1,2,-\frac{1}{2},0\right)	&  \left(1,2,-\frac{1}{2}\right):=L					& M_{\rm PQ}\\
  \hline
 \end{array} 
 \end{align*}
 \caption{\label{tab:allfermions}Decomposition of the fermion multiplets according to the various subgroups in our breaking chains.
 All SM fermions have masses set by the Higgs mechanism, the heavy right handed neutrinos acquire their mass at the BL breaking scale from the coupling to the $\overline{126}_H$. Fermions in the $10_F$ representation can obtain a mass from a Yukawa coupling to the $45_H$ or to a scalar singlet (if present). 
 }
\end{table}

Despite the lack of viability of the model, it is instructive to construct the axion explicitly using the techniques outlined in Section \ref{axiongeneralities}; this will serve as a simple example 
that will pave the way to the computations in viable models. Again, the axion involves the fields charged under PQ and getting nonzero VEVs, which are contained in  the  $\overline{126}_H$ and $10_H$ multiplets. As detailed in table \ref{tab:allvevs}, the PQ fields getting VEVs are the Higgses $H_u,H_d\supset (1,2,2)_{10}$, $\Sigma_u,\Sigma_d\supset (15,2,2)_{126}$, and the SM singlet $\Delta_R\supset({10},1,3)_{126}$. To simplify the notation as much as possible, we will  denote the VEVs with $v$ and the phases with $A$, with appropriate subindices, as in equation \eqref{eq:scalar_param}. 
We define
\begin{equation}
\label{eq:fields_not}\begin{aligned}
 \phi_1\equiv &\Sigma_u, & \phi_2\equiv &\Sigma_d, & \phi_3\equiv &H_u, & \phi_4\equiv &H_d, & \phi_5\equiv &\Delta_R.
\end{aligned}\end{equation}
For simplicity, and as was anticipated in Section \ref{thecaseforapqsymmetry}, we consider a zero VEV $v_L$ for the $(\overline{10},3,1)_{126}$ multiplet, in order to avoid $B-L$ violation at low energies, and to realise the simplest version of the seesaw mechanism (see equations \eqref{eq:VEVS_10_126} through \eqref{seesawmatrix}).  
We will also denote $v_{\rm BL} \equiv v_R$ as this is now the only B-L breaking VEV.
A general parameterisation of the axion, without knowledge of the PQ charges, can now be written as in \eqref{eq:axionsimple}.
As detailed in Section \ref{axiongeneralities}, we may constrain the previous coefficients by imposing orthogonality with respect to the Goldstone bosons of the broken gauge symmetries, as well 
as perturbative masslessness. For the gauge constraints, the choice of nonzero VEVs is such that, as commented in \ref{subsec:axion_physical} and and shown in Appendix 
\ref{app:ferphases} the only nontrivial orthogonality conditions are those with 
respect to the Goldstones associated with the generators in the five-dimensional Cartan subalgebra of the gauge group. Since all the VEVs corresponds to colour singlets, they carry no weights under the two generators of the Cartan subalgebra of $SU(3)_C\supset SU(4)_C$. By assumption, the fields also carry no electric charge, which eliminates another combination of Cartan generators (see equation \eqref{eq:charge} for the relation between the electric charge and the weights corresponding to the Cartan generators  of the group $SU(4)_C\times SU(2)_L\times SU(2)_R\supset SO(10)$). This leaves two independent Cartan generators giving rise to two nontrivial orthogonality constraints. We may use the generators $U(1)_{B-L}\supset SU(4)_C$ and $U(1)_R\supset SU(2)_R$ --see Appendix \ref{app:roots} for how $B-L$ is embedded into the Cartan algebra of $SU(4)_C\times SU(2)_L\times SU(2)_R$). The charges of our fields $\phi_i=\{H_{u,d},\Sigma_{u,d},\Delta_R\}$ under these symmetries are given in table \ref{tab:vevs}. The orthogonality constraints \eqref{eq:orthogauge2} yield
 \begin{equation}\label{d1}
 \begin{aligned}
  {c_1}{v_1}-{c_2}{v_2}+{c_3}{v_3}-{c_4}{v_4}&=0,\\
  c_5&=0.
 \end{aligned}
 \end{equation}
 Moving on to impose perturbative masslessness, we note that in the scalar potential the term $10_H\,10_H\,\overline{126}^\dagger_H \overline{126}^\dagger_H+h.c.$ is allowed by both the gauge and PQ-symmetries. After symmetry breaking, these terms induce masses for some combinations of phase fields. Denoting gauge-invariant contractions by ``$\rm inv$'', we have:
 \begin{align*} 
  10_H\,10_H\,\overline{126}^\dagger_H\, \overline{126}^\dagger_H|_{\rm inv}+h.c.
  &\supset (1,2,2)(1,2,2,)(15,2,2)(15,2,2)|_{\rm inv}+h.c.\\
  &\supset (H_u+H_d)(H_u+H_d)(\Sigma^\dagger_u+\Sigma^\dagger_d)(\Sigma^\dagger_u+\Sigma^\dagger_d)|_{\rm inv}+h.c.\\
  &\supset -{v_3^2v_1^2}
  \left(\frac{A_3}{v_3}-\frac{A_1}{v_1}\right)^2- {v_4^2 v_2^2}
  \left(\frac{A_4}{v_4}-\frac{A_2}{v_2}\right)^2.
 \end{align*}
 %
The orthogonality conditions as in equations \eqref{eq:mAs},\eqref{eq:orthomass} yield
 \begin{equation}\label{d4}
 \begin{aligned}
-\frac{c_1}{v_1}+\frac{c_3}{v_3}=&0\\
-\frac{c_2}{v_2}+\frac{c_4}{v_4}=&0.
\end{aligned}\end{equation}
More massive combinations can be found under closer inspection of the scalar potential, but they cannot give additional constraints on the axion as we already identified four constraints which, together with the requirement
for a canonical normalisation of the axion, fix the five
independent coefficients $c_i$. Proceeding in this way  we can finally conclude that the axion is given, up to a minus sign\footnote{We choose the sign that gives a positive value for the $f_{A,k}$, see \eqref{eq:fAs}.}, by
\begin{align}
 \label{gutaxion}
 A=-\frac{(A_4v_4+A_2v_2)(v_3^2+v_1^2)+(A_3v_3+A_1v_1)(v_4^2+v_2^2)
 }
 { \sqrt{ v^2(v_4^2+v_2^2)(v_3^2+v_1^2) }},\quad v^2\equiv \sum_{i=1}^4 v^2_i.
\end{align}
We remind the reader that the above parameters $v_i,A_i$ are defined by equations \eqref{eq:scalar_param} and \eqref{eq:fields_not}.
The axion couplings to matter can be calculated using  the results of section \ref{subsec:effL}. Equations \eqref{eq:Lint} and \eqref{eq:csfA} imply that the effective Lagrangian can be simply derived from
the ratios $q_a/f_{\rm PQ}$ corresponding to the fermions. The ones corresponding to the scalars can be obtained from the scalar ratios $q_i/f_{\rm PQ}$, which can be immediately derived from the axion coefficients $c_i$ by using \eqref{eq:csqs}. Applying the latter identity to the axion combination \eqref{gutaxion}, it follows that
\begin{equation}\label{eq:charges}\begin{aligned}
 \frac{q_1}{f_{\rm PQ}}= \frac{q_3}{f_{\rm PQ}}=&\,-\frac{\sqrt{v_4^2+v_2^2}}{v\sqrt{(v_3^2+v_1^2)}},\quad
 \frac{q_2}{f_{\rm PQ}}= \frac{q_4}{f_{\rm PQ}}=&\,-\frac{\sqrt{v_3^2+v_1^2}}{v\sqrt{(v_2^2+v_4^2)}},\quad
 q_5=&\,0.
\end{aligned}\end{equation}
From these we may derive the PQ charges $q_a/f_{\rm PQ}$ of the Weyl fermions by identifying the appropriate combination of global symmetries in the Lagrangian that gives rise to the charges in \eqref{eq:charges}.
The physical symmetry ${\rm PQ}_{\rm phys}$ can be expressed as a combination of the global ${\rm PQ},U(1)_R$ and $U(1)_{B-L}$ --as anticipated in \ref{subsec:axion_physical}, the modification of ${\rm PQ}$ involves  the symmetries within the Cartan algebra of the group:
\begin{align}
\label{eq:U1s}
 {\rm PQ}_{\rm phys}=s_1 \,{\rm PQ}+s_2\, U(1)_R+s_3\, U(1)_{B-L}.
\end{align}
From the conventions in \eqref{eq:fields_not}, the PQ charges in \eqref{pgsymmetry_yukawas} and the $U(1)_R,U(1)_{B-L}$ charges given in table \ref{tab:vevs} one deduces:
\begin{equation}\label{eq:s}\begin{aligned}
 \frac{s_1}{\fPQ}=&\frac{v}{4 \sqrt{\left(v_1^2+v_3^2\right) \left(v_2^2+v_4^2\right) }}\,,\quad
 \frac{s_2}{\fPQ}=&\,\frac{v_1^2-v_2^2+v_3^2-v_4^2}{v\sqrt{\left(v_1^2+v_3^2\right) \left(v_2^2+v_4^2\right) }},\\
 \frac{s_3}{\fPQ}=&\,\frac{v_1^2-3 v_2^2+v_3^2-3 v_4^2}{4 v\sqrt{ \left(v_1^2+v_3^2\right) \left(v_2^2+v_4^2\right)}}.
\end{aligned}\end{equation}
Finally, the values of $q_a/\fPQ$ for the fermions follow from \eqref{eq:U1s} and \eqref{eq:s}, and the charge assignments in table \ref{tab:allfermions}:
\begin{equation}\label{eq:fermionq}\begin{aligned}
 \frac{q_q}{\fPQ}=&\,\frac{1}{3v}\sqrt{\frac{v_1^2+v_3^2}{\left(v_2^2+v_4^2\right) }},
 &\frac{q_u}{\fPQ}=&\,\frac{-v_1^2+3 v_2^2-v_3^2+3 v_4^2}{3v \sqrt{ \left(v_1^2+v_3^2\right) \left(v_2^2+v_4^2\right)}},\\
 \frac{q_d}{\fPQ}=&\,\frac{2}{3v}\sqrt{\frac{v_1^2+v_3^2}{\left(v_2^2+v_4^2\right) }},
 &\frac{q_l}{\fPQ}=&\,\frac{1}{v}\sqrt{\frac{v_2^2+v_4^2}{\left(v_1^2+v_3^2\right) }},\\
 \frac{q_e}{\fPQ}=&\,\frac{v_1^2-v_2^2+v_3^2-v_4^2}{v\sqrt{\left(v_1^2+v_3^2\right) \left(v_2^2+v_4^2\right)}},
  &\frac{q_n}{\fPQ}=&\,0.
\end{aligned}\end{equation}
From the above we may obtain the $f_{A,k}$ using \eqref{eq:csfA}:
\begin{align}
\label{eq:fAs}
 f_{A,3_C}=f_{A,2_L}=\frac{5}{3}f_{A,Y}=\frac{1}{3} \sqrt{\frac{\left(v_1^2+v_3^2\right) \left(v_2^2+v_4^2\right)}{v^2}}.
\end{align}
As explained in  \ref{subsec:effL}, the value of $f_{A,3_C}$ only depends on the scalar PQ charges, and can be also obtained from equation \eqref{eq:fA3}.
The simple relations above reproduce exactly the result of equation \eqref{eq:LintGUT}, which was derived in the grand unified theory. Calling $f_A\equiv f_{A,3}$ we obtain the effective Lagrangian for 
the axion
\begin{align}
\label{lowenergylag}
 \mathcal{L}_{\rm int}=& \frac{1}{2}\partial_\mu A \partial^\mu A +\frac{\alpha_s}{8\pi} \frac{A}{f_A} \, G^b_{\mu\nu} \tilde G^{b\mu\nu} +\frac{\alpha}{8\pi} \frac{8}{3}\frac{A}{f_A} \,F_{\mu\nu} \tilde F^{\mu\nu}+\partial_\mu A\sum_{f=q,u,d,l,e} \frac{q_f}{f_{\rm PQ}} (f^\dagger \bar\sigma^\mu f),
\end{align}
where the $q_f/\fPQ$ factors (which are the same across generations) are given in equation \eqref{eq:fermionq}. 

As stated at the end of section \ref{subsec:effL},  one 
may obtain a physically equivalent effective Lagrangian ${\cal L}'_{\rm int}$ by starting from the usual fermion kinetic terms and Yukawa interactions and perform different phase rotations
\eqref{eq:arbitrot} that remove the scalar phases in the Yukawa terms. This does not affect the coupling of the axion to the photon; see Appendix \ref{app:ferphases} for more details.
At low energy,  incorporating the QCD effects from the 
axion-meson mixing in equation \eqref{eq:photon} and the nucleon interactions in \eqref{eq:nucleon}, \eqref{eq:nucleon2}, and expressing the electron interactions in the axial basis as in eq. \eqref{eq:axial2},
the Lagrangian involving the axion, the photon, nucleons and electrons is:
\begin{equation}
\label{eq:DFSZ}
\begin{aligned}
 \mathcal{L}^{\rm QCD}_{\rm int}=&\frac{1}{2}\partial_\mu A \partial^\mu A 
- \frac{1}{2} m_A^2 A^2 
+ \frac{\alpha}{8\pi} \,\frac{C_{A\gamma}}{f_A}\,A\,F_{\mu\nu} {\tilde F}^{\mu\nu} \\
 &-\partial_\mu A\left[ \frac{C_{AP}}{2f_A} \overline P^\dagger \gamma^\mu \gamma_5 P+\frac{C_{AN}}{2f_A} \overline N^\dagger \gamma^\mu \gamma_5 N+\frac{C_{AE}}{2f_A} \overline E^\dagger \gamma^\mu \gamma_5 E\right],
 \\
C_{A\gamma}=&\, \frac{8}{3}-1.92(4), \\
 C_{AP}=&\,-0.62+0.43\cos^2\beta\pm0.03,\\
 C_{AN}=&\,0.26-0.41\cos^2\beta\pm0.03,\\
 C_{AE}=&\,\frac{1}{3}\sin^2\beta,
\end{aligned}
\end{equation}
where we defined
\begin{align}\label{eq:gamma}
 \tan^2\beta\equiv \frac{v_1^2+v_3^2}{v_2^2+v_4^2}.
\end{align}
The couplings to fermions coincide with those in the usual DFSZ model \cite{Zhitnitsky:1980tq,Dine:1981rt,diCortona:2015ldu}, although the relation between the parameter $\beta$ 
and the scalar VEVs now involves additional fields. As commented in \ref{subsec:axial_basis}, potential differences with respect to DFSZ models could come from the axion interactions with the weak bosons, 
which in the axial basis leading to \eqref{eq:DFSZ} will contain the information of the ${\rm PQ}_{\rm phys}$ charges of the fermions.

The domain-wall number of the model can be calculated from \eqref{eq:NDW}. We may first consider the ``naive'' domain wall number obtained by using the PQ charges of equation \eqref{pgsymmetry_yukawas}, without imposing orthogonality conditions. In this case $\hat{N}=12$ (see \eqref{eq:anomaly}) is an integer --note that the value of $\hat{N}$ is the same for the GUT group and all its non-Abelian subgroups, as follows
from the fact that $\hat{N}$ in \eqref{eq:anomaly} can be expressed as a single trace over all fermions, which fall into GUT representations.
On the other hand, the scalar charges have $k=2$ as a maximum common divisor. In this situation, as discussed in section \ref{subsec:ND}, the domain wall number would be $\hat{N}/k=6$, as corresponds to a DFSZ axion model. On the other hand, using the physical PQ charges in \eqref{eq:charges}, the calculation is a bit more involved. Starting from equation \eqref{eq:NDW}, the quantity in brackets is a rational function of the $v_i$. In order to have an integer result, we must demand that the numerator is proportional to the denominator.
This gives a system of equations, as many as there are independent monomials in the denominator. Denoting the minimum integer as $n_{\rm min}$ (which will be the domain-wall number) one has:
\begin{equation}\begin{aligned}
n_{\rm min}+3(n_1+n_2)=&\,0, &
n_{\rm min}+3(n_2+n_3)=&\, 0,\\
n_{\rm min}+3(n_1+n_4)=&\,0, &
n_{\rm min}+3(n_3+n_4)=&\,0.
\end{aligned}\end{equation}
Since the $n_i$ are integers, clearly one has $N_{\rm DW}=n_{\rm min}=3.$ That is, the domain wall number is half of the naive estimate with the unphysical PQ symmetry in \eqref{pgsymmetry_yukawas}. As discussed around equation \eqref{eq:NDphys}, this is due to the fact that the naive estimate is not taking into account the need to quotient the remnant discrete symmetry by the center $Z_2$ of the gauge group .

As anticipated at the beginning of this section, all VEVs appearing in $f_A\equiv f_{A,3_C}$ in equation \eqref{eq:fAs} have to be at the electroweak scale, as follows
from the conventions in \eqref{eq:fields_not} and equation \eqref{nonsusysumrule}. Hence, the axion described in this model is visible, being just a GUT-embedded variant of the original Peccei-Quinn-Weinberg-Wilczek model which is
phenomenologically unacceptable (for a review, see  \cite{Patrignani:2016xqp}).

There are several ways to lift the axion decay constant to higher values, which, as follows from the discussion in \ref{subsec:axion_physical}, must involve additional scalars with PQ charges. The simplest way is to  give a PQ charge to the Higgs field
responsible for the GUT symmetry breaking~\cite{Mohapatra:1982tc}.
An alternative way is to introduce a new scalar multiplet, e.g. a
$45_H$,  also charged under the PQ symmetry \cite{Holman:1982tb,Altarelli:2013aqa}. A third 
way is to introduce an $SO(10)$ singlet complex scalar field responsible for the $U(1)_{\rm PQ}$ symmetry breaking~\cite{Babu:2015bna}.  We will consider 
in this paper benchmark models from all these three categories. 

\subsection{Models with axion decay constants at the unification scale}
\label{gutaxionmodel}

As follows from the arguments in Section \ref{subsec:axion_physical}, in order to have a heavy axion one needs at least two fields charged under PQ and getting large VEVs. 
In the model of the previous section, the scalar $210_H$, which was needed to ensure the breaking of the GUT group, was not charged under PQ. Thus the most minimal way to decouple the axion decay constant from the electroweak scale is to extend the PQ symmetry 
 \eqref{pgsymmetry_yukawas} to the $210_H$, 
\begin{equation}
\label{eq:PQGUT}
\begin{array}{cc}
\begin{aligned}
 &16_F\rightarrow16_F e^{i\alpha}, \\
{\bf Model\ 1:}\hspace{6ex} & 10_H\rightarrow10_H e^{-2i\alpha},  \\
 & \overline{126}_H\rightarrow\overline{126}_H e^{-2i\alpha}, \\
 & 210_H \rightarrow 210_He^{4i\alpha}. 
\end{aligned}
\end{array}
\end{equation} 
The PQ charge of $210_H$ follows from the requirement of allowing gauge invariant cubic interactions between the $210_H$ multiplet  the other scalars. The only possibility is $210_H\,\overline{126}_H\, 10_H$, which fixes the above PQ charge. 

The construction of the axion field in this model goes along the same lines as in the previous section, yet with an added extra phase associated with the $\phi=(1,1,1)$ component of the $210_H$ multiplet whose VEV $v_{\rm U}\equiv v_\phi=v_6$ breaks $SO(10)$ to $4_C\times2_L\times2_R$ (see table \ref{tab:vevs}). We now define
\begin{equation}
\label{eq:fields_not3}\begin{aligned}
 \phi_1\equiv &\Sigma_u, & \phi_2\equiv &\Sigma_d, & \phi_3\equiv &H_u, & \phi_4\equiv &H_d, & \phi_5\equiv &\Delta_R, &  \phi_6\equiv &\phi.
\end{aligned}\end{equation}
The general parameterisation of the axion is now $ A=\sum_{i=1}^6 c_i A_i$.
When imposing perturbative masslessness, we have the same constraints \eqref{d4} as before, plus new ones coming from the new interaction $210_H\,\overline{126}_H\,10_H$:
 \begin{align*} 
  210_H\,\overline{126}_H\, 10_H|_{\rm inv} +h.c. &\supset (1,1,1)[(10,1,3)+(15,2,2)](1,2,2)|_{\rm inv}+h.c.\\
 \textstyle
 &=\phi\Sigma_uH_d+\phi \Sigma_d H_u +h.c. \\ 
 &\supset-\frac{v_6 v_1 v_4}{2\sqrt{2}}\left(\frac{A_6}{v_6}+\frac{A_1}{v_1}+\frac{A_4}{v_4}\right)^2-
 \frac{v_6 v_2 v_3}{2\sqrt{2}}\left(\frac{A_6}{v_6}+\frac{A_2}{v_2}+\frac{A_3}{v_3}\right)^2,
 \end{align*}
 where ``inv'' denotes a projection into gauge-invariant contractions.
 Masslessness of the axion requires then 
 \begin{equation}\begin{aligned}
 \frac{c_6}{v_6}+\frac{c_2}{v_2}+\frac{c_3}{v_3}&=0,\\
  \frac{c_6}{v_6}+\frac{c_1}{v_2}+\frac{c_4}{v_3}&=0.
  \end{aligned}\end{equation}
 In addition, since $\phi$ is a singlet under $U(1)_R$ and $U(1)_{B-L}$, we still have the same constraints from orthogonality as before, equations \eqref{d1} and \eqref{d4}. 
 Solving the linear system of equations and normalising, we construct the axion for this model:
\begin{align}
 \label{gutaxion2}
 A=-\frac{(A_4v_4+A_2v_2)(v_3^2+v_1^2)+(A_3v_3+A_1v_1)(v_4^2+v_2^2)
 -A_6v_6v^2}
 { \sqrt{ v^2((v_4^2+v_2^2)(v_3^2+v_1^2)+v_6^2 v^2) }},\quad v^2\equiv \sum_{i=1}^4 v^2_i.
\end{align}
Note that in the limit $M_Z\ll M_{\rm U}$ the axion is just $A=A_6$. This follows from the field assignments in \eqref{eq:fields_not2} and the scales in table \ref{tab:vevs}. Therefore,
the dominant contribution to the axion field comes from the $210_H$.\footnote{Note that the $210_H$ is not charged under $U(1)_R$ and $U(1)_{B-L}$ --see table \ref{tab:allvevs}-- so that the axion can be aligned with the phase of a field getting a large VEV, like $\phi\supset 210_H$, without violating any orthogonality condition. This is in contrast to the model \ref{subsec:fAEW}. } The PQ$_{\rm phys}$ charges of the scalars are now:
\begin{equation}\label{eq:charges2}\begin{aligned}
 \frac{q_1}{f_{\rm PQ}}= \frac{q_3}{f_{\rm PQ}}=&\,-\frac{v_2^2+v_4^2}{v\sqrt{ \left(v^2 v_6^2+\left(v_1^2+v_3^2\right) \left(v_2^2+v_4^2\right)\right)}},\\
 \frac{q_2}{f_{\rm PQ}}= \frac{q_4}{f_{\rm PQ}}=&\,-\frac{v_1^2+v_3^2}{v\sqrt{ \left(v^2 v_6^2+\left(v_1^2+v_3^2\right) \left(v_2^2+v_4^2\right)\right)}},\\
 q_5=&\,0,\\
 q_6=&\,\frac{v}{\sqrt{v^2 v_6^2+\left(v_1^2+v_3^2\right) \left(v_2^2+v_4^2\right)}}.
\end{aligned}\end{equation}

Once more, the global symmetry ${\rm PQ}_{\rm phys}$ can be expressed as a combination of ${\rm PQ}$ and the Cartan generators $U(1)_R$ and $U(1)_{B-L}$, as in equation
\eqref{eq:s}, but with coefficients $s_i$ that now take  the values
\begin{equation}\label{eq:s2}\begin{aligned}
 \frac{s_1}{\fPQ}=&\frac{v}{4 \sqrt{\left(v_1^2+v_3^2\right) \left(v_2^2+v_4^2\right) +v^2 v^2_6}}\,,\quad
 \frac{s_2}{\fPQ}=&\,\frac{v_1^2-v_2^2+v_3^2-v_4^2}{v\sqrt{\left(v_1^2+v_3^2\right) \left(v_2^2+v_4^2\right)+v^2 v^2_6}},\\
 \frac{s_3}{\fPQ}=&\,\frac{v_1^2-3 v_2^2+v_3^2-3 v_4^2}{4 v\sqrt{ \left(v_1^2+v_3^2\right) \left(v_2^2+v_4^2\right)+v^2 v^2_6}}.
\end{aligned}\end{equation}
The ${\rm PQ}_{\rm phys}$ charges of the fermions can be obtained from \eqref{eq:U1s} and \eqref{eq:s2}, and the charges of table \ref{tab:allfermions}:
\begin{equation}\label{eq:fermionq2}\begin{aligned}
 \frac{q_q}{\fPQ}=&\,\frac{v_1^2+v_3^2}{3v\sqrt{\left(v_2^2+v_4^2\right)(v_1^2+v_3^2)+v^2v^2_6 }},
 &\frac{q_u}{\fPQ}=&\,\frac{-v_1^2+3 v_2^2-v_3^2+3 v_4^2}{3v \sqrt{ \left(v_1^2+v_3^2\right) \left(v_2^2+v_4^2\right)+v^2v^2_6 }},\\
 \frac{q_d}{\fPQ}=&\,\frac{2(v_1^2+v_3^2)}{3v\sqrt{(v_1^2+v_3^2)\left(v_2^2+v_4^2\right) +v^2v^2_6 }},
 &\frac{q_l}{\fPQ}=&\,\frac{v_2^2+v_4^2}{v\sqrt{\left(v_1^2+v_3^2\right)(v_2^2+v_4^2)+v^2v^2_6 }},\\
 \frac{q_e}{\fPQ}=&\,\frac{v_1^2-v_2^2+v_3^2-v_4^2}{v\sqrt{\left(v_1^2+v_3^2\right) \left(v_2^2+v_4^2\right)+v^2v^2_6}},
  &\frac{q_n}{\fPQ}=&\,0.
\end{aligned}\end{equation}
The $f_{A,k}$, following from \eqref{eq:csfA}, satisfy again the GUT relations in \eqref{eq:LintGUT}, and are given by:
\begin{align}
\label{eq:fAGUT}
 f_{A,3_C}=f_{A,2_L}=\frac{5}{3}f_{A,Y}=\frac{1}{3} \sqrt{\frac{v_6^2v^2+(v_4^2+v_2^2)(v_3^2+v_1^2)}{v^2}}.
\end{align}
In the limit $M_Z\ll M_{\rm U}$, $f_A\sim\frac{v_6}{3}=\frac{v_{\rm U}}{3}$, so that the axion decay constant is dominated by the GUT-breaking scale. The effective Lagrangian 
for the axion is as in equation \eqref{lowenergylag}, with the ${\rm PQ}_{\rm phys}$ charges in \eqref{eq:fermionq2}. At low energies, incorporating QCD effects and going into 
the axial basis, one gets the DFSZ-like interactions in \eqref{eq:DFSZ}, with the the parameter $\beta$ in \eqref{eq:gamma}.

As in the previous model, the original PQ symmetry in \eqref{eq:PQGUT} involves scalar charges with a maximum common divisor $k=2$, and one has integer $\hat{N}=12$ (common to the GUT group and its non-Abelian subgroups). Thus the naive domain-wall number --without imposing orthogonality of the axion with respect to the gauge fields-- is again $\hat{N}/k=6$. To get the physical domain-wall number we may use \eqref{eq:NDW}. As was done for the previous model, \eqref{eq:NDW} can be converted into a system of equations involving $n_{\rm min}$ and integer $n_i$:
\begin{equation}\begin{aligned}
n_{\rm min}+3(n_1+n_2)=&\,0, &
n_{\rm min}+3(n_2+n_3)=&\,0,\\
n_{\rm min}+3(n_1+n_4)=&\,0, &
n_{\rm min}+(n_3+n_4)=&\,0,\\
n_{\rm min}-3n_6=&\,0.
\end{aligned}\end{equation}
Once again, one has $N_{\rm DW}=n_{\rm min}=3$, half of the naive estimate\footnote{As already pointed out in \cite{Geng:1990dv}, a DFSZ model featuring $N_{\rm DW}=3$ can also be constructed without reference to a bigger gauge group. The defining criterion is a Peccei-Quinn that allows a dimension three coupling between the PQ charged scalars. This is common to the models 2.1, 2.2, 3.1 and 3.2 considered in this paper.} .

\subsection{Models with an intermediate scale axion decay constant}
\subsubsection{Additional $45_H$}
\label{altarellimodel}
As was mentioned before, lifting the axion from the electroweak scale requires a scalar other than the  $\overline{126}_H$ having a  nonzero PQ charge and a large VEV. In the previous section,
this scalar was chosen as the one responsible for the first stage of GUT breaking. This linked $f_A$ and the GUT scale. Choosing nonzero VEVs along other components of the $210_H$ multiplet 
which are not involved in the first-stage breaking and can thus have smaller values does not help in lowering $f_A$, as there is always the GUT-scale VEV.
However, one may consider
an additional scalar with a lower-scale vacuum expectation value. To motivate the choice of representation under the unified group,
we can be guided by minimality and predictivity. We would like to constrain the axion mass by the requirement of gauge coupling unification, which is only possible if the PQ-breaking VEV of the new
scalar is also 
related to the breaking of a gauge group. It should therefore be a singlet under an intermediate symmetry group between $4_C2_L2_R$ and $3_C2_L1_Y$. In other words, it should break the Pati-Salam group, but not to the Standard Model. There are few multiplets in the $SO(10)$ representations up to $210_H$ which fulfill this criterion -the lowest ones being the $(1,1,3)$ and the $(15,1,1)$ of the $45_H$, denoted by their Pati-Salam quantum numbers,
\begin{equation}
45_H = (1,1,3) \oplus (1,3,1) \oplus (6,2,2) \oplus (15,1,1)\,.
\end{equation}
The only other option would be the $(15,1,1)$ of the $210_H$, which, as mentioned above does not help in lowering $f_A$, as the multiplet contains a GUT-scale VEV. One could consider an additional $210_H$, independent of the GUT breaking, but minimality favours a smaller multiplet like the $45_H$. We will adopt this choice, and to comply with existing literature \cite{Altarelli:2013aqa,Holman:1982tb}, we choose to use the field $\sigma=(1,1,3)\supset 45_H$, which breaks $SU(2)_R$ down to $U(1)_R$ when it acquires its VEV $v_{\rm PQ}\equiv \langle \sigma \rangle$. Aside from the GUT scale $M_{\rm U}$, the theory will now have two additional physical scales $M_{\rm BL}$ and $M_{\rm PQ}$ related with the
VEVs of the $\overline{126}_H$ and $45_H$, respectively (see table \ref{tab:vevs}).
In this model, the $210_H$ does not carry PQ-charge, as again this would lift $f_A$ to the GUT scale. For the $45_H$ we can choose different PQ charges, depending on the interactions we want to allow with
the other scalars. As  opposed to the case of the $210_H$ in the previous section, there are no cubic interactions of the $45_H$ with the other scalars that are compatible with a nonzero PQ charge for the $45_H$. On the other hand, one can allow the quartic couplings $210_H\times\overline{126}_H\times\overline{126}_H\times45_H$, $210_H\times10_H\times{\overline{126}}_H\times45_H$, which enforce a PQ charge of four units for the new scalar:
\begin{equation}
\label{eq:PQ45}
\begin{array}{cc}
\begin{aligned}
  &16_F\rightarrow16_F e^{i\alpha},  \\
  &10_H\rightarrow10_H e^{-2i\alpha}, \\
{\bf Model\ 2.1:}\hspace{6ex} &  \overline{126}_H\rightarrow\overline{126}_H e^{-2i\alpha}, \\
  &210_H \rightarrow 210_H,   \\ 
  &45_H \rightarrow 45_H e^{4 i\alpha}.
\end{aligned} 
\end{array}
\end{equation}

The construction of the axion goes analogous to Section \ref{gutaxionmodel}. The VEV  $v_{\rm PQ}$ of the $45_H$ now plays the role of the VEV of the $210_H$, so we define:
\begin{equation}
\label{eq:fields_not2}\begin{aligned}
 \phi_1\equiv &\Sigma_u, & \phi_2\equiv &\Sigma_d, & \phi_3\equiv &H_u, & \phi_4\equiv &H_d, & \phi_5\equiv &\Delta_R, &  \phi_6\equiv &\sigma.
\end{aligned}\end{equation}
The masslessness conditions now arise from the interactions $210_H\times\overline{126}_H\times\overline{126}_H\times45_H$ --which includes terms going as $\phi\sigma \Sigma_u \Sigma_d$, see table \ref{tab:vevs}-- and $210_H\times10_H\times{\overline{126}}_H\times45_H$, which includes $\phi\sigma (H/ \Sigma)_u (\Sigma/H)_d$. Since $\sigma$ is not charged under $U(1)_R$ and $U(1)_{B-L}$, the orthogonality conditions are as in Section
\ref{subsec:fAEW}. Despite the different masslessness conditions, the formulae \eqref{gutaxion2}, \eqref{eq:charges2}, \eqref{eq:fermionq2} and \eqref{eq:fAGUT}  of the previous section apply to this model, although with $v_6$ and $A_6$ now referring to the field $\sigma$. In the limit $M_Z\ll M_{\rm PQ}$, the axion is dominated by the VEV of the $45_H$ and we have 
\begin{align}
\label{eq:fA45}
 f_A\sim \frac{v_{6}}{3}=\frac{v_{\rm PQ}}{3}.
\end{align}

Once more, the initial PQ symmetry in \eqref{eq:PQ45} has scalar charges with a maximum common divisor $k=2$; on the other hand, for the GUT group and its non-Abelian subgroups one has integer $\hat{N}=12$, giving a naive domain-wall number of 6. The physical domain-wall number follows from  \eqref{eq:NDW}, which is equivalent to the following system of equations:
\begin{equation}\begin{aligned}
n_{\rm min}+3(n_1+n_2)=&\,0, &
n_{\rm min}+3(n_2+n_3)=&\,0,\\
n_{\rm min}+3(n_1+n_4)=&\,0, &
n_{\rm min}+3(n_3+n_4)=&\,0,\\
n_{\rm min}-3n_6=&\,0.
\end{aligned}\end{equation}
Once again, one has $N_{\rm DW}=n_{\rm min}=3$, half of the naive estimate. The effective Lagrangian for the axion is as in \eqref{lowenergylag}, with the values of  $f_A$ and $q_i/\fPQ$ given in equations \eqref{eq:charges2} and \eqref{eq:fAGUT}.  Accounting for QCD effects in
the axial basis, one recovers again the DFSZ-like interactions in \eqref{eq:DFSZ}, \eqref{eq:gamma}.

In \cite{Altarelli:2013aqa} $v_{\rm PQ}$ was chosen to lie at the same scale as the VEV of the \hts. In principle, there is no reason for this equality, so we will not use it in our analysis. Generically, as mentioned before  we have two physical scales $\MBL$ and $\MPQ$ associated with the VEVs of  $\overline{126}_H$ and $45_H$. We will now distinguish between two cases:
\newcommand{\DelR}{\langle \Delta_R\rangle}
\newcommand{\sig}{\langle \sigma\rangle}

\paragraph{Case A: $M_{\rm PQ}>M_{\rm BL}$.} 

If the $45_H$ acquires its VEV before the $126_H$, it takes part in the gauge symmetry breaking. This is because, as said before, $v_{\rm PQ}$ breaks the Pati-Salam group to $SU(4)_C\times SU(2)_L\times U(1)_R$
(see table \ref{tab:vevs}).
We are therefore confronted with the following three-step symmetry breaking chain:
\begin{eqnarray}
\label{chain2}
SO(10)&\stackrel{M_{\rm U}-210_H}{\longrightarrow}
4_{C}\, 2_{L}\, 2_{R}\, \stackrel{M_{\rm PQ}-45_H}{\longrightarrow}
4_{C}\, 2_{L}\, 1_{R}\, \stackrel{M_{\rm BL}-126_H}{\longrightarrow}
3_{C}\, 2_{L}\, 1_{Y}\, \stackrel{M_Z-10_H}{\longrightarrow} \ 3_{C}\,1_{\rm em}.
\end{eqnarray}
Both $v_{\rm PQ}$, related to $M_{\rm PQ}$, and  the VEV $v_{\rm BL}$, related to  $M_{\rm BL}$, have to be compatible with gauge coupling unification at $M_{\rm U}$. Since $v_{\rm PQ}\sim 3 f_A$ (see \eqref{eq:fA45}), this 
constrains the axion decay constant. Such constraints will  be analysed in Section \ref{gaugecouplingunification}.

\paragraph{Case B: $M_{\rm BL}>M_{\rm PQ}$.} 

In this case the $45_H$ does not take part in the gauge symmetry breaking, because  $v_{\rm BL}$  breaks the Pati-Salam group to the SM, which is preserved by the VEV $v_{\rm PQ}$ of $\sigma$. 
Hence, in these scenarios one cannot constrain the axion-decay constant $f_A$ from  unification requirements. The only limit on $v_{\rm PQ}$ is set by the requirement $v_{\rm BL}>v_{\rm PQ}$.

\subsubsection{Additional $45_H$ and extra fermions}
\label{hlsmod}
All the models analyzed so far feature $N_{\rm DW}=3$ and are therefore troubled by a domain-wall number problem, if the 
topological defects are not diluted by inflation. A variant of the model in Section \ref{altarellimodel} which does not suffer from the domain wall problem was originally proposed in \cite{Lazarides:1982tw}. It additionally contains two generations of fermions in the $10_F$ representation which become massive via Yukawa interactions with the $45_H$, 

\begin{equation}
\label{eq:PQ45fer}
\begin{array}{cc}
\begin{aligned}
  &16_F\rightarrow16_F e^{i\alpha},  \\
 & 10_H\rightarrow10_H e^{-2i\alpha},  \\
{\bf Model\ 2.2:}\hspace{6ex} & \overline{126}_H\rightarrow\overline{126}_H e^{-2i\alpha},  \\
 & 210_H \rightarrow 210_H,   \\ 
 & 45_H \rightarrow 45_H e^{4 i\alpha},  \\
  & 10_F\rightarrow10_F e^{-2i\alpha}.
\end{aligned} 
\end{array}
\end{equation}

The axion is given by  the same combination of phases as in section \ref{altarellimodel}, as the construction only depends on the scalar PQ charges. The axion decay constant can be obtained from 
equation \eqref{eq:csfA} --which, applied to models with $N_{10}$ extra fermion multiplets in the $10_F$, gives \eqref{eq:fA3}-- substituting the values of $q_i/\fPQ$ in \eqref{eq:charges2}, and using $N_{10}=2$. This gives
\begin{align}
\label{eq:fA45frt}	
f_A\equiv{f_{A,3_C}}= \sqrt{\frac{v_6^2v^2+(v_4^2+v_2^2)(v_3^2+v_1^2)}{v^2}}.
\end{align}

Due to the extra fermions, the PQ symmetry in \eqref{eq:PQ45fer} has now $\hat{N}=4$, as opposed to the previous value of 12. Again, the scalar PQ charges have a maximum common divisor of 2, so that 
the naive domain wall number is 2. When taking the quotient of the discrete symmetry group with respect to the center $Z_2$ of the gauge group, one expects then a physical domain wall number $N_{\rm DW}=1$,
which gets rid of the domain-wall problem. This can be explicitly checked using equation \eqref{eq:NDW}, which now implies the following system of equations for the $n_i$ and the minimum integer $n_{\rm min}$ that gives the domain-wall number:
\begin{equation}\begin{aligned}
n_{\rm min}+(n_1+n_2)=&\,0, &
n_{\rm min}+(n_2+n_3)=&\,0,\\
n_{\rm min}+(n_1+n_4)=&\,0, &
n_{\rm min}+(n_3+n_ 4)=&\,0,\\
n_{\rm min}-n_6=&\,0.
\end{aligned}\end{equation}

As expected, we have $N_{\rm DW}=n_{\rm min}=1$. In order to obtain an axion effective Lagrangian we need the ${\rm PQ}_{\rm phys}$ charges of the extra fermions $\tilde D, D,\tilde L,L$ in the $10_F$ (see table \ref{tab:allfermions}). 
As the scalar content of the theory is as in the previous section, we have that, as before, ${\rm PQ}_{\rm phys}$ is given by  \eqref{eq:U1s} with the $s_i$ in \eqref{eq:s2}. Using the  $1_R$ and $1_{B-L}$ assignments in table \ref{tab:allfermions}, the charges of the extra fermions  are:
\begin{equation}\label{eq:charges3}\begin{aligned}
 \frac{q_{\tilde D}}{\fPQ}=&\,-\frac{2(v_1^2+v_3^2)}{3v\sqrt{\left(v_2^2+v_4^2\right)(v_1^2+v_3^2)+v^2v^2_6 }},
 &\frac{q_D}{\fPQ}=&\,-\frac{v_1^2 +3v_2^2+v_3^2+3 v_4^2}{3v \sqrt{ \left(v_1^2+v_3^2\right) \left(v_2^2+v_4^2\right)+v^2v^2_6 }},\\
 \frac{q_{\tilde L}}{\fPQ}=&\,-\frac{(v_2^2+v_4^2)}{v\sqrt{(v_1^2+v_3^2)\left(v_2^2+v_4^2\right) +v^2v^2_6 }},
 &\frac{q_L}{\fPQ}=&\,-\frac{v_1^2+v_3^2}{v\sqrt{\left(v_1^2+v_3^2\right)(v_2^2+v_4^2)+v^2v^2_6 }}.\\
\end{aligned}\end{equation}
The axion interactions are then given by 
\begin{align}
\label{lowenergylag2}
 \mathcal{L}_{\rm int}=& \frac{1}{2}\partial_\mu A \partial^\mu A +\frac{\alpha_s}{8\pi} \frac{A}{f_A} \, G^b_{\mu\nu} \tilde G^{b\mu\nu} +\frac{\alpha}{8\pi} \frac{8}{3}\frac{A}{f_A} \,F_{\mu\nu} \tilde F^{\mu\nu}+\partial_\mu A\sum_{f=\tiny{\begin{array}{c}q,u,d,l,e,\\
         \tilde D,D,\tilde L,L                                                                                                                                                                                                                                                                                                                                                                                                                                                                        \end{array}}
} \frac{q_f}{f_{\rm PQ}} (f^\dagger \bar\sigma^\mu f),
\end{align}
with the values of  $f_A$ and $q_i/\fPQ$ given in equations \eqref{eq:fAGUT}, \eqref{eq:charges2}, and \eqref{eq:charges3}. Including QCD effects and going to the axial basis, the Lagrangian for axion, photon, nucleon and electrons now has a different relative weight between axion and fermion couplings than in the previous DFSZ-like result of \eqref{eq:DFSZ}. In the current domain-wall-free model one has now an extra factor of three in the fermionic couplings:
\begin{equation}\label{eq:DFSZ2}\begin{aligned}
 \mathcal{L}^{\rm QCD}_{\rm int}=&\frac{1}{2}\partial_\mu A \partial^\mu A 
- \frac{1}{2} m_A^2 A^2 
+ \frac{\alpha}{8\pi} \,\frac{C_{A\gamma}}{f_A}\,A\,F_{\mu\nu} {\tilde F}^{\mu\nu} \\
 &-\partial_\mu A\left[ \frac{C_{AP}}{2f_A} \overline P^\dagger \gamma^\mu \gamma_5 P+\frac{C_{AN}}{2f_A} \overline N^\dagger \gamma^\mu \gamma_5 N+\frac{C_{AE}}{2f_A} \overline E^\dagger \gamma^\mu \gamma_5 E\right],
 \\
C_{A\gamma}=&\, \frac{8}{3}-1.92(4), \\
 C_{AP}=&\,-0.91+1.30\cos^2\beta\pm0.05,\\
 C_{AN}=&\,0.81-1.24\cos^2\beta\pm0.05,\\
 C_{AE}=&\,\sin^2\beta,
\end{aligned}\end{equation}
with $\beta$ as in \eqref{eq:gamma}. Since the scalar content of the models is unchanged with respect to the previous section, the symmetry breaking chains are the same as before. A slight difference occurs in case B: If $M_{\rm BL}>M_{\rm PQ}$, the extra fermions acquire masses only below the scale $M_{\rm PQ}$. In the analysis of gauge coupling unification, one has to take into account the extra contributions of these fermions between $M_{\rm U}$ and $M_{\rm PQ}$. 

\subsection{Models with decay constants independent of the gauge symmetry breaking}
\label{independentaxion}

A third way to lift the PQ-breaking scale from the electroweak scale, also relying on a new scalar charged under the PQ symmetry and getting a large VEV, relies in the introduction of a complex 
gauge singlet scalar $S$ and exploits thus an even 
more minimal scalar sector than the previous intermediate scale axion models exploiting an additional $45_H$. However, this choice lacks the  predictivity of the previous approaches since the singlet $S$ does not participate in the gauge symmetry breaking.  As in the models containing a $45_H$, the minimal model has a domain wall problem which can be avoided by introduction of two generations of heavy fermions. Under the Peccei Quinn symmetry, the scalar fields transform as follows for the two models:
%
%

\begin{equation}
\label{eq:justindepmod}
\begin{array}{cc}
\begin{aligned}
  &16_F\rightarrow16_F e^{i\alpha},  \\
  &10_H\rightarrow10_H e^{-2i\alpha},\\
 {\bf Model\ 3.1:}\hspace{6ex} &\overline{126}_H\rightarrow\overline{126}_H e^{-2i\alpha}, \\
  &210_H \rightarrow 210_H\,,  \\
  &S \rightarrow S e^{4i\alpha}, 
\end{aligned} 
\end{array}
\end{equation}
which features $N_{\rm DW}=3$, and 
%
%

\begin{equation}
\label{eq:scalarand10}
\begin{array}{cc}
\begin{aligned}
  &16_F\rightarrow16_F e^{i\alpha},  \\
  &10_H\rightarrow10_H e^{-2i\alpha},\\
 {\bf Model\ 3.2:}\hspace{6ex} &  \overline{126}_H\rightarrow\overline{126}_H e^{-2i\alpha},  \\
  &210_H \rightarrow 210_H\,,  \\
  &S \rightarrow S e^{4i\alpha},  \\
   & 10_F\rightarrow10_F e^{-2i\alpha},
\end{aligned}
\end{array}
\end{equation}
for the model with $N_{\rm DW}=1$.

In both of these models the construction is analogous to \ref{altarellimodel} and \ref{hlsmod}, where the role of the $45_H$ is now played by the singlet $S$. Like $\sigma$ in Sections \ref{altarellimodel} and \ref{hlsmod}, $S$ is not charged under $U(1)_R$ and $U(1)_{B-L}$, so that the orthogonality conditions are unchanged. The massive combinations which the axion needs to be orthogonal to only occur at dimension 6 in the shape of the operator $\overline{126}_H\,\overline{126}_H\,10_H\,10_H\,S\,S$. The resulting system of equations however yields the same formulae as in equations \eqref{gutaxion2}, \eqref{eq:charges2}, as well as \eqref{eq:fA45} --with no additional fermions in the $10_F$-- and \eqref{eq:fA45frt} --with fermions in the $10_F$-- yet with $v_6$ and $q_6$ corresponding now to the field $S$. The calculation of the domain wall number is identical to that in Sections \ref{altarellimodel} and \ref{hlsmod}, and so is the axion effective Lagrangian. In particular, at low energies we get the DFSZ interactions in \eqref{eq:DFSZ} for the $N_{\rm DW}=3$ model, and the result in \eqref{eq:DFSZ2} for the $N_{\rm DW}=1$ case.

%% file: axion_mass_so10_constraints_unification.tex
\section{Constraints on axion properties from gauge coupling unification}
\label{gaugecouplingunification}

In this section we analyse the constraints put on the axion mass by the requirement of gauge coupling unification in the models introduced in the previous section. In each case, we take into account the running of the gauge couplings at two-loop order. A consistent analysis to this order needs to take into account one-loop threshold corrections. The general and model-specific 
$\beta$-functions and matching conditions are given in Appendix \ref{rgeevolution}.\footnote{Our analysis has been performed using \emph{Mathematica} \cite{Mathematica}. In the calculation of the beta functions, we have employed the \emph{LieART} package\cite{Feger:2012bs}, as well as our own code. 
The appearance of our plots was enhanced using Szabolcs Horvat's \emph{MaTeX} package.} 
 Since the scalar masses 
depend strongly on the parameters of the scalar potential, which are not known a priori, 
the scalar threshold corrections due to the Higgs scalars cannot be calculated exactly. Instead, 
we have assumed that the scalar masses are distributed randomly in the interval $[\frac{1}{10}M_{T},10M_{T}]$, where $M_{\mathrm T} (\mathrm{T} \in \{\mathrm{U,BL,PQ}\})$ is the threshold at which these particles acquire their masses. For the RG running, we have employed a modified version of the \emph{extended survival hypothesis} \cite{delAguila:1980qag}. According to the latter, scalars get masses of
the order of their VEVs, so that the scalars remaining active in the RG at a given scale are those whose VEVs lie below that scale. We consider however an exception \cite{Babu:2015bna,Babu:1992ia}: in order to have a 2HDM at low energies,
we will assume that $\Sigma_u$ and $\Sigma_d$, the SM doublets in the $(15,2,2)_{PS}$ of the $\overline{126}_H$, have masses of the order of $\MBL$. Such a choice is not arbitrary. First, as commented in Section \ref{thecaseforapqsymmetry}, realistic fermion masses require
VEVs $v^{126}_{u,d}$ of the order of the electroweak scale for the previous fields. Small VEVs for massive fields can be achieved through mixing with the light doublets $H_u,H_d$ in the $10_H$, which themselves must acquire electroweak VEVs $v^{10}_{u,d}$. The mixing can be induced by a PQ invariant operator such as  $10_H\, \,\overline{126}_H^\dagger \,\overline{126}_H \,  \overline{126}_H^\dagger$, which gives VEVs $v^{126}_{u,d}$ of the order of $v_{\rm BL}^2/M^2_{(15,2,2)} v^{10}_{u,d}$, where  $M_{(15,2,2)}$ is the mass
 of the $(15,2,2)_{\rm PS}$ multiplet \cite{Babu:2015bna,Babu:1992ia}. If the mass is of the order of $v_{\rm BL}$, the desired electroweak-scale VEVs are achieved.
 Taking into account the scalar content of each model, the surviving multiplets can be read from table \ref{tab:allvevs}. The RG equations also take into account the fermionic representations --three generations of fermions in the $16_F$ representations, and two additional generations in the $10_F$ in the models defined in equations \eqref{eq:PQ45fer}  and \eqref{eq:scalarand10}.

To sharpen the predictions of our models, we take into account constraints from the non-observation of proton decay, bounds on the B-L breaking scale obtained from fits to fermion masses, as well as black hole superradiance and stellar cooling constraints.

In regards to proton decay, we use a naive estimate for its lifetime, considering  only the decay mediated by superheavy gauge bosons \cite{DiLuzio:2011my}. We approximate the lifetime of the proton by $\tau \sim \frac{\MU^4}{m_p^5 \alpha_\mathrm{U}^2}$ (for $m_p=0.94 \GeV$) and compare it to the current experimental limits \cite{Miura:2016krn} $\tau(p\rightarrow\pi^0e^+)>1.6\times 10^{34}{\rm y}$. 
In subsequent plots, constraints imposed by current limits from proton decay will be shown in blue.  

The constraints on the B-L scale in $SO(10)$ models can be obtained by fitting the observed values of  fermion masses and mixing angles to the relationships implied by the gauge symmetry (eq. \eqref{nonsusysumrule}). Such fits have been performed for example in \cite{Dueck:2013gca} and \cite{Joshipura:2011nn}. In the former the fit was performed at the weak scale, while in  the latter it was done at the GUT scale. As in the models in our analysis, \cite{Joshipura:2011nn} considered a two-Higgs-doublet model at low scales above $M_Z$. Both studies only considered the scalar fields contributing to the Yukawa interactions  --in our model  the $10_H$ and the $\overline{126}_H$-- since these are largely model independent. In both cases the analysis yielded an upper bound on the B-L breaking scale of about $3\times 10^{15}\GeV$.  The final formula for the B-L breaking VEV can be derived from  \eqref{nonsusysumrule} and the seesaw formula, and it includes two mixing angles $\beta$ and $\gamma$:
\begin{align}
v^R_{126}=v_{\mathrm{BL}}=3\times  \sin \gamma  \cos \beta \times 10^{15} \GeV, 
\end{align}
where we have defined
\begin{equation}\begin{aligned}
\tan\beta&=\frac{v_u}{v_d}=\frac{\sqrt{(v_u^{10})^2+(v_u^{126})^2}}{\sqrt{(v_d^{10})^2+(v_d^{126})^2}},\\
\tan \gamma&=\frac{v_d^{126}}{v_d^{10}}.
\end{aligned}\end{equation}

Since the fits only determine the ratios $\frac{v_u^{126}}{v_d^{126}}$ and $\frac{v_u^{10}}{v_d^{10}}$, the two factors $\sin \gamma$  and $\cos \beta$ are not constrained. Allowing for some fine tuning -- as it is customary in $SO(10)$ models-- the B-L breaking scale can be lowered to  $10^9 \GeV$. For each of our models we have considered different levels of fine tuning in this sector, allowing $v_{\rm BL}$ to be within windows with an upper value of $10^{15} \GeV$ and a lower value of either $10^9,10^{11}$ or $10^{13}\,\GeV$. In the figures of the rest of the section, constraints imposed by the B-L scale will be shown in green.  

Finally, black hole superradiance constraints arise from the fact that axion condensates around black holes can affect their rotational dynamics and the emission of gravitational waves \cite{Arvanitaki:2009fg,Arvanitaki:2010sy,Arvanitaki:2014wva}. We will show the associated constraints in black. Bounds from stellar cooling arise from taking into account the loss of energy by axion emission due to photon axion-conversion in helium-burning horizontal branch stars in globular clusters  \cite{Ayala:2014pea}. Such constraints will be shown in gray.

\subsection{Running with one intermediate scale}
\label{runninggutscale}
Let us first consider Model 1 described by \eqref{eq:PQGUT} with PQ charged scalars in the $210_H$, $\overline{126}_H$ and $10_H$ representations. 
 \begin{figure}[h]
\begin{centering}
\includegraphics[width=0.7\textwidth]{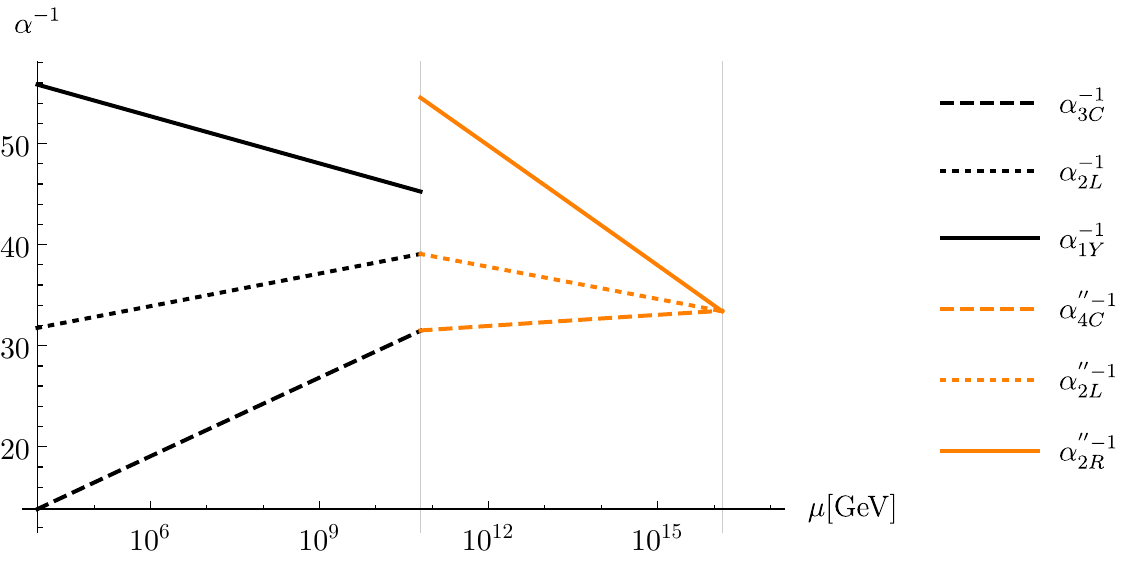}\\
\end{centering}
\caption{\label{fig:gcuplot} Running and gauge coupling unification in Model 1 in the case of minimal threshold corrections.}
\end{figure}
Figure \ref{fig:gcuplot} shows the predicted running of the gauge couplings for the case of 
minimal threshold corrections, in which all scalar masses are degenerate with the corresponding gauge boson masses.
\begin{figure}[h]
\begin{centering}
\begin{minipage}[t]{0.31\textwidth}
\includegraphics[width=\textwidth]{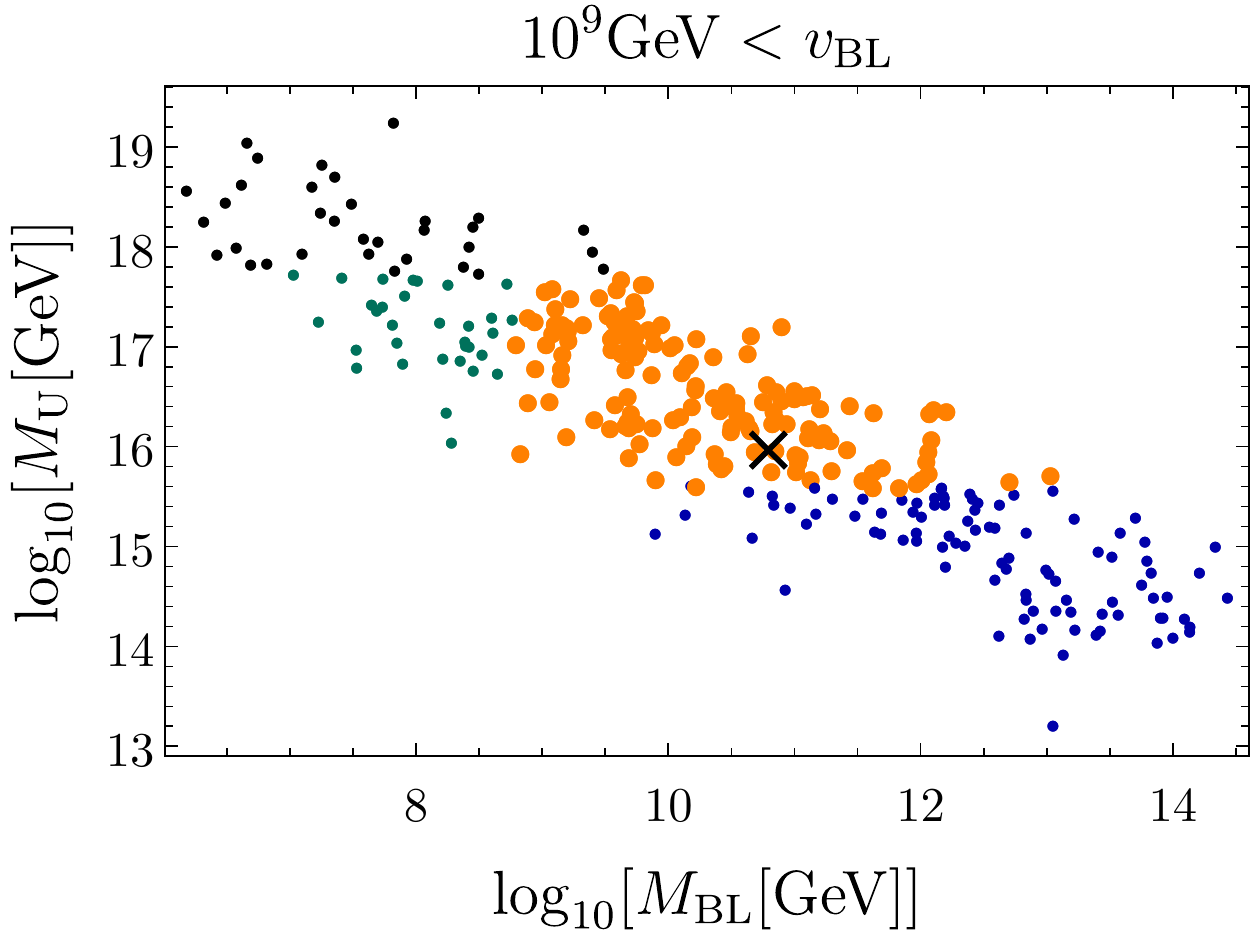}
\end{minipage}
\begin{minipage}[t]{0.31\textwidth}
\includegraphics[width=\textwidth]{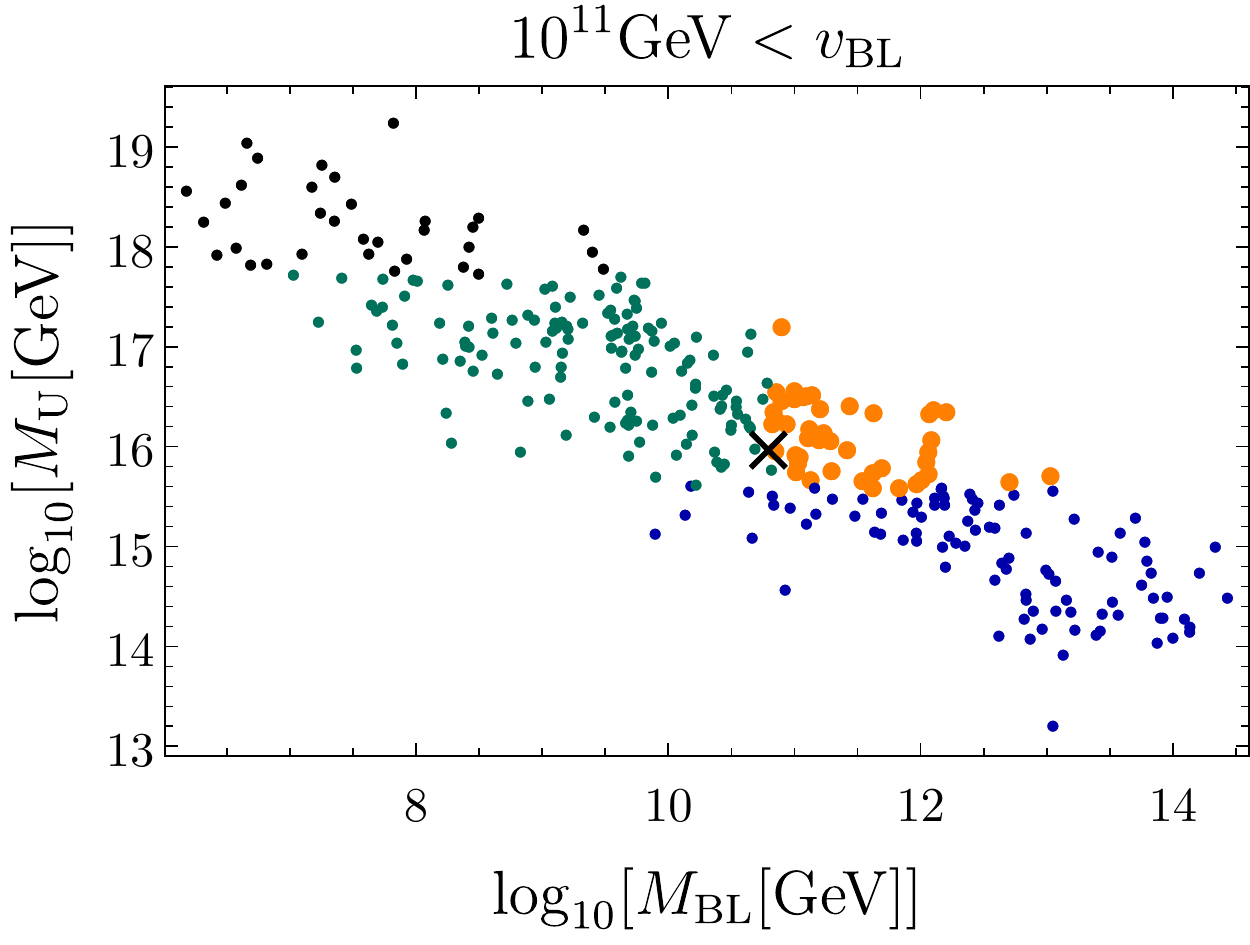}
\end{minipage}
\begin{minipage}[t]{0.31\textwidth}
\includegraphics[width=\textwidth]{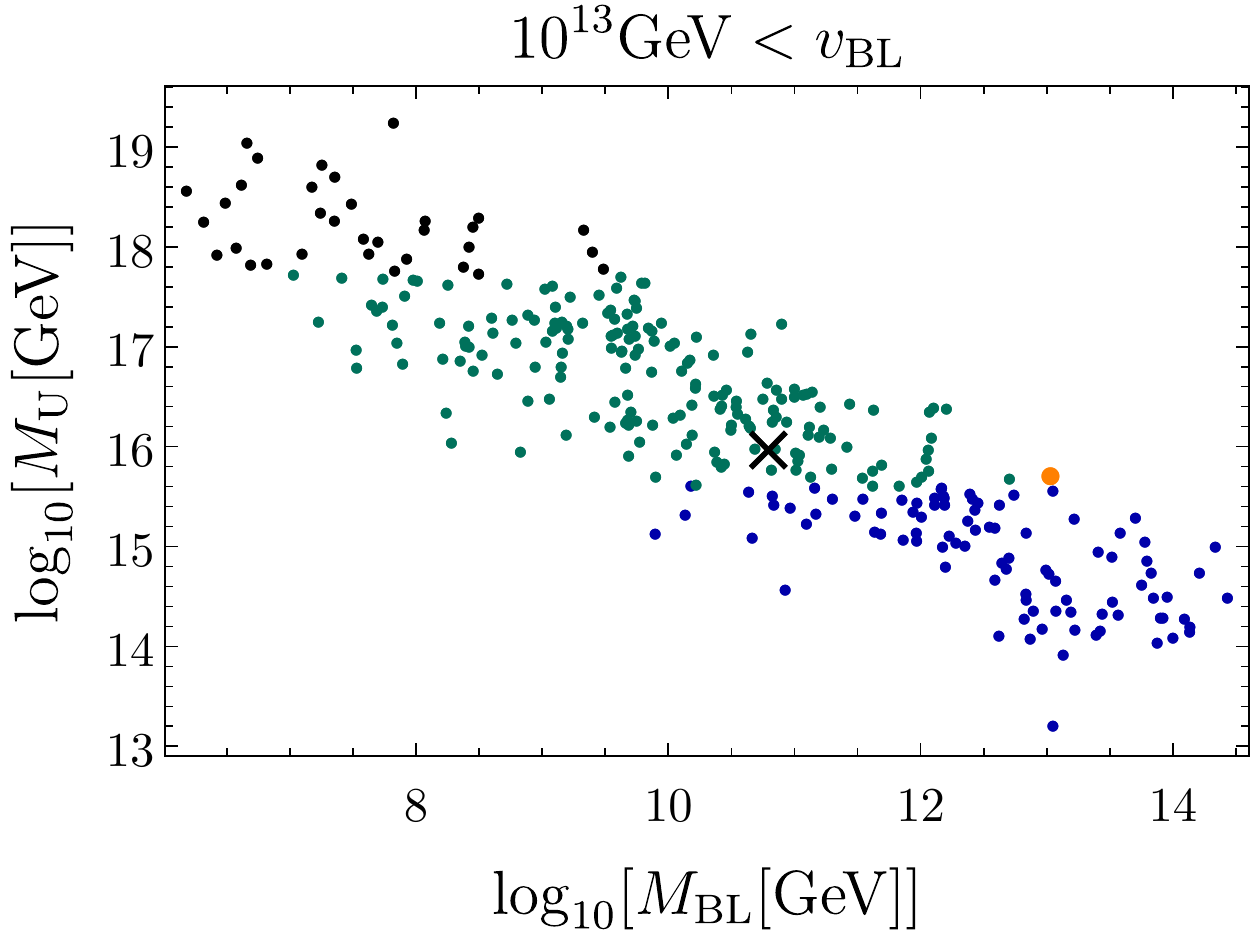}
\end{minipage}
\end{centering}
\caption{\label{fig:mumiplot210} Intermediate and unification scale for randomised scalar threshold corrections in Model 1.
Only the large orange points are not excluded by our constraints. Points in blue are excluded by proton decay limits, points in black are excluded by the limits from black hole superradiance constraints. Points in green are allowed by black hole superradiance and proton decay, but forbidden by the chosen range of B-L breaking. The black cross indicates the minimal threshold case, i.e. the case when all scalar masses are degenerate and at the corresponding unification scales for which the running of the gauge couplings is illustrated in figure \ref{fig:gcuplot}. We have performed the scan for 400 sets of initial conditions, 310 of which yielded unification of the gauge couplings.}
\end{figure}
 Gauge coupling unification fixes the different scales in this case to 
\begin{equation}
\MU=\MPQ=1.4\times 10^{16}\GeV,\  \alpha_\mathrm{U}(\MU)^{-1}=33.6,\   \MBL= 6.3\times 10^{10}\GeV.
\end{equation}
The unification scale is well above constraints from proton decay. 
\begin{figure}[h]
\begin{centering}
\includegraphics[width=\textwidth]{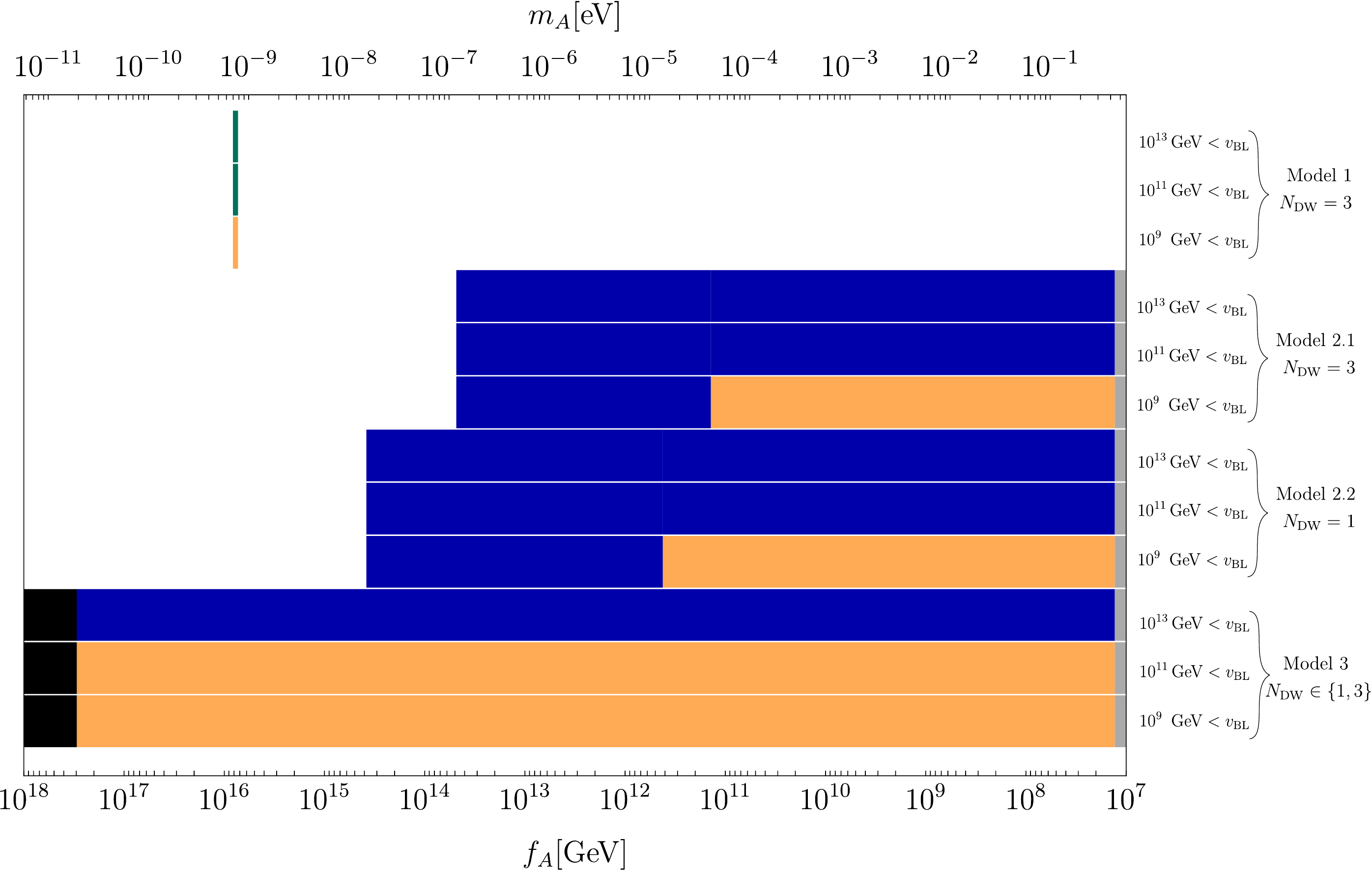}\\
\end{centering}
 \caption{\label{fig:summarylongnothresholds} 
Possible ranges of the axion mass and decay constant consistent with gauge coupling unification in our models,  
for the case where all heavy scalars are degenerate at their various threshold scales. Regions in black are excluded by constraints from black hole superradiance, regions in blue by proton stability constraints. Regions in green are disfavoured depending on the allowed range of the B-L breaking scale. Regions in gray are excluded by stellar cooling constraints. The width of the region in Model 1 is exaggerated to make the bar visible. Note that for the Models 2.1, 2.2 and 3 the exclusion of the higher B-L breaking scales comes from an interplay of the proton stability constraint and the limit on $v_{\rm BL}$. In  all these models, a higher B-L breaking scale corresponds to a lower GUT unification scale, which leads to an instability of the proton and is therefore excluded. 
  }
\end{figure}
Exploiting the relation 
\begin{equation}
\MU=g_Uv_{\rm U} , 
\end{equation}
between the mass of the superheavy gauge bosons and the VEV $v_{\rm U}$ and the relation  \eqref{eq:fAGUT} 
between the axion decay constant and the VEVs, we obtain  
\begin{align}
\label{eq:fA_MU_model1}
   f_A\simeq \frac{1}{3}\,v_{\rm U} =\frac{\MU}{3 g_U}=\frac{\sqrt{\alpha_U(\MU)^{-1}}}{3 \sqrt{4\pi}}\MU=7.7\times 10^{15} \GeV,
\end{align}
yielding, via \eqref{zeroTma}, an axion mass  
\begin{align}
   m_A= 7.4\times 10^{-10}\,\eV.
\end{align}
This result is illustrated in the first three lines of figure \ref{fig:summarylongnothresholds}, which summarises our results for the case of vanishing threshold corrections.

As illustrated in figure \ref{fig:mumiplot210} and as already pointed out in \cite{Dixit:1989ff}, taking 
into account the possibility of scalar threshold corrections induces large uncertainties in the prediction of the GUT scale, which result in corresponding large uncertainties in the prediction of the axion mass. 
Including constraints from proton decay limits and the non-observation of black hole superradiance, the allowed
range is
\begin{equation}\begin{aligned}
2.6 \times 10^{15} \GeV <f_A<3.0\times 10^{17}\text{GeV},\\
1.9 \times 10^{-11}\eV < m_A<2.2\times 10^{-9} \eV.
\end{aligned}\end{equation}
Finally, we have considered the various constraints imposed by the B-L breaking scale. As shown in figure \ref{fig:mumiplot210}, varying the allowed range of $v_{\mathrm{BL}}$ changes the viable range of $v_{\mathrm{PQ}}$and therefore of $f_A$.  For $\vbl>10^{11}(10^{13})\GeV$, the upper bound on $f_A$ is lowered to $ 1.0\times 10^{17}(3.5\times 10^{15})\GeV$. In the latter case, our random sample contains only two viable points (cf. figure \ref{fig:mumiplot210}).
\begin{figure}[h]
\begin{centering}
\includegraphics[width=\textwidth]{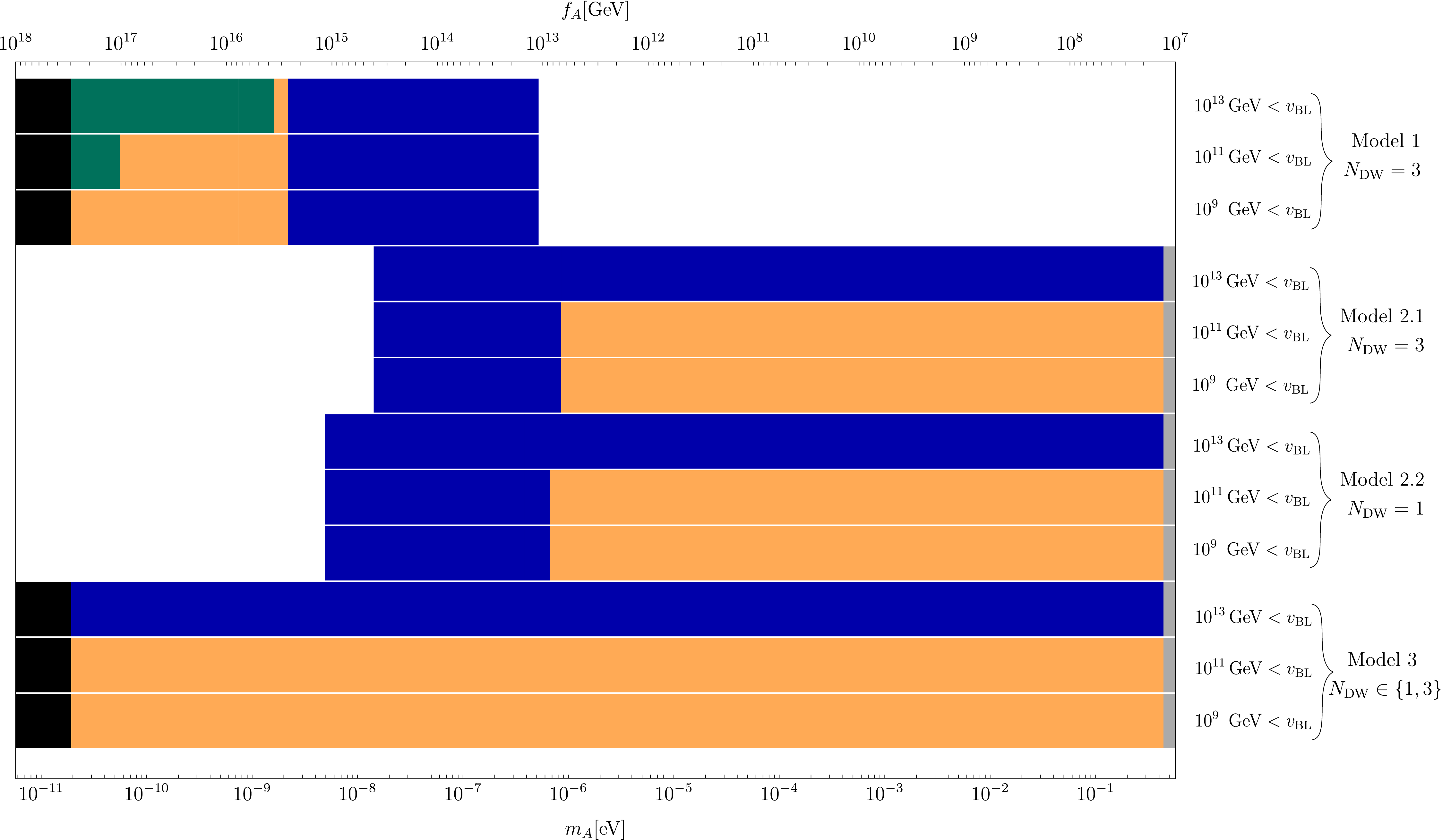}\\
\end{centering}
 \caption{\label{fig:summarylong} 
Possible ranges of the axion mass and decay constant consistent with gauge coupling unification in our  models,  
for the case where scalar threshold corrections have been taken into account. 
Regions in black are excluded by constraints from black hole superradiance, regions in blue by proton stability constraints. Regions in green are disfavoured depending on the allowed range of the B-L breaking scale. 
Regions in gray are excluded by non-observation of excessive cooling of helium burning stars by axion emission.
  }
\end{figure}
These findings are 
summarised in the first three lines of figure \ref{fig:summarylong}.

\subsection{Running with two intermediate scales}
\subsubsection{An extra multiplet}
In the Model 2.1 described in \eqref{eq:PQ45}, the requirement of gauge coupling unification does not sufficiently constrain the system of differential equations to uniquely fix both intermediate scales - we can only infer a relationship between the three unification scales $\MPQ,\MBL$ and $\MU$. We have calculated this relationship, and also imposed the aforementioned limits on the unification scale and on the B-L breaking scale. 

Depending on which VEV is bigger, the RG running is different. In  case B, i.e. $\MPQ$<$\MBL$, the Peccei-Quinn breaking VEV does not break any gauge symmetries, and thus  it is unconstrained by the  evolution. In this case, we have essentially a two-step breaking model, in which the two symmetry breaking scales $\MU$ and $\MBL$ are fixed. In case A, with $\MPQ$>$\MBL$, $v_{\mathrm{PQ}}$ breaks the $SU(2)_R$ gauge symmetry and is therefore constrained by the  evolution. 
\begin{figure}
\begin{centering}
\includegraphics[width=0.7\textwidth]{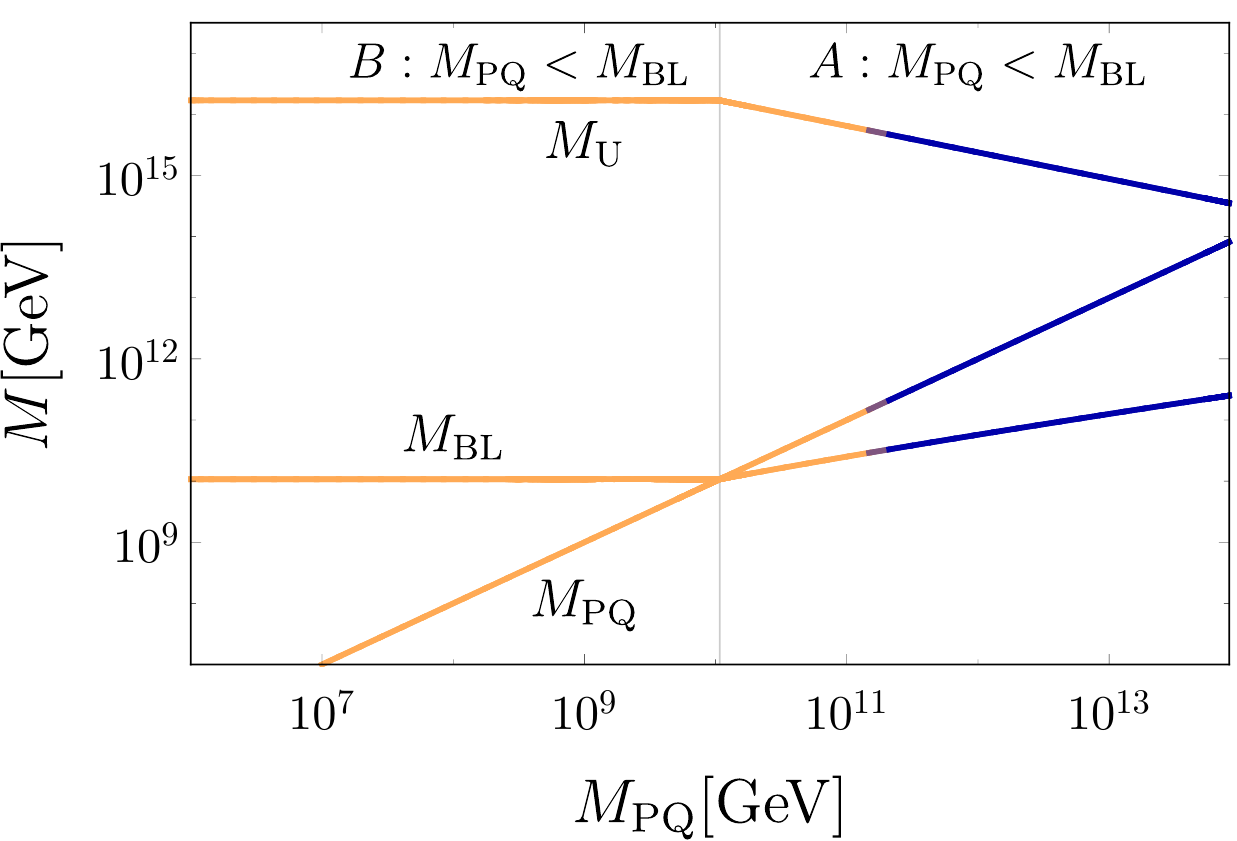}\\
\end{centering}
 \caption{\label{fig:m21nothresh} Relationship between the three unification scales for the Model 2.1 described in \eqref{eq:PQ45} in the case of minimal threshold corrections. The regions in blue are excluded by the non-observation of proton decay. In  case B, $\MPQ<\MBL$, $\MPQ$ is unconstrained - it can take any value, while the B-L breaking scale $\MBL$ is fixed at $\sim 10^{10}\GeV$. }
\end{figure}
Both cases are indicated in figure \ref{fig:m21nothresh}.

In the case of minimal threshold corrections, in which all scalars are assumed to be degenerate in mass with the gauge bosons that get masses at the corresponding threshold scale, gauge coupling unification and limits from proton decay constrain the intermediate scale $\MBL$ between
\begin{equation}
10^{10}\GeV\lesssim \MBL\lesssim  2.3 \times 10^{10}\GeV
\end{equation} 
and put an upper bound on  $\MPQ$ of order 
\begin{equation}
\label{upper_bound_mpq}
\MPQ < 1.3\times10^{11}\GeV,
\end{equation} 
cf. figure \ref{fig:m21nothresh}. 
 \begin{figure}[h]
\begin{centering}
\includegraphics[width=0.7\textwidth]{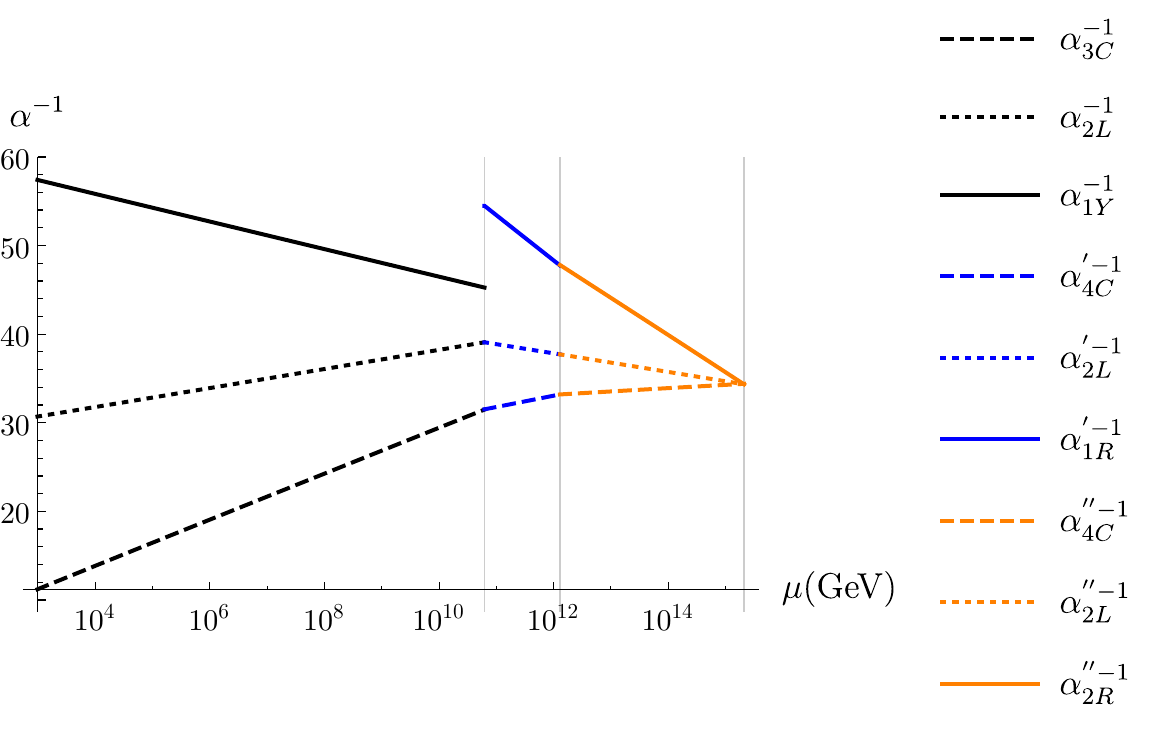}\\
\end{centering}
 \caption{\label{fig:3stepbreaking} \noindent Running gauge couplings for $M_{\rm BL}=6.3\times 10^{10}\GeV$ in Model 2.1. The corresponding higher unifications scales are $M_{\rm PQ}=1.3\times 10^{12} \GeV$ and $M_{\rm U}=2.1 \times 10^{15} \GeV$.
  Threshold corrections due to non-degenerate scalars are not included. Beta functions for this model are given in Appendix \ref{rgeevolution}. }
\end{figure}
An example of the evolution of the gauge couplings in this case is shown in figure \ref{fig:3stepbreaking}. 
For completeness, let us also mention the special case in which the PQ  and the B-L scales are taken to coincide \footnote{See also  \cite{Altarelli:2013aqa}, which considered the same model, however only taking into account only one-loop running and 
a single Higgs doublet at the weak scale. They find in this case $\MPQ=1.3\times 10^{11} \GeV$ and $\MU=1.9\times 10^{16} \GeV$.}, $\MPQ=\MBL$. We find in this case, for minimal threshold corrections, 
\begin{equation} 
\MPQ= \MBL=1.1\times 10^{10} \GeV, \hspace{3ex} \MU=1.6 \times 10^{16} \GeV.
\end{equation}

The upper limit \eqref{upper_bound_mpq} on $\MPQ$ -- derived by proton decay constraints in the case of minimal threshold corrections -- can be turned 
into an upper limit on the axion decay constant and a corresponding lower limit on the axion mass as follows.  
Since $\MPQ$ is the mass of the gauge bosons that become heavy by the $SU(2)_R \rightarrow U(1)_R$ breaking, we get
\begin{equation}
\label{mpq_vsigma}
\MPQ= g_R v_{\rm PQ}.
\end{equation} 
The corresponding limit on the axion mass follows straightforwardly from 
  \begin{equation}\begin{aligned}
   f_A&=\frac{1}{3}v_{\rm PQ}=\frac{\MPQ}{3g_R}= \frac{ \MPQ}{3\sqrt {4\pi}}\sqrt{\alpha_R^{-1}}\,<\,1.4\times 10^{11}\,\GeV\\
   m_A&>4.1 \times 10^{-5} \,\eV.
 \end{aligned}\end{equation}
This limit is illustrated in lines 4 to 6 of figure \ref{fig:summarylongnothresholds}.
This constraint however is still subject to potentially large corrections from scalar threshold effects. 
\begin{figure}[h]
\begin{centering}
\begin{minipage}[t]{0.31\textwidth}
\includegraphics[width=\textwidth]{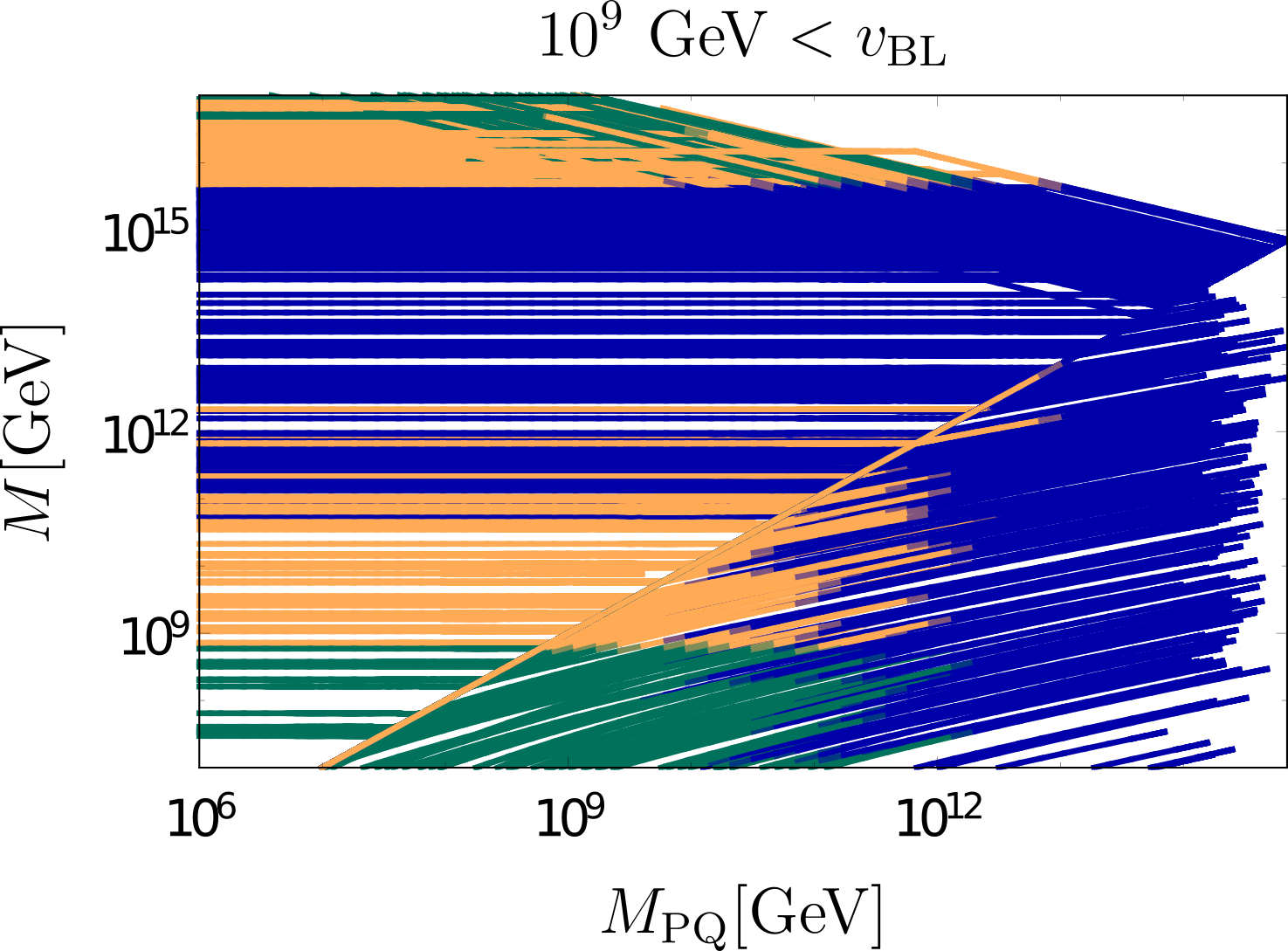}

\end{minipage}
\begin{minipage}[t]{0.31\textwidth}
\includegraphics[width=\textwidth]{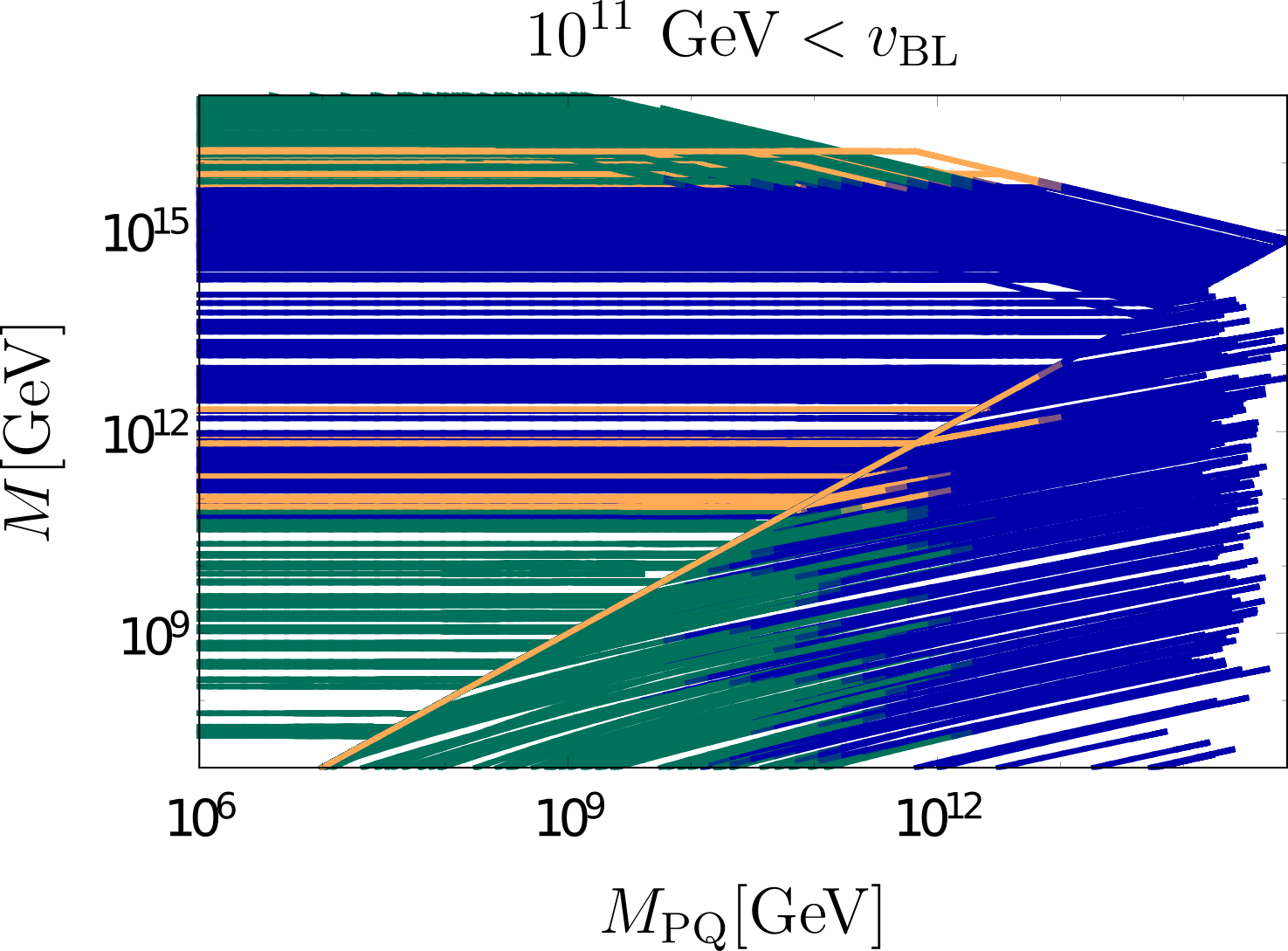}

\end{minipage}
\begin{minipage}[t]{0.31\textwidth}
\includegraphics[width=\textwidth]{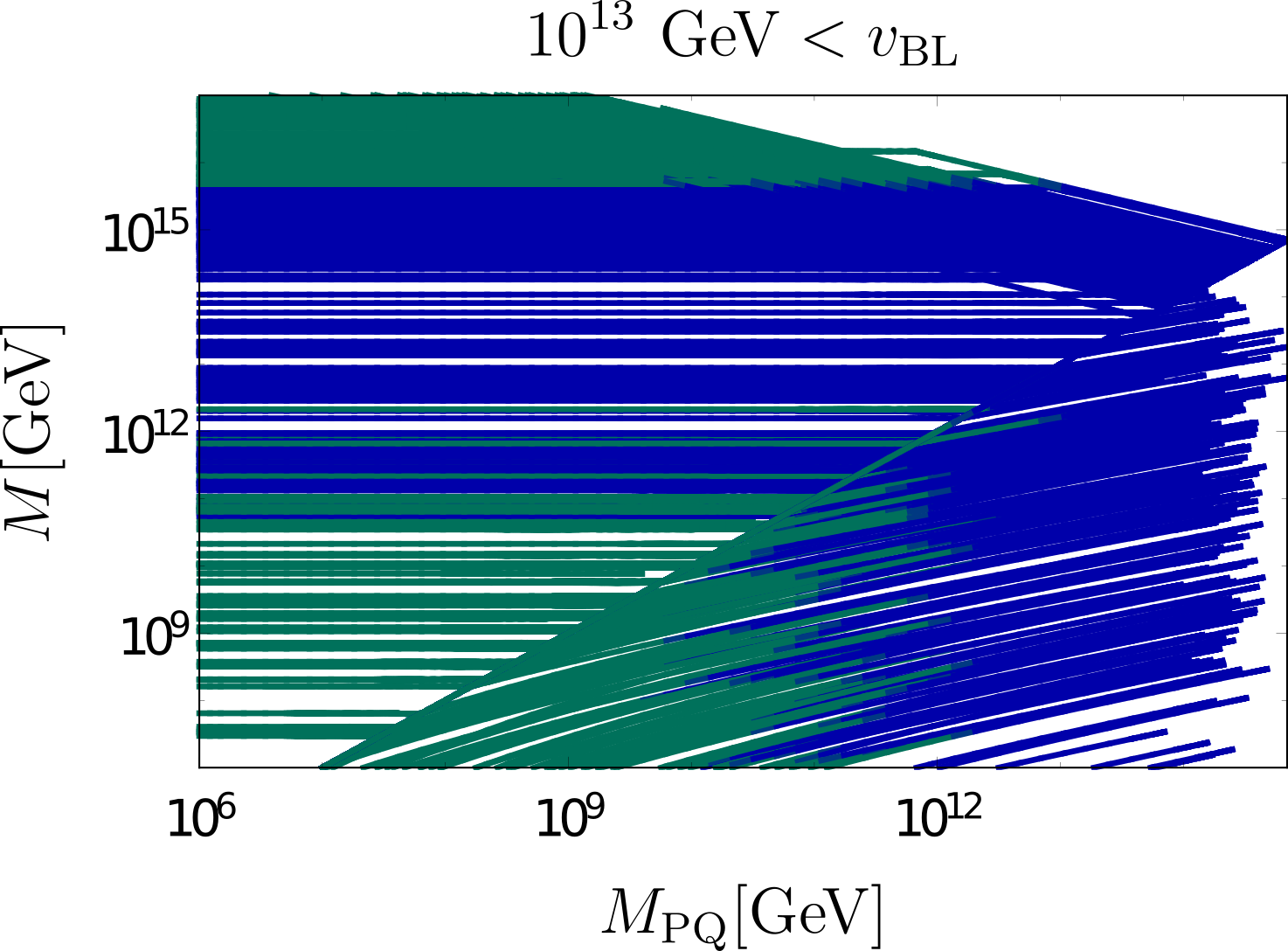}

\end{minipage}
\end{centering}
 \caption{\label{fig:threescalerunning} Intermediate scales $\MPQ$ and $\MBL$ and GUT-scale $\MU$ for different threshold corrections in Model 2.1.  Refer to figure \ref{fig:m21nothresh} for clearer view of the different scales.  The curves in blue are excluded by gauge-mediated proton decay limits, the curves in green by the limit on the B-L breaking scale. We have considered three different ranges of allowed B-L breaking scales. The threshold corrections are randomised in the following way: All scalars take masses among the values $\{ \frac{1}{10},1,10\}$ times the corresponding threshold scale, where we have taken care not to make proton decay mediating scalars contained in the $(6,2,2)$ (Pati-Salam) multiplet light. We have chosen this discrete set of masses in order to focus on the largest possible corrections coming from the mass degeneracies. The scan was performed using 240 different sets of threshold corrections. Allowing the scalars to take masses in the whole interval $[\frac{1}{10},10]$ times the threshold scale, one could "fill the gaps" and find even more compatible solutions. These would however not significantly increase the allowed region of $M_{\rm PQ}$, whose upper and lower limits we are interested in.}
\end{figure}
In fact, in figure \ref{fig:threescalerunning} we display the relation between the different unification scales --as in figure \ref{fig:m21nothresh}-- but now for randomised scalar threshold corrections for different ranges of $v_{\rm BL}$. Obviously, 
the threshold corrections can increase the bound on $\MPQ$ and thus on $f_A$.  
For $10^9(10^{11}) \GeV< \vbl<10^{15} \GeV$, we get $f_A<6.7 \times 10^{12}\GeV$ and $m_A> 8.5 \times 10^{-7} \eV $, while for $10^{13} \GeV< \vbl<10^{15} \GeV$ no allowed range of $f_A$ remains - in this case, the model is excluded. These findings are summarised in lines 4 to 6 of figure \ref{fig:summarylong}.

\subsubsection{An extra multiplet, and additional fermions}
Model 2.2 with PQ charges given in \eqref{eq:PQ45fer} contains additional quarks which acquire masses at the scale $\MPQ$. Above $\MPQ$, they contribute to the running of the coupling constants (cf. Appendix \ref{rgeevolution}). Correspondingly, in this model we obtain a relation between $\MPQ$ and $\MBL$ even in the case where $\MPQ$ does not break a gauge symmetry. 
\begin{figure}[h]
\begin{centering}
\begin{minipage}[t]{0.31\textwidth}
\includegraphics[width=\textwidth]{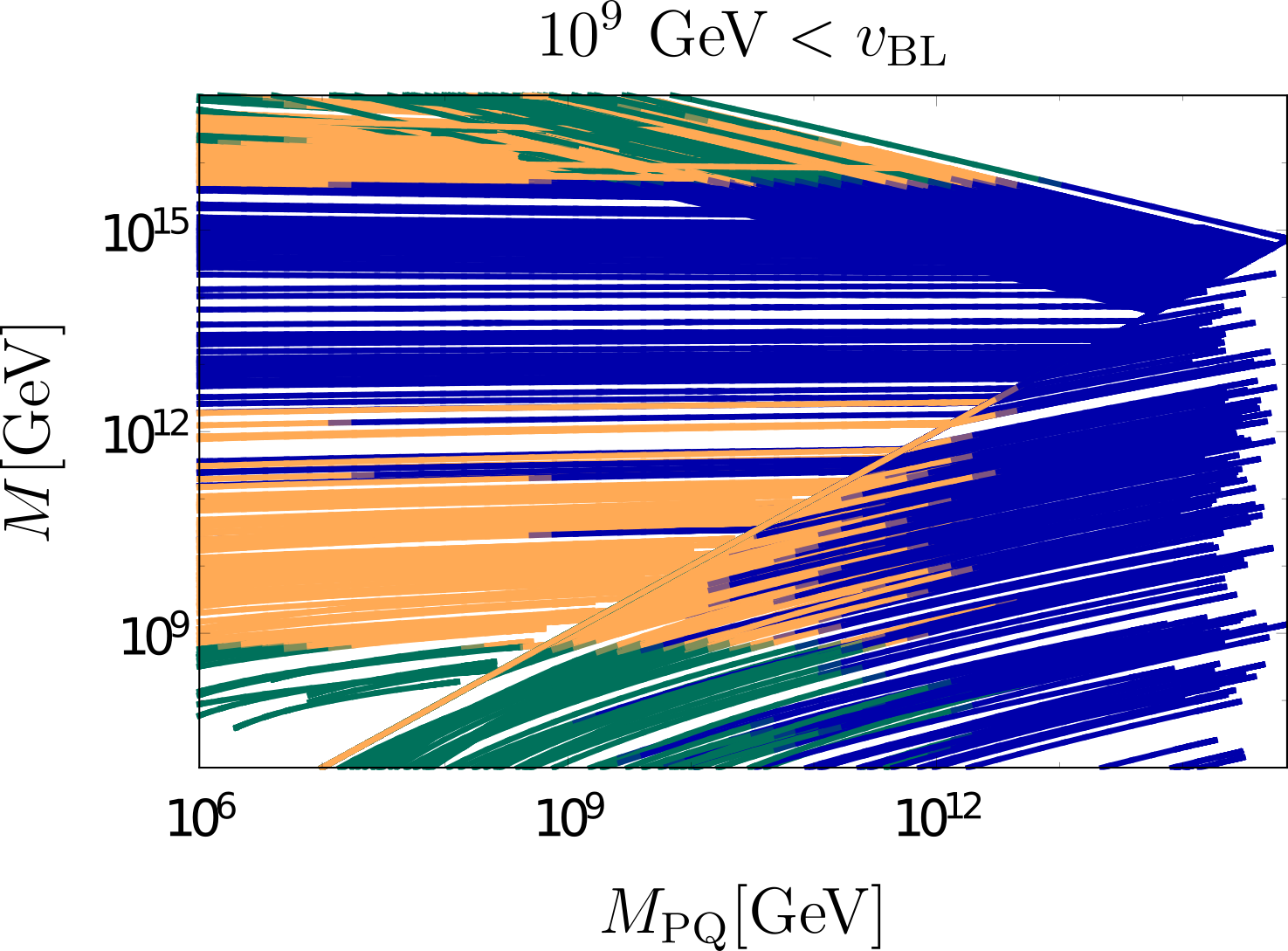}

\end{minipage}
\begin{minipage}[t]{0.31\textwidth}
\includegraphics[width=\textwidth]{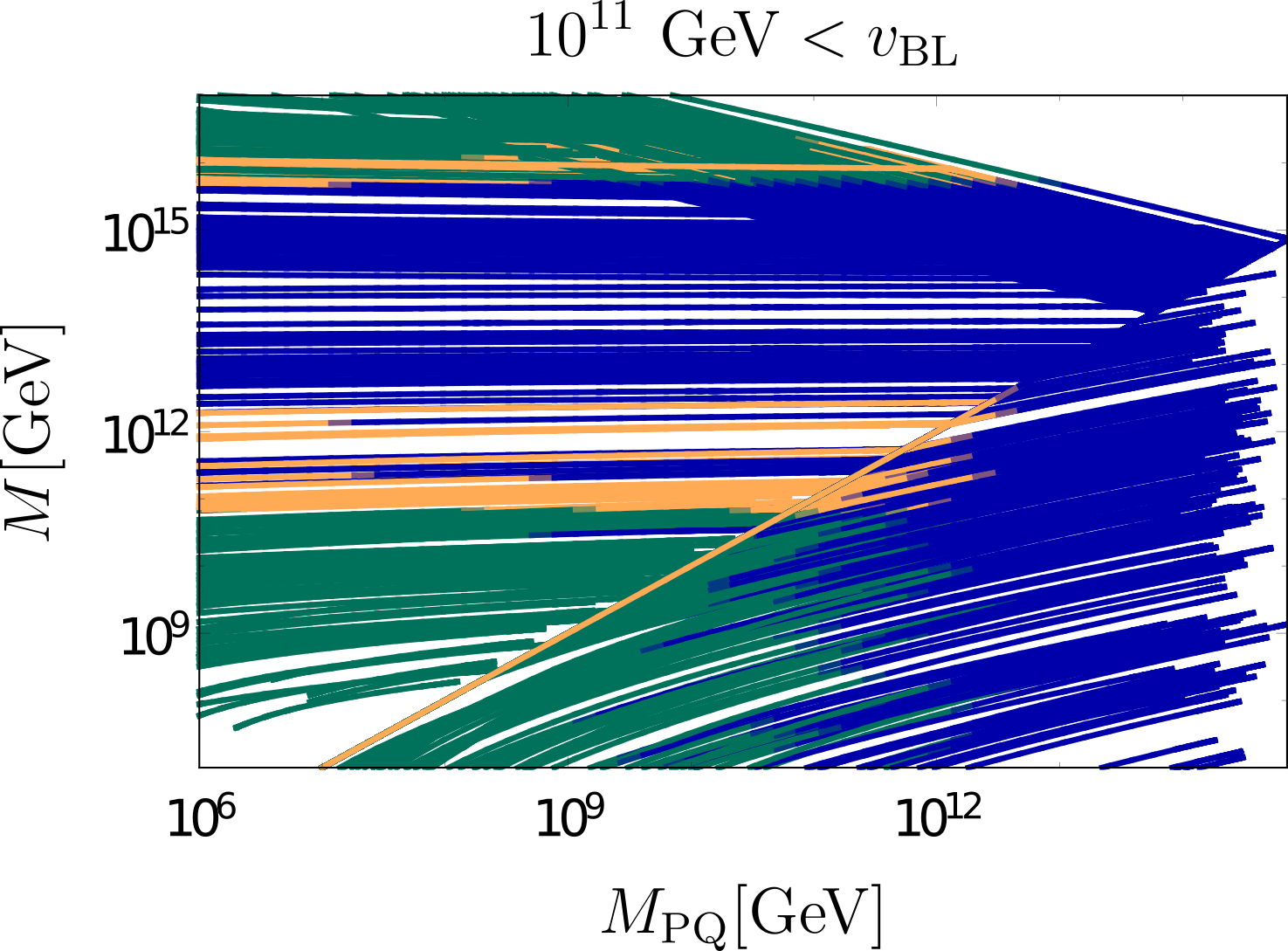}

\end{minipage}
\begin{minipage}[t]{0.31\textwidth}
\includegraphics[width=\textwidth]{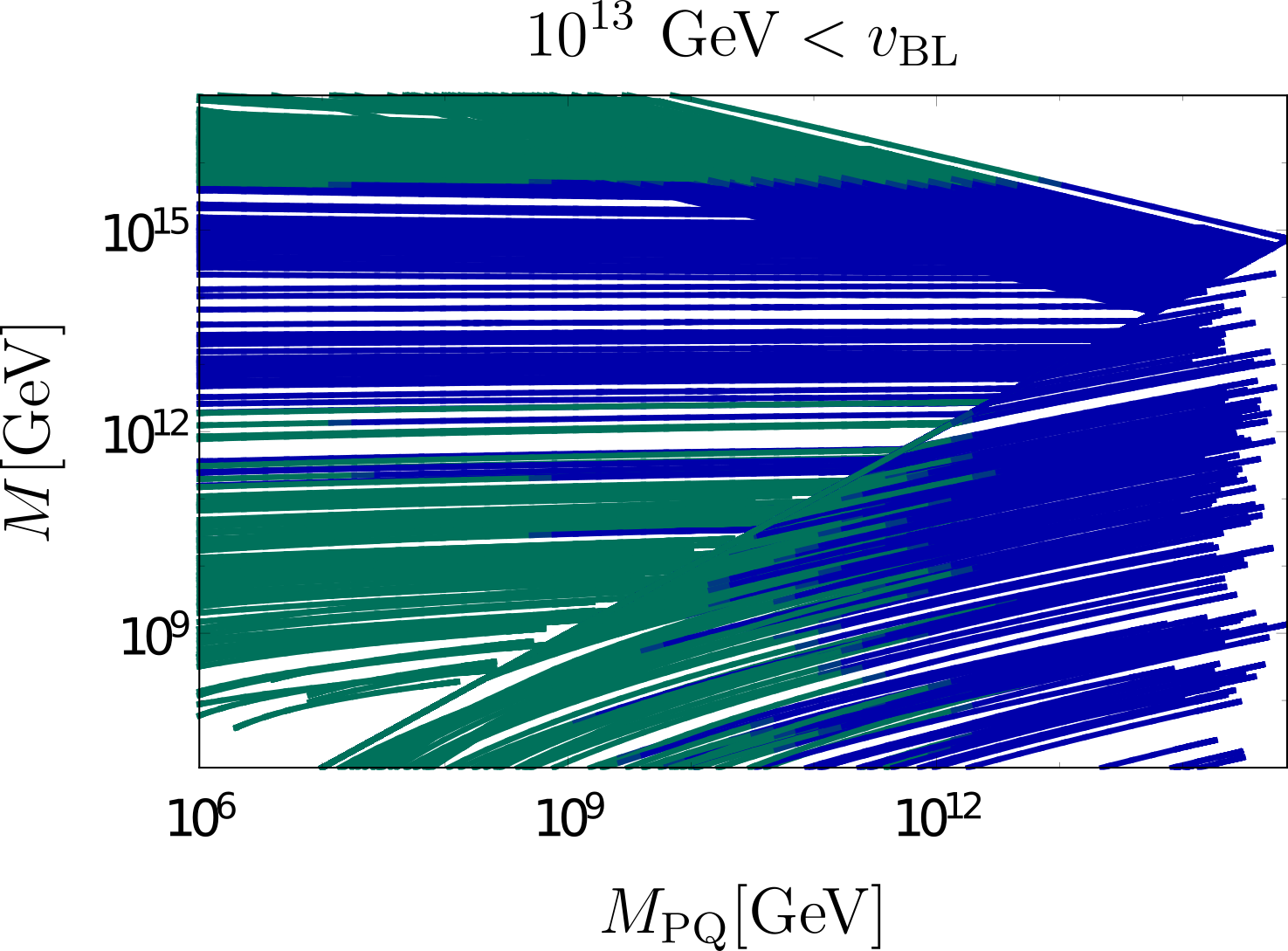}

\end{minipage}
\end{centering}
 \caption{\label{fig:hlsthreescale} Intermediate scales $\MPQ$ and $\MBL$ and GUT-scale $\MU$ for Model 2.2 for different threshold corrections.  The curves in blue are excluded by gauge-mediated proton decay limits, the curves in green by the limit on the B-L breaking scale. We have considered three different ranges of allowed B-L breaking scales. The threshold corrections are randomised in the following way: All scalars take masses among the values $\{ \frac{1}{10},1,10\}$ times the corresponding threshold scale, where we have taken care not to make proton decay mediating scalars contained in the $(6,2,2)$ (Pati-Salam) multiplet light.
}
\end{figure}
The corresponding plots are shown in figure \ref{fig:hlsthreescale}. After constraining the B-L breaking scale we obtain an upper limit on the axion mass and decay constant in this model. 
The minimal threshold case is only allowed if $\MBL$ can be as low as $10^9 \GeV$, in this case the maximal allowed $f_A$ is $4.2\times10^{11} \GeV$ -- this is also shown in lines 7 to 9 of figure \ref{fig:summarylongnothresholds}.

Including all threshold corrections in our random sample, 
for $10^9(10^{11}) \GeV< \vbl<10^{15} \GeV$, $f_A$ is constrained to be smaller than $8.6\times 10^{12} \GeV$. For $10^{13} \GeV< \vbl<10^{15} \GeV$, no viable solutions were found in the sample - the model is strongly disfavoured in this case.  The results on the axion decay constant are summarised for this model in lines 7 to 9 of figure \ref{fig:summarylong}.

\subsection{Models with a scalar singlet}

In the simplest Model 3.1, described in \eqref{eq:justindepmod}, the Peccei-Quinn breaking is driven by a scalar singlet and the axion mass is unconstrained, 
cf. line 12 in figure \ref{fig:summarylong}. There is no relation between the PQ breaking scale and the two other scales. The possible ranges for $\MU$ and $\MBL$ however can be read from figure \ref{fig:mumiplot210} --the extra scalar singlet in this model does not change the  running. Also in this model a lower B-L breaking scale is preferred, and the model is excluded if $v_{\rm BL}>10^{13} \GeV$ is imposed.

If, however, we employ the mechanism of reference \cite{Lazarides:1982tw} to reduce the domain wall number and introduce additional heavy fermions (Model 3.2, \eqref{eq:scalarand10}), one has to account for how the latter change the running of the gauge couplings above the scale $M_{\rm PQ}$ at which they acquire their masses, if $M_{\rm PQ}<M_{\rm U}$. 
\begin{figure}[h]
\begin{centering}
\begin{minipage}[t]{0.31\textwidth}
\includegraphics[width=\textwidth]{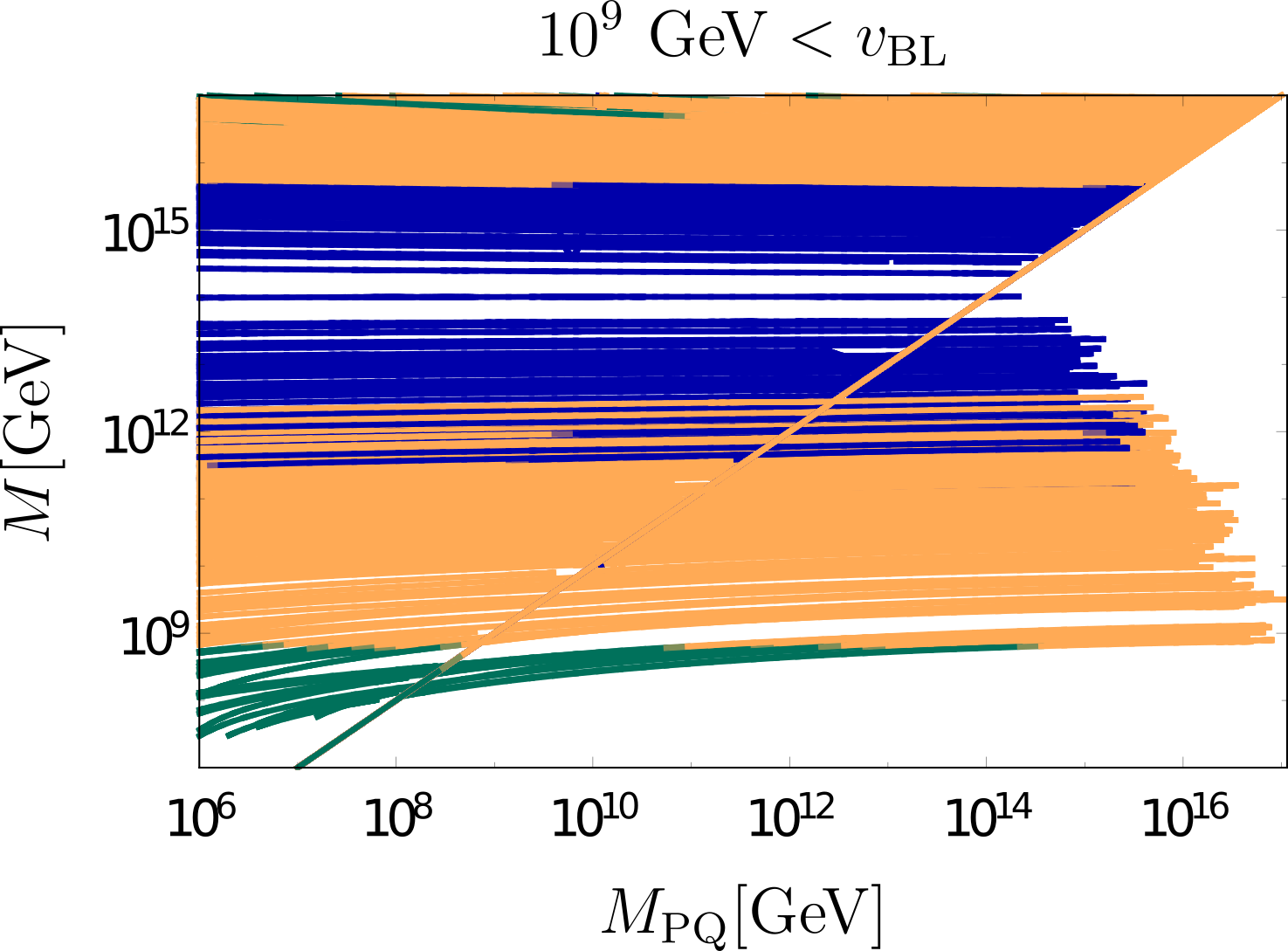}

\end{minipage}
\begin{minipage}[t]{0.31\textwidth}
\includegraphics[width=\textwidth]{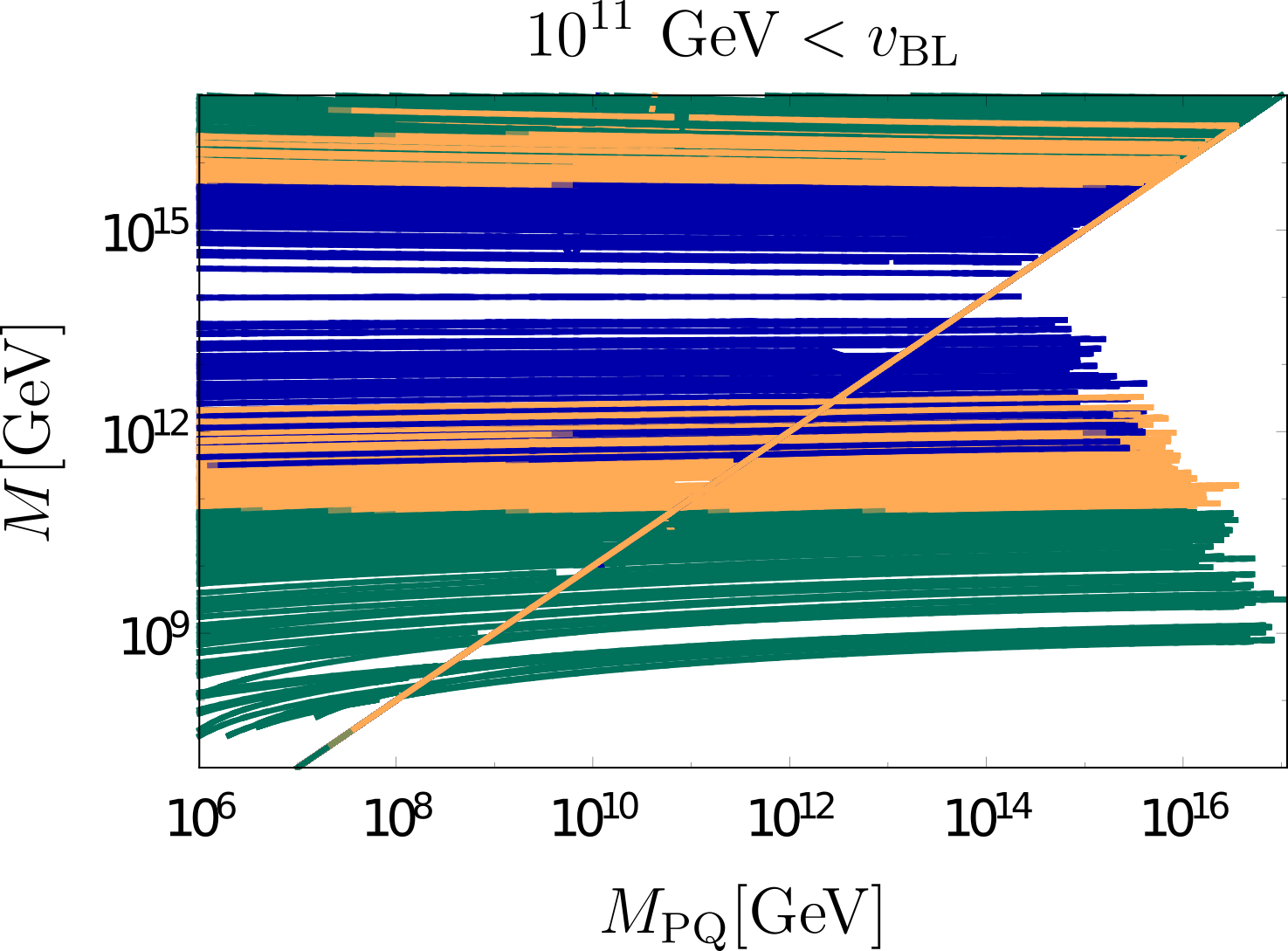}

\end{minipage}
\begin{minipage}[t]{0.31\textwidth}
\includegraphics[width=\textwidth]{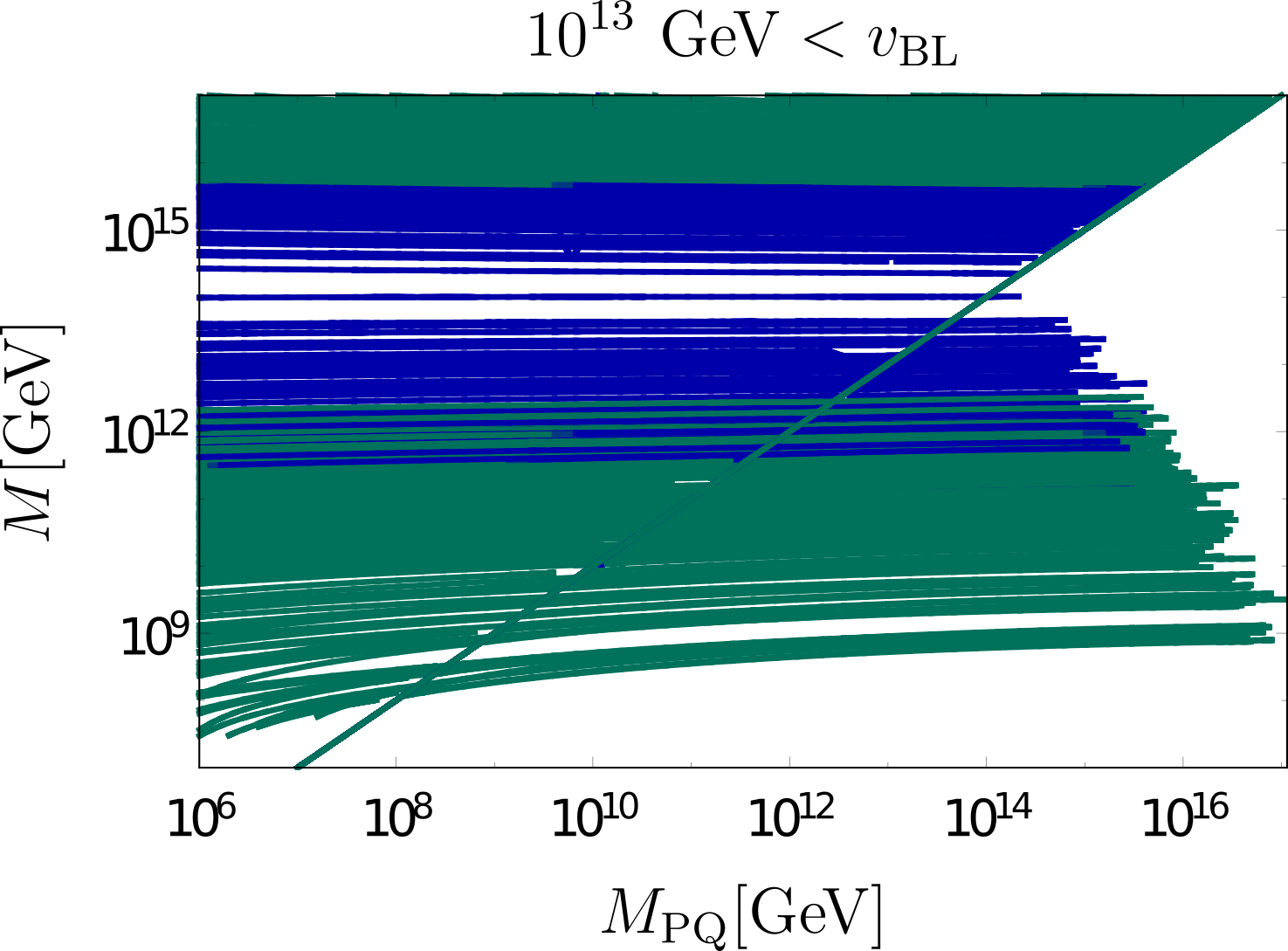}

\end{minipage}
\end{centering}
 \caption{\label{fig:singletthreescale} Intermediate scales $\MPQ$ and $\MBL$ and GUT-scale $\MU$ for Model 3.2 for different threshold corrections.  The curves in blue are excluded by gauge-mediated proton decay limits, the curves in green by the limit on the B-L breaking scale. We have considered three different ranges of allowed B-L breaking scales. The threshold corrections are randomised in the following way: All scalars take masses among the values $\{ \frac{1}{5},1,5\}$ times the corresponding threshold scale, where we have taken care not to make proton decay mediating scalars light. We have reduced the range of possible threshold corrections since the bigger range did not yield enough viable solutions.
}
\end{figure}
Correspondingly we obtain a relation between the scales $\MU$, $\MPQ$ and $\MBL$ also in this model. However, this dependence --plotted in figure \ref{fig:singletthreescale} for different sets of threshold corrections-- is very weak, and the additional fermions do not change the beta functions enough to introduce additional constraints.  We have verifed that the model is still allowed in the entire parameter space of $v_{\rm PQ}$.\\

For $M_{\rm PQ}>M_{\rm U}$, the extra fermions are integrated out above the GUT scale and do not change the running of the three gauge couplings. This case is always allowed, as long as the model without the additional fermions is not ruled out. The only constraint on both Models 3.1 and 3.2 -- which we summarise as Model 3-- comes then from the B-L breaking scale. For degenerate scalars at the thresholds, we need to allow for $v_{\rm BL}$ as low as $10^9 \GeV$, as indicated in lines 10 to 12 of figure \ref{fig:summarylongnothresholds}. If we allow variations in the masses of the heavy scalars, values of $v_{\rm BL}$ of order $10^{11}-10^{12}\GeV$ are still allowed. For $v_{\rm BL}>10^{13} \GeV$, Model 3 is excluded. This is illustrated in in lines 10 to 12 of figure \ref{fig:summarylong}.

\subsection{Dependence on the proton lifetime}

\begin{figure}[h]
\begin{centering}
\includegraphics[width=0.45 \textwidth]{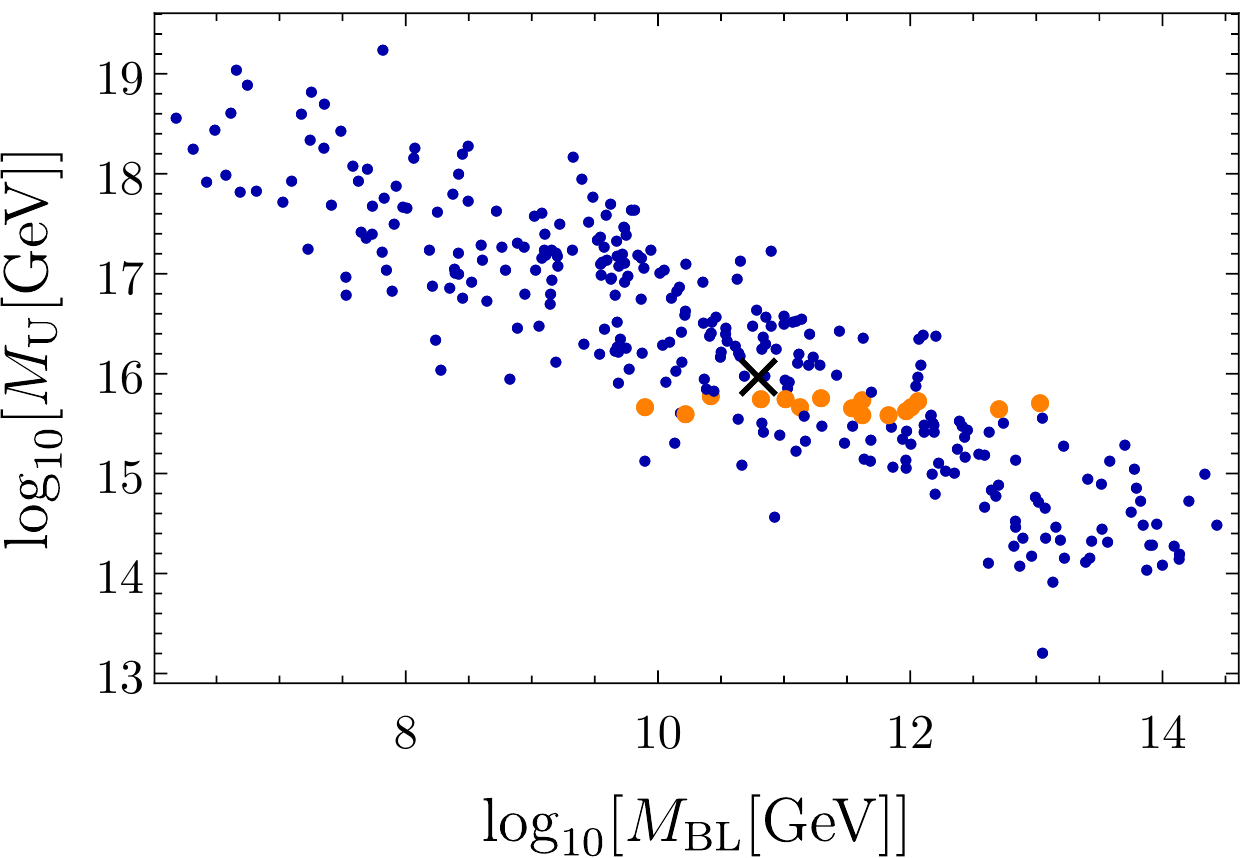}\\
\end{centering}
\begin{minipage}[t]{0.45\textwidth}
\includegraphics[width=\textwidth]{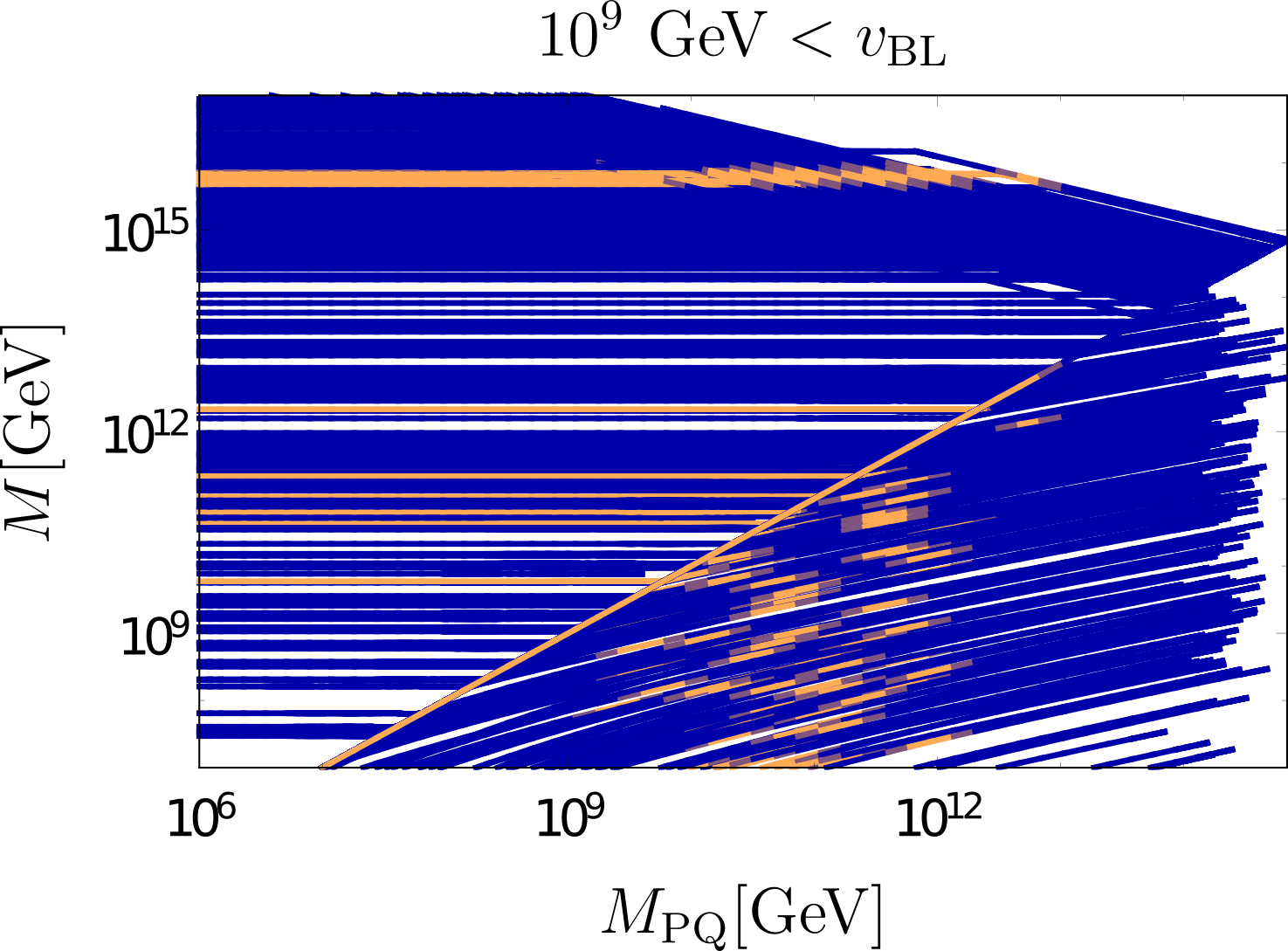}
\end{minipage}
\begin{minipage}[t]{0.45\textwidth}
\includegraphics[width=\textwidth]{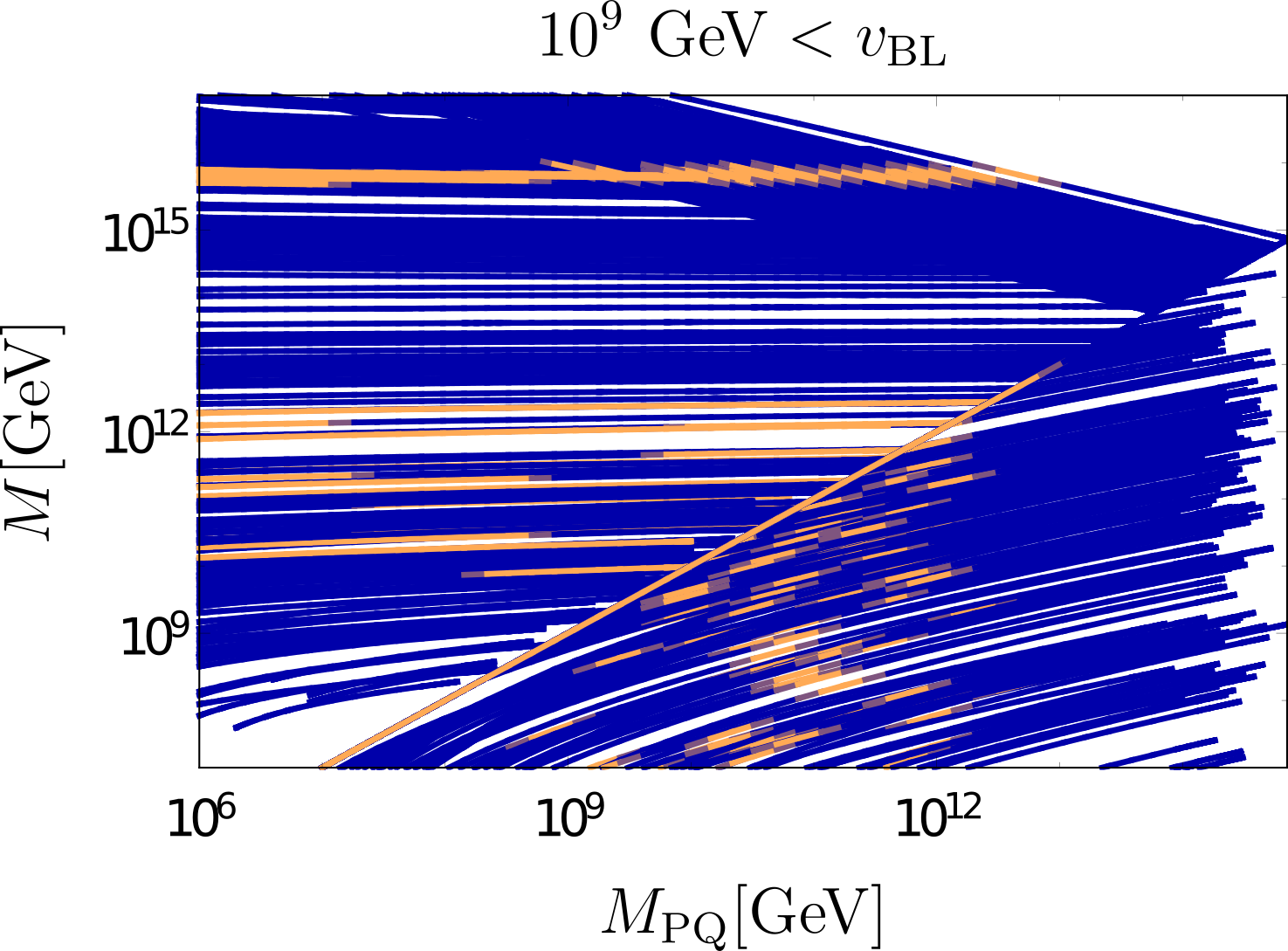}
\end{minipage}
\caption{\label{fig:hypprot} Intermediate and unification scale for randomised scalar threshold corrections in Models 1 (top), 2.1 (bottom left) and 2.2 (bottom right) including the limit coming from a hypothetical observation of proton decay at Hyper-Kamiokande. Only the orange points/regions are not excluded by the constrained proton lifetime. For models 2.1 and 2.2, the threshold corrections induce large uncertainties in the viable $f_A$ ranges even in the case where the unification scale is known.}
\end{figure}

\begin{figure}[h]
\begin{centering}
\includegraphics[width=\textwidth]{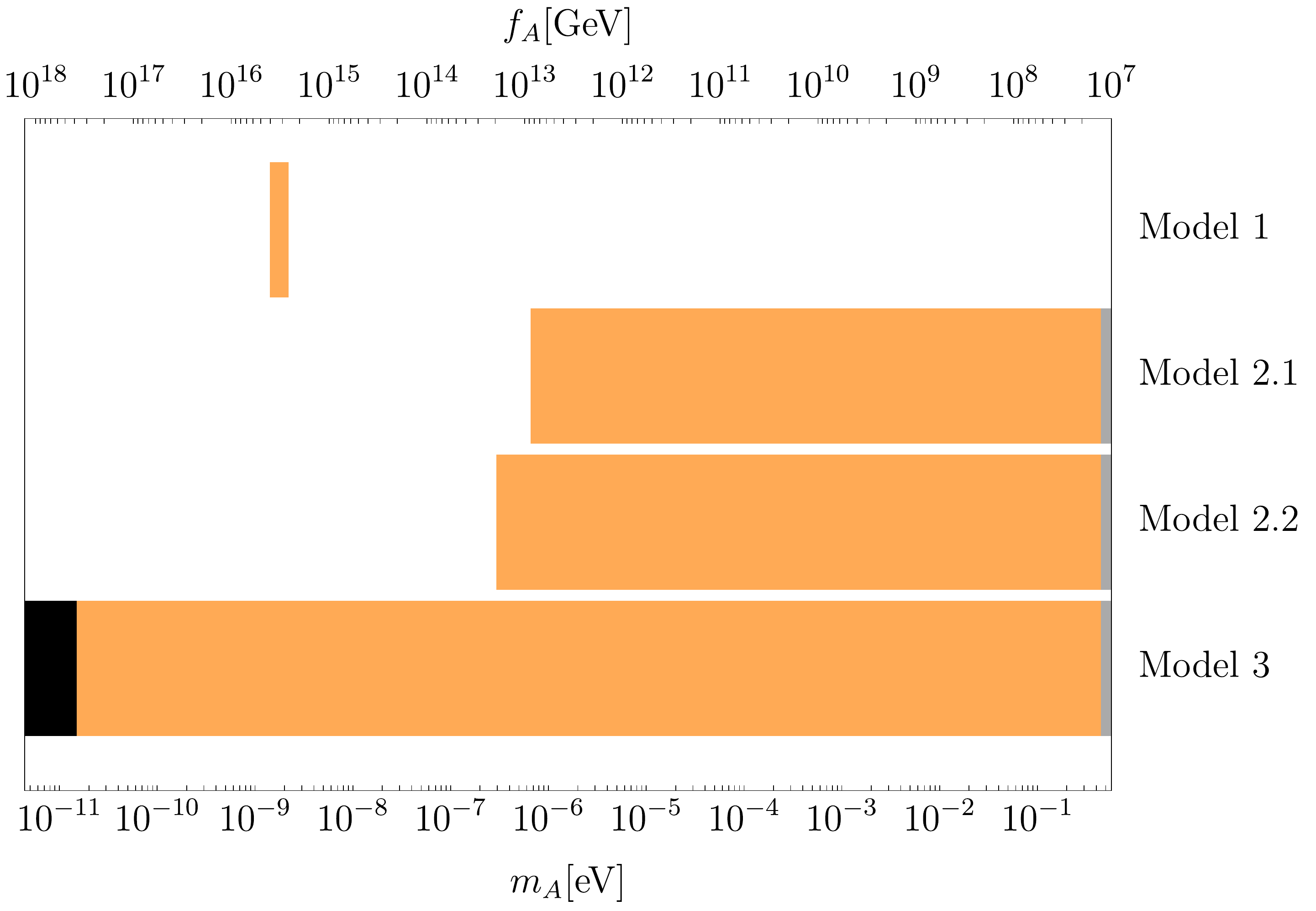}

\end{centering}
\caption{\label{fig:hypsummary} Viable ranges of $f_A$ in the hypothetical case of a known proton lifetime between $1.6 \times 10^{34}$ and $1.3\times 10^{35}$ years. Allowed regions are plotted in orange. Regions in black are excluded from black hole superradiance constraints, regions in gray from the non-observation of excessive stellar cooling.}
\end{figure}

In our analysis we are dealing with three different scales, none of which have been observed so far. Apart from the axion mass, one could also hope to constrain the unification scale in the future  by detecting proton decay. The projected sensitivity of the Hyper-Kamiokande Cherenkov detector to the channel $p\rightarrow e^+\pi^0$  after 10 years of measurement is $1.3 \times 10^{35} {\rm yr}$ at $90\%$CL \cite{Abe:2014oxa}. Assuming that proton decay is observed during the first decade of Hyper-Kamiokande, we can place further (hypothetical) constraints on the axion decay constant in each of our models. In figure \ref{fig:hypprot} we illustrate how an upper bound
 on the proton decay scale constrains the scale of Peccei-Quinn breaking and ergo the axion decay constant and mass. 
We make a naive analysis and assume that proton decay is only mediated by the heavy gauge bosons. As a lower bound for the proton lifetime we use the present limit $1.6\times 10^{34} {\rm yr}$ \cite{Miura:2016krn}. As shown in figure \ref{fig:hypsummary}, an observation of proton decay is very constraining only for our Model 1 --here we obtain $2.6\times 10^{15}\GeV< f_A<4.0 \times 10^{15} \GeV$- , while in the other models the allowed ranges of $f_A$ are still rather large.

%% file: axion_mass_so10_conclusions.tex
\section{Summary and discussion}
\label{discussionandconclusions}

We have analysed various non-supersymmetric Grand Unified $SO(10)\times U(1)_{\rm PQ}$ models for their predictions on the axion mass, the domain wall number, and the low-energy couplings to SM particles. 
The basic field content of all the considered models consisted of three spinorial $16_F$ representations 
of $SO(10)$ representing the fermionic matter content and three Higgs representations: $210_H$, 
$\overline{126}‚_H$ and $10_H$, see table \ref{tab:PQassignments}. 
\begin{table}[h]
\begin{align*}
\begin{array}{|l|c|c|c|c|c|c|c||c|}
\hline
 & 16_F & \overline{126}_H & 10_H & 210_H  & 45_H & S & 10_F & N_{\rm DW} \\
\hline
{\bf Model\ 1}  & 1 & -2 &-2 & 4  & - & -& -  & 3  \\
\hline
{\bf Model\ 2.1}  & 1 & -2 &-2 & 0  & 4 & -& -  & 3   \\
\hline
{\bf Model\ 2.2}  & 1 & -2 &-2 & 0  & 4 & -& -2 & 1   \\
\hline
{\bf Model\ 3.1}  & 1 & -2 &-2 & 0  & - & 4 & -  & 3 \\
\hline
{\bf Model\ 3.2}  & 1 & -2 &-2 & 0  & - & 4 & -2  & 1
\\
\hline
\end{array}
 \end{align*}
 \caption{\label{tab:PQassignments}Field content, PQ charge assignments, and resulting domain wall number $N_{\rm DW}$ in the various $SO(10)\times U(1)_{\rm PQ}$ models considered in this paper.}
\end{table}
The latter have been assumed to take VEVs in such a way that $SO(10)$ is 
broken along the symmetry breaking chain 
\begin{eqnarray*} 
SO(10)&\stackrel{M_{\rm U}-210_H}{\longrightarrow}
&4_{C}\, 2_{L}\, 2_{R}\ \stackrel{M_{\rm BL}-\overline{126}‚_H}{\longrightarrow}3_{C}\, 2_{L}\, 1_{Y}\ \stackrel{M_Z-10_H}{\longrightarrow} \ 3_{C}\,1_{\rm em}\,.
\end{eqnarray*}
In some of the models, this basic field content was extended by further scalar and fermion representations. This includes an additional scalar in the $45_H$, in which case we have considered a further symmetry breaking chain,
\begin{eqnarray*}
SO(10)&\stackrel{M_{\rm U}-210_H}{\longrightarrow}
4_{C}\, 2_{L}\, 2_{R}\, \stackrel{M_{\rm PQ}-45_H}{\longrightarrow}
4_{C}\, 2_{L}\, 1_{R}\, \stackrel{M_{\rm BL}-126_H}{\longrightarrow}
3_{C}\, 2_{L}\, 1_{Y}\, \stackrel{M_Z-10_H}{\longrightarrow} \ 3_{C}\,1_{\rm em}.
\end{eqnarray*}

\begin{figure}[h]
\begin{centering}
\includegraphics[width=\textwidth]{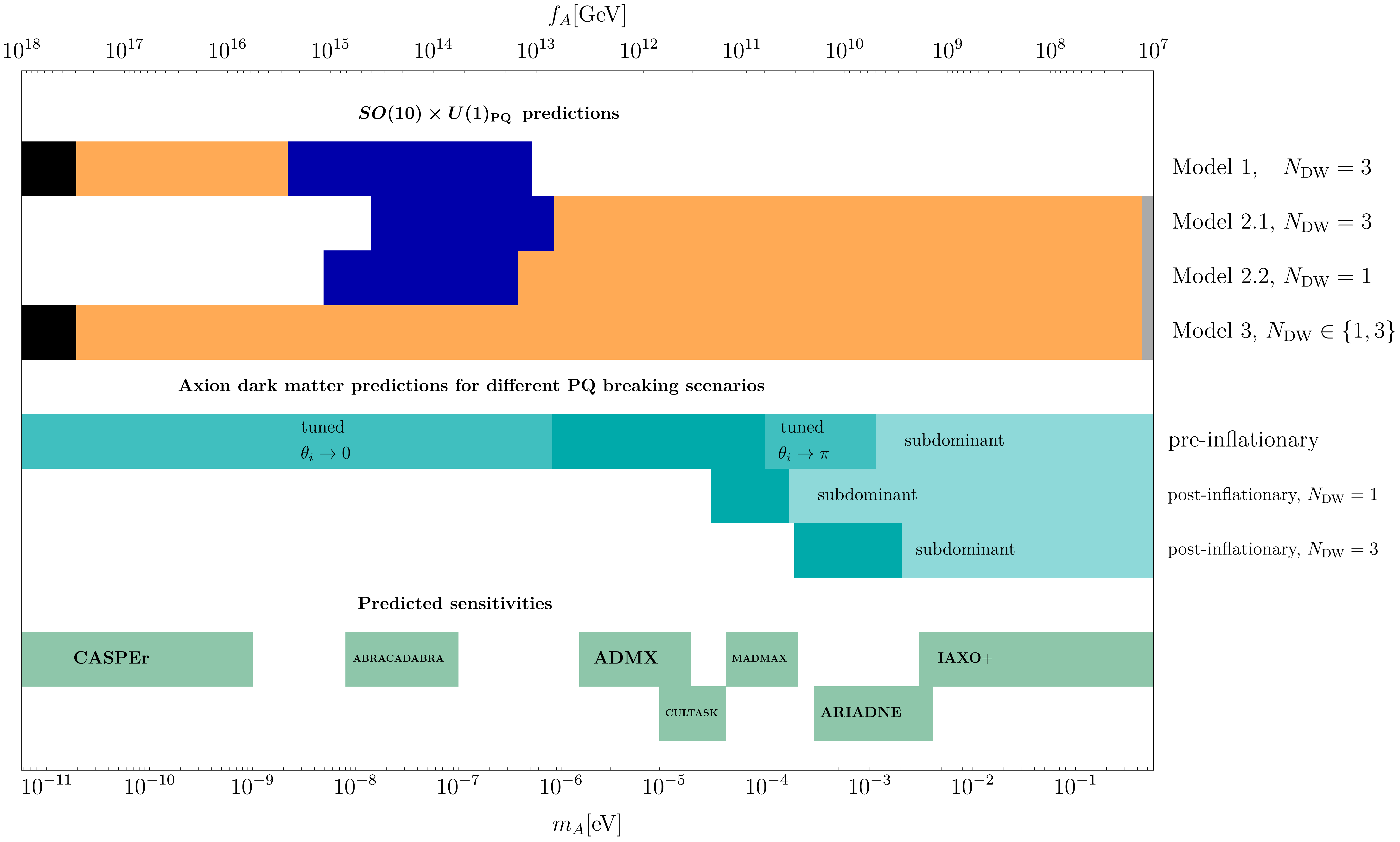}\\
\end{centering}
  \caption{\label{fig:summaryshort} 
Possible ranges of the axion mass and decay constant consistent with gauge coupling unification in our four models.
Regions in black are excluded by constraints from black hole superradiance, regions in dark blue by proton stability constraints.
Regions in gray are excluded by stellar cooling constraints from horizontal branch stars in globular clusters  \cite{Ayala:2014pea}. For comparison, we show also the mass regions preferred by axion dark matter (DM) (lines 5 to 7), cf. \cite{Saikawa:2017lzn}. Here, the dark regions indicate the ranges where the axion can make up the main part of the observed DM, with the possibility of fine tuning the initial misalignment angle in the scenario where the PQ symmetry is broken before the end of inflation and not restored thereafter (pre-inflationary PQ symmetry breaking scenario). In the light regions, axions could still be DM, but not the dominant part. The remaining regions are not allowed - axions in this mass range would be overabundant. 
Note that the region in the $N_{\rm DW}=3$ case has been derived under the assumption that the PQ symmetry is protected by a discrete symmetry, so that Planck scale suppressed PQ violating operators are allowed at dimension 10 or higher \cite{Ringwald:2015dsf}. 
In the last two lines the projected sensitivities of various experiments are indicated \cite{Budker:2013hfa,Stern:2016bbw, Chung:2016ysi,Kahn:2016aff,TheMADMAXWorkingGroup:2016hpc, Arvanitaki:2014dfa,Armengaud:2014gea,Giannotti:2017hny}. }
\end{figure}

In all models one can choose a basis of fermion fields for which the phenomenologically most important couplings to photons ($\gamma$), electrons ($f=e$), protons ($f=p$) and 
neutrons ($f=n$)  read, at energies lower than $\Lambda_{\rm QCD}$,
\begin{equation}
{\cal L} = \frac{1}{2} \partial_\mu A \partial^\mu A - \frac{1}{2} m_A^2 A^2 
+ \frac{\alpha}{8\pi} \,\frac{C_{A\gamma}}{f_A}\,A\,F_{\mu\nu} {\tilde F}^{\mu\nu} 
-
\frac{1}{2}\, \frac{C_{Af}}{f_A} \, 
\partial_\mu A\ \overline{\Psi}_f \gamma^\mu 
\gamma_5   \Psi_f \,,
\end{equation}
with
\begin{equation}
m_A= 
{57.0(7)\,   \left(\frac{10^{11}\rm GeV}{f_A}\right)\mu \text{eV}, }
\end{equation}
and with the couplings $C_{Ax}$ given by 
\begin{eqnarray}
&& C_{A\gamma}= \frac{8}{3}-1.92(4) \,, \qquad  C_{Ae}= \frac{1}{N_{\rm DW}} \sin^2\beta\,,  \nonumber \\
&& C_{Ap}=-0.47(3)+\frac{3}{N_{\rm DW}}[0.29\cos^2\beta-0.15\sin^2\beta\pm 0.02]\,,  \\
&& C_{An}=-0.02(3)+\frac{3}{N_{\rm DW}}[-0.14\cos^2\beta+0.28\sin^2\beta\pm 0.02]\,, \nonumber
\end{eqnarray}
where $\tan^2\beta = ((v_u^{10})^2+(v_u^{126})^2)/((v_d^{10})^2+(v_d^{126})^2)$, and $N_{\rm DW}$ is the domain-wall number, which in the models considered is either 3 or 1. For $N_{\rm DW}=3$ one recovers the results for the DFSZ axion \cite{Zhitnitsky:1980tq,Dine:1981rt,Srednicki:1985xd,diCortona:2015ldu}, although the 
microscopic origin of the parameter $\beta$ differs (as it is determined by the VEVs of four Higgses, as opposed to two in DFSZ models). The fermion fields for which the above interactions are valid are obtained after special axion-dependent rotations of the fermion fields that carry charges under the
global symmetry  ${\rm PQ}_{\rm phys}$  compatible with the GUT symmetry. These fermion rotations do not act in the same way over all the components of the $16_F$ multiplets, and thus will ``hide'' the GUT symmetry, and moreover modify the axion couplings to the weak gauge bosons. A possible 
measurement of the latter couplings would open up the possibility of recovering the GUT compatible charges under ${\rm PQ}_{\rm phys}$, and discriminate these models from e.g. simpler DFSZ scenarios.

The overall phenomenologically viable range in the axion decay constant of these models spans 
a very wide range, $10^{7}\,\GeV\lesssim  f_A\lesssim 10^{17}\, \GeV$, corresponding to an axion mass $m_A$ between  $10^{-10}$\,eV and  $10^{-1}$\,eV (see the orange regions in figure \ref{fig:summaryshort}).
These predictions survive constraints from gauge coupling unification, from black hole superradiance (black), from proton decay (blue),  and stellar cooling\footnote{For the stellar cooling bound, we took the one on the photon coupling from horizontal branch stars in globular clusters  \cite{Ayala:2014pea}. The one on the nucleon coupling from supernova 1987A  is presumably stronger, $f_A> 3\times 10^8\,\GeV$, corresponding to $m_A< 0.02\,$meV \cite{Giannotti:2017hny}, but suffers still from nuclear physics uncertainties.} (gray).  The features of the different models are summarised next.

{\bf Model 1} -- employing just the basic field content mentioned above and assuming that all these fields are charged under the PQ symmetry, cf. equation \eqref{eq:PQGUT} and table \ref{tab:PQassignments} -- appears to be most predictive.
In fact, we infer from the first line in figure \ref{fig:summaryshort} that the aforementioned phenomenological constraints result in an axion parameter region 
\begin{equation}
2.6 \times 10^{15} \GeV <f_A<3.0\times 10^{17}\text{GeV}, \hspace{3ex}
1.9 \times 10^{-11}\eV < m_A<2.2\times 10^{-9} \eV,
\end{equation} 
if we allow the seesaw scale to get as low as $v_{\rm BL}\simeq 10^9\,\GeV$. 
The allowed axion mass range moves towards the upper end, $m_A \simeq 2.2\times 10^{-9}\,\eV$, if we 
restrict the seesaw scale to higher values, $v_{\rm BL}\gtrsim 10^{13}\,\GeV$, 
cf. figure \ref{fig:summarylong} (first and second line). 

The small axion mass predicted in Model 1 implies that the PQ symmetry has to be broken before and during inflation and must not be restored thereafter \cite{Borsanyi:2016ksw} (pre-inflationary PQ symmetry breaking scenario). In fact, in the 
opposite case (post-inflationary PQ symmetry breaking scenario), the axion mass is bounded from below  
by about $23\,\mu$eV \cite{Borsanyi:2016ksw,Klaer:2017ond}, cf. figure \ref{fig:summaryshort}. In order for axion cold dark matter not to become overabundant, the initial value of the axion field 
in the causally connected patch which contains the present universe had to be small, 
$10^{-3}\lesssim |\theta_i| = |A(t_i)/f_A|\lesssim 10^{-2}$  \cite{Borsanyi:2016ksw}. Furthermore,  the 
Hubble expansion rate during inflation must have been small, $H_I\lesssim 10^9\,\GeV$, to avoid constraints from the non-observation 
of traces of isocurvature fluctuations in the cosmic microwave background radiation \cite{Turner:1990uz,Beltran:2006sq,Hertzberg:2008wr,Hamann:2009yf}.  
The latter constraint disqualifies Model 1 as a possible GUT SMASH candidate, since in Higgs portal inflation the Hubble expansion rate during inflation is of order 
$10^{13}\,\GeV\lesssim H_I\lesssim 10^{14}\,\GeV$ \cite{Ballesteros:2016xej}. 

Remarkably, the predicted axion mass range of Model 1 will be probed in the 
upcoming decade by the CASPEr experiment \cite{Budker:2013hfa}, cf. figure \ref{fig:summaryshort}, which aims to detect directly axion dark matter by precision nuclear  magnetic resonance techniques. If successful and interpreted in terms of Model 1, one may translate, via \eqref{eq:fA_MU_model1}, the measurement of the axion mass into an indirect determination of the mass of the superheavy gauge bosons, i.e. the unification scale, 
\begin{align}
\label{eq:MU_determination_model1}
 \MU \simeq   3 g_U \sqrt{\chi}/m_A ,
\end{align}
where $\chi$ is the topological susceptibility in QCD, $\chi = [75.6(1.8)(0.9) {\rm MeV}]^4$ \cite{diCortona:2015ldu,Borsanyi:2016ksw}. 
The unification scale 
can be probed complementarily by 
the next generation of experiments searching for signatures of proton decay, such as Hyper-Kamiokande \cite{Abe:2011ts} or DUNE \cite{Kemp:2017kbm}. 

Models featuring an axion with a larger mass, $m_A\gtrsim 23\,\mu$eV, compatible with the post-inflationary 
PQ symmetry breaking scenario, can only be obtained if the $210_H$ responsible for
the first gauge symmetry breaking at $\MU$ has no PQ charge. But in addition -- as emphasised already in references 
\cite{Holman:1982tb,Bajc:2005zf} -- the scalar sector has to be extended by yet another PQ charged field obtaining 
a VEV. Otherwise the axion decay constant is predicted to be as low as the electroweak scale, which is firmly 
excluded by laboratory experiments and stellar astrophysics. 
Correspondingly, we considered also models which have additional scalar fields 
such as a PQ charged $45_H$ (Models 2.1 and 2.2) or a PQ charged $SO(10)$ singlet complex scalar field $S$ (Models 3.1 and 3.2),
cf. table \ref{tab:PQassignments}. 
Crucially, in these extended models, the PQ symmetry breaking scale $v_{\rm PQ}$ and the seesaw scale 
$v_{\rm BL}$ are independent parameters -- in fact, the axion field has to be orthogonal to the Goldstone field 
which is eaten by the gauge bosons getting mass by B-L breaking. This is different in the original SMASH 
model \cite{Ballesteros:2016euj,Ballesteros:2016xej}, where the PQ symmetry is at the same time also a B-L symmetry.

{\bf Model 2.1} features PQ charges for the $16_F$, $\overline{126}_H$ and $10_H$ multiplets and an additional PQ charged $45_H$, cf. \eqref{eq:PQ45} and table \ref{tab:PQassignments}. 
The range in the axion decay-constant/mass is predicted to be quite distinct from the one of Model 1, 
see figure \ref{fig:summaryshort}. 
Just accounting for gauge coupling unification with scalar threshold corrections, we have found an 
upper bound on the decay-constant $f_A< 4.0\times 10^{14}\,\GeV$, and a corresponding lower bound on the axion mass, $m_A>1.4\times 10^{-8}\,{\rm eV}$.  
Imposing in addition constraints from proton decay, the upper limit on $f_A$ comes further down, while constraints on the photon coupling from stellar 
cooling introduce also a lower limit on $f_A$, 
\begin{equation}
1.3 \times 10^{7} \GeV <f_A<6.7\times 10^{12}\text{GeV}, \hspace{3ex}
8.5 \times 10^{-7}\eV < m_A<0.5\, \eV .
\end{equation}
Furthermore, the model features an upper limit on the scale of B-L breaking: $v_{\rm BL}<10^{13}\GeV$, cf. figure \ref{fig:summarylong}. 

In this model, both the pre-inflationary as well as the post-inflationary PQ symmetry breaking scenarios are possible, 
see figure \ref{fig:summaryshort}. As far as the latter case is concerned, 
it is important to note that the possibly inherent domain wall problem 
can be circumvented if the PQ symmetry is only 
a low energy remnant of an exact discrete symmetry, so that Planck scale suppressed PQ violating operators -- which have been argued to be induced in any case by quantum gravity effects \cite{Ghigna:1992iv,Barr:1992qq,Kamionkowski:1992mf,Holman:1992us}, and which render string-wall systems with $N_{\rm DW}>1$
unstable -- occur at dimension ten\footnote{For explicit examples of such discrete symmetries and more details 
see table \ref{tab:Z10} and Appendix \ref{discretesymmetry}.} \cite{Ringwald:2015dsf}. 
\begin{table}[h]
\begin{align*}
\begin{array}{|c|c|c|c|c|c|}
\hline
&16_F&10_H&\overline{126}_H& 45_H& S\\
  \hline
 \text{ Model 2.1}  & -\frac{1}{40} & \frac{1}{20} & \frac{1}{20}&-\frac{1}{10}  & -\\
   \hline
 \text{ Model 3.1} & -\frac{1}{40} & \frac{1}{20} & \frac{1}{20}&-& -\frac{1}{10}  \\
 \hline
 \end{array} 
 \end{align*}
 \caption{\label{tab:Z10}
Charges of the fermionic and scalar fields under a PQ-protecting discrete $\mathcal{Z}_{40}$ symmetry for Models 2.1 and 3.1. The lowest dimensional PQ violating operators allowed by these symmetries are $45_H^{10}$ for Model 2.1 and $S^{10}$ for Model 3.1. 
 }
\end{table}
In this case, the preferred mass range for dark matter is a bit higher than the 
one singled out in the post-inflationary $N_{\rm DW}=1$ scenario. In particular, for $N_{\rm DW}=3$, as in Model 2.1,  it 
is given by $0.18\,{\rm meV}\lesssim m_A\lesssim 2.0\, {\rm meV}$, cf. figure \ref{fig:summaryshort}. 
Intriguingly, for $m_A= {\mathcal O}(10)$\,meV and $\tan\beta \gtrsim 0.3$, the axion in this model may explain 
at the same time recently observed stellar cooling excesses observed from helium burning stars, red giants and white dwarfs  \cite{Giannotti:2017hny}.

Fortunately, 
there are a number of running (ADMX \cite{Stern:2016bbw}, HAYSTAC \cite{Brubaker:2016ktl}, ORGAN \cite{McAllister:2017lkb}), presently being assembled (CULTASK \cite{Chung:2016ysi}, QUAX \cite{Barbieri:2016vwg}), or planned (ABRACADABRA \cite{Kahn:2016aff}, KLASH \cite{Alesini:2017ifp}, MADMAX \cite{TheMADMAXWorkingGroup:2016hpc}, ORPHEUS \cite{Rybka:2014cya}) axion dark matter experiments, which cover a large portion of the mass range of interest for axion dark matter in the pre-inflationary PQ symmetry scenario in Model 2.1, see figure \ref{fig:summaryshort}. 
Furthermore, in the meV mass range of interest for the post-inflationary PQ symmetry breaking scenario, the model can be probed by the presently being build fifth force experiment ARIADNE \cite{Arvanitaki:2014dfa} and the proposed helioscope IAXO \cite{Armengaud:2014gea}, cf. figure \ref{fig:summaryshort}.

{\bf Model 2.2} adds to the field content of Model 2.1 two PQ charged vector representations $10_F$ of $SO(10)$, getting their mass from the $45_H$ and cancelling two units of the domain wall number. Correspondingly, 
Model 2.2 has no domain wall problem whatsoever. 
Allowing $v_{\rm BL}$ to be as small as $10^9\GeV$, the allowed mass range in this model is very similar to the one of Model 2.1. Taking into account additional limits from gauge coupling unification, proton decay and the limit from stellar cooling, the preferred ranges in this model are
\begin{equation}
1.3 \times 10^7\GeV <f_A<1.5\times 10^{13} \GeV, \hspace{3ex}
3.8 \times 10^{-7}\eV < m_A<0.5\, \eV .
\end{equation}

This model can be probed by the same experiments as Model 2.1.
Similar to Model 2.1, this model allows for axions in the post-inflationary DM window.  
Remarkably, the model features a second potential DM candidate: the lightest stable combination of the additional fermions \cite{Mambrini:2015vna,Nagata:2015dma,Arbelaez:2015ila,Boucenna:2015sdg}. Therefore we do not need to insist on the axion being $100\%$ of the observed dark matter and can allow for bigger axion masses (cf. the region labelled "subdominant" in the $N_{\rm DW}=1$ bar of figure \ref{fig:summaryshort}).

{\bf Model 3.1} contains PQ-charged fermions in the $16_F$ representation, as well as PQ charged scalars in the $10_H$, $\overline{126}_H$ and a singlet $S$. 
The axion decay constant in this model is set by the VEV of the scalars singlet. It is a free parameter not constrained by gauge coupling unification, since it does not break any gauge symmetries. 
In a sense, Model 3.1 is the most minimal GUT model with an invisible axion with decay constant possibly in the intermediate range between the electroweak scale and the unification scale, see  
figure \ref{fig:summaryshort}. Similar to Model 2.1, its possible domain wall problem in the 
 post-inflationary symmetry breaking case  can be avoided if the PQ symmetry is only an accidental symmetry of a discrete symmetry which forbids PQ-violating operators up to dimension 10. For an example of such a discrete symmetry, we refer to table \ref{tab:Z10}.

{\bf Model 3.2} adds to the field content of Model 3.1  two vectorial $10_F$ representations of fermions 
getting their masses by the VEV $v_S$ of the singlet $S$. Despite the fact that the fermions affect the running of gauge couplings at scales above $v_S$, we found that, as in Model 3.1, the axion decay constant cannot be constrained by the running of the gauge coupling. Both models 3.1 and 3.2 feature a B-L breaking scale lower than $10^{13}\GeV$.

{\bf Variations of these models} can be obtained by:
{\em (i)} Changing the PQ-charges of the scalar that sets the axion decay constant while keeping the other PQ charges constant.  E.g., a lower PQ charge 2 of the $S$ in our model 3.1 will result in an increased domain wall number  $N_{\rm DW}=6$. 
{\em (i)} Choosing a different scalar sector and therefore different breaking chains. For example, one can choose $SU(3)_C\times SU(2)_L\times SU(2)_R\times U(1)_{B-L}$ as the intermediate gauge group by assigning a VEV to the corresponding singlet in the $210_H$. Or one may replace the $210_H$ by a $45_H$. 
For both these model variations -- {\em (i)} and {\em (ii)} --  we expect similarly large ranges of viable axion masses. 
{\em (iii)} Employing a different gauge group at the unification scale. If the breaking chain does not go via $SO(10)$, the analysis can be very different from the one in this paper.  

Finally, we return back to the question posed in the introduction, concerning whether the field content of the considered models may allow for a self-contained description of particle physics and cosmology on the same level as the minimal SMASH model.
For Model 3, this question can almost certainly be answered in the affirmative, since nearly all the considerations and calculations done in minimal SMASH can be translated to Model 3 if one exploits the (in general non-minimally gravitationally coupled) complex singlet $S$ (or a linear combination with the Higgs) as the inflaton.  Note that, although $f_A$ in Model 3 is poorly constrained for $v_{\rm BL}<10^{13}$ GeV, there can be further bounds coming from demanding stability in the Higgs direction, as analyzed for the SMASH theory in refs.  \cite{Ballesteros:2016euj,Ballesteros:2016xej}. Such study was beyond the scope of this paper.
Although less minimal, Model 2 may also be a good candidate for a GUT SMASH model, and features a more constrained axion mass.

%% file: app_phases.tex
\section{Invariance of axion-neutral gauge boson couplings under fermionic rephasings\label{app:ferphases}} 

As mentioned in Section \ref{subsec:effL}, one may obtain the axion effective Lagrangian by performing redefinitions of the fermionic phases which eliminate the dependence 
of the Yukawa interactions on the axion  field. Although rephasings fixed by the fermionic PQ charges suffice, one may choose alternative redefinitions --all cancelling the axion dependence coming from the Yukawas-- which will give rise to different effective actions. 
These are  physically equivalent, as they only differ by field redefinitions whose effects vanish on-shell. In this appendix we show explicitly that,
in keeping with these expectations, the $SU(3)_C$ axion decay constant $f_{A,3_C}$ --and with it the axion mass-- as well as the coupling of the axion to photons are not
sensitive to rephasings of the fermion fields. Although the interactions of the axion with fermions and 
massive gauge boson remain sensitive to the choice of fermionic PQ charges, the different effective Lagrangians should yield identical on-shell results. 

To prove the assertions about the couplings of the axion to the massless bosons, we will consider the  Yukawa interactions for the Weyl fermion fields $q,u,d,l,e,n$ (see table \ref{tab:fermions}) generated by the $SO(10)$ invariant couplings in \eqref{SO10Yukawacomplex}, with $\tilde Y_{10}=0$ due to the assumed PQ symmetry \eqref{pgsymmetry_yukawas}. For completeness, we will also add the Yukawas
of additional $N_{10}$ fermion multiplets in the $10_F$ of $SO(10)$ coupled to a scalar in the $45_H$, as needed in some of the models of section \ref{so10xu1pqmodels}:
\begin{equation}
\label{eq:Yuk45}
\mathcal{L}_Y = 16_F \left( Y_{10} 10_H   + Y_{126} \overline{126}_H \right) 16_F + Y_{45} 10_F 45_H 10_F + {\rm h.c.}.
\end{equation}
Using the decompositions of the scalars in the  $10_H$, $\overline{126}_H$ and $45_H$ into SM  representations given in table \ref{tab:vevs}, as well as the decompositions of the fermion representations
in table \ref{tab:fermions}, we may write
\begin{equation}\label{eq:Yuk452}\begin{aligned}
 {\cal L}_Y\supset&\, Y_{10}(qH_u+qH_d d+ l H_d e+l H_u n)+Y_{126}(q\Sigma_u u+q\Sigma_d d+ l \Sigma_d e+l \Sigma_u n)\\
 &+Y_{45}\,\sigma\,(\tilde D D +\tilde L L).
\end{aligned}\end{equation}
Since they couple to the same fermion fields, $H_u$ and $\Sigma_u$ must have identical $PQ_{\rm phys}$ charges $q_1/\fPQ$; similarly,  $H_d$ and $\Sigma_d$ must have a common charge $q_2/\fPQ$. We also allow a charge
$q_6/\fPQ$ for the field $\sigma$.\footnote{The notation is chosen for compatibility with sections \ref{altarellimodel}, \ref{hlsmod}.} Then 
we may remove the axion contributions to the Yukawa couplings by performing any of the following fermion rotations, parameterised by arbitrary $\hat q_q,\hat q_l,\hat q_D,\hat q_L$:
\begin{equation}\label{eq:arbitrot}\begin{aligned}
\psi_a\rightarrow& \,e^{i\hat q_a/\fPQ A} \psi_a, \quad \psi_a=\{q,u,d,l,e,n,D,\tilde D,L,\tilde L\},\\
\hat q_u=&\,-q_1-\hat q_q, & \hat q_d=&\,-q_2-\hat q_q, \\
\hat q_e=&\,-q_2-\hat q_l, & \hat q_n=&\,-q_1-\hat q_l,\\
\hat q_{\tilde D}=&\,-q_6-\hat q_{D}, & \hat q_{\tilde L}=&\,-q_6-\hat q_L
\end{aligned}\end{equation}
Under such anomalous transformations, after redefining $A\rightarrow -A$, the axion-gauge boson interactions become:
\begin{equation}\label{eq:axiongauge}\begin{aligned}
 \delta{\cal L}\supset&\,\frac{A}{\fPQ}\sum_k \left(2\sum_a q_a T_k(R_a)\right)\frac{g_k^2}{16\pi^2} \,{\rm \bar{T}r}\,\tilde F^k_{\mu\nu} F^{k,\mu\nu}=\\
 &- \, A\,\left(\frac{3q_1}{\fPQ}+\frac{3q_2}{\fPQ}+\frac{N_{10}q_6}{\fPQ}\right)\,\frac{g_3^2}{16\pi^2} \,{\rm \bar{T}r}\,\tilde F^3_{\mu\nu} F^{3,\mu\nu}\\
 &+\, A\,\left(\frac{3\hat q_l}{\fPQ}+\frac{9\hat q_q}{\fPQ}-\frac{N_{10}q_6}{\fPQ}\right)\frac{g_2^2}{16\pi^2} \,{\rm \bar{T}r}\,\tilde F^2_{\mu\nu} F^{2,\mu\nu}\\
 &- \, A\,\left(\frac{3\hat q_l}{\fPQ}+\frac{9\hat q_q}{\fPQ}+\frac{8q_1}{\fPQ}+\frac{8q_2}{\fPQ}+\frac{5N_{10}q_6}{3\fPQ}\right)\,\frac{g_1^2}{16\pi^2} \,{\rm \bar{T}r}\,\tilde F^1_{\mu\nu} F^{1,\mu\nu}\\
 &\supset - \, A\,\left(\frac{3q_1}{\fPQ}+\frac{3q_2}{\fPQ}+\frac{N_{10}q_6}{\fPQ}\right)\,\frac{\alpha_s}{8\pi}\, {\tilde G}^a_{\mu\nu} G^{a\mu\nu}\\
 &-8 \, A\,\left(\frac{q_1}{\fPQ}+\frac{q_2}{\fPQ}+\frac{N_{10}q_6}{3\fPQ}\right)
 \,\frac{\alpha}{8\pi}\tilde F_{\mu\nu}F^{\mu\nu},
\end{aligned}\end{equation}
where $\alpha_s=g_s^2/(4\pi)$, $\alpha=e^2/(4\pi)=g_2^2g_1^2/(g_1^2+g_2^2)/(4\pi)$ are the strong and electromagnetic fine-structure constants, and $G^{a\mu\nu}, F^{\mu\nu}$ denote the components of the
strong and electromagnetic field strengths, respectively. The previous result shows explicitly that, although the fermionic PQ charges appear in the effective Lagrangian,
the interactions between the axion and the massless bosons only depend on the PQ charges of the scalars, and thus are independent of possible rephasings of the fermions in the low-energy theory. From our results we may obtain an expression for  $f_A\equiv f_{A,3_C}$ in terms of the scalar PQ charges:
\begin{align}
\label{eq:fA3}
 f^{-1}_{A}=f^{-1}_{A,3_C}=-3 \left(\frac{q_1}{\fPQ} + \frac{q_2}{\fPQ}\right) - N_{10} \frac{q_6}{\fPQ}.
\end{align}
Including the axion-fermion interactions arising from the fermion rephasings, we may finally write the effective Lagrangian defined for the general fermion rotations in \eqref{eq:arbitrot} and the above $f_A$ value:
\begin{equation}\label{eq:axionLaggen}\begin{aligned}
  {\cal L}_{\rm eff}=&\, \partial_\mu A  \sum_a \frac{\hat q_a}{f_{\rm PQ}} (\psi_a^\dagger \bar\sigma^\mu \psi_a)+\,\frac{ A}{f_A}\,\,\frac{\alpha_s}{8\pi}\, {\tilde G}^a_{\mu\nu} G^{a\mu\nu}+\,\frac{ A}{f_A}\,\frac{8}{3}
 \,\frac{\alpha}{8\pi}\tilde F_{\mu\nu}F^{\mu\nu}.
\end{aligned}\end{equation}

%% file: app_roots.tex
\section{Roots and weights of $SO(10)$ and $4_C\times2_L\times2_R$\label{app:roots}} 

In this appendix we review some basic properties of the roots and weights of $SO(10)$ --for more details, see \cite{Slansky:1981yr}-- and prove that for our choice of VEVs in table \ref{tab:vevs} the axion is automatically orthogonal to the Goldstone bosons associated with the broken, non-diagonal generators of the gauge group.

The Lie algebra of $SO(10)$ has rank five, as it contains a subspace of mutually commuting generators --the Cartan subalgebra-- of dimension five. The latter is spanned by generators $H_i,i=1\dots5$. 
As the $H_i$ are hermitian and commute with each other, in any representation of the algebra one can find a basis of orthonormal eigenstates of the $H_i$, with real eigenvalues $\lambda$ called ``weights''. This applies
in particular to the adjoint representation, whose weights are called the roots of the Lie algebra. The roots are defined in a basis of generators in which the adjoint action of
the $H_i$ is diagonal. Aside from the $H_i$, the Lie algebra is then spanned by generators $E_\alpha$ satisfying
\begin{align}
\label{eq:commutatora}
 [H_i,E_\alpha]=\alpha_i E_\alpha.
\end{align}
The $E_\alpha$ are not hermitian --in fact \eqref{eq:commutatora} implies $E_\alpha^\dagger=E_{-\alpha}$-- but one can always recover hermitian generators from the combinations 
$1/2(E_\alpha-E_{-\alpha})$,  $1/{2i}(E_\alpha-E_{-\alpha})$.
In an arbitrary representation, one can label the states with well-defined weights as $|\lambda\rangle$, which satisfy
 \begin{align}
 H_i |\lambda\rangle=\lambda_i|\lambda\rangle, \quad \langle\lambda|\lambda'\rangle = \delta_{\lambda\lambda'}.
\end{align}
The commutation relations \eqref{eq:commutatora} imply 
\begin{align}
 E_\alpha |\lambda\rangle= N_{\alpha,\lambda}|\lambda+\alpha\rangle
\end{align}
for some normalisation constant $N_{\alpha,\lambda}$. The roots $\alpha=\{\alpha_i\}$ and weights $\lambda=\{\lambda_i\}$ can be seen as vectors in an Euclidean space of dimension equal to the rank of the group.
There is a convenient choice of
non-orthonormal basis called the Dynkin basis in which the $\alpha_i$ and  $\lambda_i$ can be represented with integer numbers. One starts by identifying a set of linearly independent roots $\alpha^p=\{\alpha^p_i\}$ called ``simple 
roots'',  such that any root can be expressed as a linear combination of the simple roots, with coefficients that are all positive or zero, or alternatively all negative or zero. The scalar products
of the simple roots --defined as  $(\alpha^p,\alpha^q)=\sum_i \alpha^p_i \alpha^q_i$-- allow to define a Cartan matrix $A_{ij}$
\begin{align}
 A_{pq}=2\frac{(\alpha^p,\alpha^q)}{(\alpha^q,\alpha^q)}.
\end{align}
Then the Dynkin basis is spanned by the following basis of roots $\tilde\alpha^p=\{\tilde\alpha^p_i\}$ 
\begin{align}
 \tilde\alpha^p=\sum_q (A^{-1})_{pq}\alpha^q.
\end{align}
In the Dynkin basis, any root or weight can be expressed as a linear combination of the $\tilde\alpha^p$ with integer coefficients:
\begin{align}
\label{eq:Dynkinas}
 \lambda_i=\sum_p a^p \tilde\alpha^p_i,\quad a^p \,\text{integer}.
\end{align}

For $SO(10)$, there are 5 zero roots corresponding to the Cartan generators, plus 20 ``positive'' roots, and 20 negative roots given by minus the positive roots. The positive roots are
\begin{equation}\label{eq:roots}\begin{array}{c}
 (0,1,0,0,0)\\
 (1,-1,1,0,0)\\
 (-1,0,1,0,0),(1,0,-1,1,1)\\
 (-1,1,-1,1,1),(1,0,0,-1,1),(1,0,0,1,-1)\\
 (0,-1,0,1,1),(-1,1,0,-1,1),(-1,1,0,1,-1),(1,0,1,-1,-1)\\
 (0,-1,1,-1,1),(0,-1,1,1,-1),(-1,1,1,-1,-1),(1,1,-1,0,0)\\
 (0,0,-1,0,2),(0,0,-1,2,0),(0,-1,2,-1,-1),(-1,2,-1,0,0),(2,-1,0,0,0).
\end{array}\end{equation}
The five simple roots $\alpha^p$ of $SO(10)$ are those at the bottom of the previous list.

The maximal subgroup  $4_C 2_L 2_R\supset SO(10)$ has rank 5, and thus its Cartan generators contain those of $SO(10)$, and 
the two bases of Cartan generators for each group are related by a linear transformation.  
Then one can map weights of representations of $4_C 2_L 2_R$ into weights of $SO(10)$. The relation is given by an invertible
projection matrix
$P$ such that
\begin{align}
\label{eq:SO10P}
 \lambda_{4_C 2_L 2_R}=P\lambda_{SO(10)},\quad P=\left[
 \begin{array}{ccccc}
  0 & 0 & 1 & 1 & 1\\
  0 & 0 & 1 & 0 & 0\\
  1 & 1 & 1 & 0 & 1\\
  0 & 1 & 1 & 1 & 0\\
  -1 & -1 & -1 & -1 & 0
 \end{array}
\right].
\end{align}
With the former choice of matrix, 
the weights under $4_C 2_L 2_R$ are of the form $\{w_i\},i=1,\dots,5$, where $w_1$ is the weight corresponding to the generator
$T_3$ of $2_L$,  $w_2$ is the weight of $T_3$ for the group $2_R$ (or, as denoted in tables \ref{tab:vevs}, \ref{tab:fermions}, the $1_R$ charge), and $w_3,w_4,w_5$ are the three weights
of the Cartan algebra of $SU(4)$, with $w_3,w_4$ the weights of the Cartan generators $T^3,T^8$ of $SU(3)$. The former assignments
can be checked by starting from the $SO(10)$ weights of the 16 of $SO(10)$, computing the  $4_C 2_L 2_R$ weights with \eqref{eq:SO10P}, and identifying 
the quantum numbers of the SM fermions and the RH neutrino, as in table \ref{tab:fermions}. One can also identify the charge $B-L$ as a combination of $SU(4)$ weights
\begin{align}
\label{eq:B-L}
 B-L=\frac{1}{3}w_3+\frac{2}{3}w_4+w_5.
\end{align}
The electric charge is
\begin{align}
\label{eq:charge}
 Q=T_3+1_R+\frac{1}{2}(B-L)=\frac{w_1}{2}+\frac{w_2}{2}+\frac{1}{6}w_3+\frac{1}{3}w_4+\frac{1}{2}w_5.
\end{align}

With the previous tools we may prove now that with the choice of fields getting VEVs in table \ref{tab:vevs}, the orthogonality conditions of the axion with respect to Goldstone bosons
associated with the off-diagonal gauge generators are always satisfied. 
The non-diagonal generators of the Lie algebra in a given representation are spanned by the $E_\alpha$. 
Let's assume a representation of scalar fields in which the nonzero VEVs $v_i$  correspond to states $|\lambda(i)\rangle$. 
Then the orthogonality constraints \eqref{eq:orthogauge0} from off-diagonal generators can be satisfied with the following sufficient conditions:
\begin{align}
 (E_\alpha)_{mn}=0 \quad\text{for $m,n$ such that }v_m\neq0,v_n\neq0, \alpha\neq0.
\end{align}
The previous conditions have to be verified within each $SO(10)$ irreducible representation, as the generators only link field components within them.
One has
\begin{align}
 (E_\alpha)_{mn}=\langle \lambda(m)| E_\alpha |\lambda(n)\rangle= N_{\alpha,\lambda(n)}\langle\lambda(m) |\lambda(n)+\alpha\rangle=N_{\alpha,\lambda(n)} \delta_{\lambda(m), \lambda(n)+\alpha}.
\end{align}
This means that $ (E_\alpha)_{mn}$ will be zero --and the orthogonality condition with all the $E_\alpha$ (and with them the non-diagonal generators) automatically satisfied-- 
if the difference of the weights associated with the  $v_m\neq0$ is not a nonzero root of the Lie algebra, that is $\lambda(m)\neq \lambda(n)+\alpha$ for all roots $\alpha\neq0$. 
This will always be the case if only one component in a given representation has a nonzero VEV, but has to be checked for more general situations. 
If the property holds, then the only nontrivial orthogonality conditions are those arising from \eqref{eq:orthogauge} applied to the diagonal Cartan generators (or their linear combinations).

In this article, we consider the following scalar representations: $210_H$, $10_H$, $45_H$, $\overline{126}_H$. In both the $210_H$ and $45_H$, only one component ($\phi$ for the $210_H$, $\sigma$ for the $45_H$, 
see table \ref{tab:vevs}) gets a VEV, so that orthogonality 
with respect to the off-diagonal generators is guaranteed. For the $10_H$ and $\overline{126}_H$, however, we have in both cases two field components getting a VEV: the neutral components of $H_u$
and $H_d$, and those of $\Sigma_u$ and $\Sigma_d$, respectively. $H_u$ has the same quantum numbers as $\Sigma_d$, meaning identical weights. A similar relation holds for $H_d$ and $\Sigma_d$. Since the orthogonality condition can be checked in terms of weights, it suffices to consider  
the $\Sigma$ components. From table \ref{tab:vevs} one gets their quantum numbers under $1_R$, $B-L$. The fact that the states are colour singlets implies $w_3=w_4=0$. The table
gives $B-L=0$, so that equation \eqref{eq:B-L} implies then  $w_5=0$.
Charge neutrality, together with \eqref{eq:charge}, 
fixes $T_3=-1_R$. In the Dynkin basis, the $SU(2)$ weights are twice the usual ones, as follows from  relation \eqref{eq:Dynkinas}, and the fact that there is a unique simple $SU(2)$ root
given by the number 2, in the conventional normalisation. Then the $4_C 2_L 2_R$ weights of the neutral $\Sigma_u$ and $\Sigma_d$ states in the Dynkin basis are
\begin{align}
 \lambda({\Sigma^0_u})=(1,-1,0,0,0)_{4_C 2_L 2_R},\quad  \lambda({\Sigma^0_d})=(-1,1,0,0,0)_{4_C 2_L 2_R}.
\end{align}
Inverting the relation \eqref{eq:SO10P}, the resulting $SO(10)$ weights in the Dynkin basis are 
\begin{align}
 \lambda({\Sigma^0_u})=(0,0,-1,1,1)_{SO(10)},\quad  \lambda({\Sigma^0_d})=(0,0,1,-1,-1)_{SO(10)}.
\end{align}
One has $ \lambda_{\Sigma^0_u}- \lambda_{\Sigma^0_d}= (0,0,-2,2,2)_{SO(10)}$, which is not a root of the Lie Algebra, as neither it nor its opposite are within the list in \eqref{eq:roots}.
This means then that the orthogonality condition \eqref{eq:orthogauge0} is satisfied for all non-diagonal generators in the $\overline{126}_H$ representation. Identical results apply for the $10_H$.

%% file: app_beta_functions.tex
\section{Coupling evolution}
\label{rgeevolution}
As usual, we can write the renormalisation group equations for the gauge couplings as

\begin{equation}
\label{2loopRG}
\dfrac{d \alpha^{-1}_i(\mu)}{d \ln \mu}=-\dfrac{a_i}{2 \pi}-\sum\limits_{j}\dfrac{b_{ij}}{8 \pi^2 \alpha^{-1}_j(\mu)} 
\end{equation}
where $i,j$ indices refer to different subgroups of the unified gauge group at the energy scale $\mu$ and 
\begin{equation}
\alpha^{-1}_i=\dfrac{4 \pi}{g_i^2}.
\end{equation}
\noindent
The $\beta$-function of a gauge coupling $g_i$ associated with the gauge group $G_i$ at two-loop order in the $\overline{\rm MS}$ scheme is given by \cite{Machacek:1983tz}
\begin{eqnarray}\label{eq:betas}
\beta g_i &=& \mu \dfrac{d g_i}{d\mu}=-\dfrac{g_i^3}{(4 \pi)^2}\left\lbrace\dfrac{11}{3}C_2 (G_i)-\dfrac{4}{3}\kappa\sum_a n_{a,i}S_i(\rho_{a})-\dfrac{1}{6}\eta \sum_m n_{m,i}S_i(\rho_{m})\right\rbrace  \nonumber\\ 
&& -\dfrac{g_i^3}{(4 \pi)^4}\left\lbrace \dfrac{34g_i^2}{3}\,  C_2(G_i)^2-\kappa\sum_a\left[4 \sum_j g_j^2 C_{2,j}(\rho_{a}) +\dfrac{20g_i^2}{3} C_2(G_i)\right]n_{a,i}S_i(\rho_{a})\right. \nonumber\\ 
&& \left.-\eta \sum_m\left[2\sum_j g_j^2 C_{2,j}(\rho_{m})+\dfrac{g^2_i}{3} C_2(G_i)\right]n_{m,i}S_i(\rho_{m})\right\rbrace. 
\end{eqnarray}
\noindent
In the above equation, the irreducible fermion and  scalar representations are labelled by $a$ and $m$, respectively. An irreducible representation of a product of groups can contain several copies of irreducible representations of the  individual groups. For a fermion representation $\rho_a$ we denote the  multiplicity of representations of the group $i$ as $n_{a,i}$; similarly, for a scalar representation $\rho_m$ we use the notation and $n_{m,i}$. $S_i(\rho_{a})$ is a shorthand for the  the Dynkin index of the irreducible representation of the group $i$ contained within a given fermion representation $\rho_a$. $S_i(\rho_{m})$ is the analogue for a scalar representation $\rho_m$.  $C_2(G_i)=S_i(\rm ad)$ designates the quadratic Casimir for the gauge fields in the adjoint representation of the gauge group $i$, while $C_{2,i}(\rho_{a})$, $C_{2,i}(\rho_{m})$ are the quadratic Casimirs of the representation of the group $i$ contained in $\rho_a$ and $\rho_m$, respectively. Finally, in equation \eqref{eq:betas} one has $\kappa=1,\frac{1}{2}$ for Dirac and Weyl fermions, respectively, and $\eta=1,2$ for real and complex scalar fields. 

At each scale, one has to take care as to which multiplets have to be included in the running.  
As described in section \ref{gaugecouplingunification}, we consider for the scalars an extended survival hypothesis, modified so as to allow for a 2HDM limit at low energies, while still having electroweak VEVs for all doublets in the 10 and $\overline{126}$, as needed to achieve realistic fermion masses.
According to the extended survival hypothesis, fields contribute to the running only if they acquire a VEV at lower scales. The exceptions are the doublets $\Sigma_u$, $\Sigma_d$ in the $(15,2,2)_{\rm PS}$ component of the $\overline{126}$, which are assumed to have a mass of the order of $\MBL$. A list of  the scalar components that get VEVs is given in table \ref{tab:vevs}. The decomposition of the fermions is given in table \ref{tab:fermions}.
With the previous assumptions, between $M_W$ and the lowest intermediate scale, the beta functions for all models mentioned in this paper are the beta functions of a two-Higgs doublet model, with gauge groups given in the order $SU(3)_C\times SU(2)_L\times U(1)_Y$: 
\begin{equation}
\label{eq:beta coefficients 2HDM}
a_{\rm 2HDM}=\left(\begin{array}{r}
-7 \\
-3 \\ 
\frac{21}{5} 
\end{array} \right);\hspace{1cm} b_{\rm 2HDM}=\left( \begin{array}{rcc}
-26 & \frac{9}{2}& \frac{11}{10} \\
12 & 8 & \frac{6}{5}\\
\frac{44}{5}& \frac{18}{5} & \frac{104}{25}
\end{array} \right),
\end{equation}

\noindent
where we used the GUT normalisation for the hypercharge gauge coupling, $g_Y=\sqrt{5/3} g'$, which ensures that the generator $T_Y$ enters the Lagrangian in the combination
$g_y\sqrt{3/5} T_Y$, with $\sqrt{3/5} T_Y$ a generator with the appropriate GUT normalisation.
%
For a consistent analysis at the two-loop order, at each symmetry breaking scale one needs to impose matching conditions for the gauge couplings that account for finite one-loop thresholds. For a symmetry
 breaking scale in which each ultraviolet  group $G^{\rm UV}_i$ is broken down to a  subgroup $G^{\rm IR}_i$, the matching conditions for the gauge couplings $g_i$ are of the form \cite{Hall:1980kf,Weinberg:1980wa}:
\begin{equation}
\label{eq:Threshold Def}
\dfrac{1}{\alpha^{\rm IR}_i(\mu)}=\dfrac{1}{\alpha^{\rm UV}_G(\mu)}-\dfrac{\lambda_i(\mu)}{12 \pi} ,
\end{equation}
where, assuming diagonal mass matrices compatible with the infrared gauge symmetries --that is, with a common mass for each IR multiplet-- one has
\begin{equation}\label{eq:lambda_threshold_1}
\begin{aligned}
\lambda_i(\mu)=& {C_2(G_i^{\rm UV})-C_2(G_i^{\rm IR})}  {-21 \sum_{j}S_i(V_{j}) \ln \frac{M_{V_j}}{\mu}} \\ 
& +{\eta\sum_{k_{\rm phys}}S_i(S_{k_{\rm phys}})\ln \dfrac{M_{S_{k_{\rm phys}}}}{\mu} } +  {8\kappa\sum_{l}S_i(F_l) \ln \dfrac{M_{F_l}}{\mu} }.
\end{aligned}\end{equation}
For each value of $i$ in the above equation, the $V_j$ designate the $G^{\rm IR}_i$ representations of gauge bosons that receive a mass at the corresponding threshold, leading to the breaking of the UV group $G^{\rm UV}_i$. $S_{k_{\rm phys}}$ designate the $G^{\rm IR}_i$ representations of 
heavy scalars that are integrated out at the threshold, omitting the unphysical Goldstone bosons. Finally, $F_l$ are the $G^{\rm IR}_i$ representations of heavy Dirac fermions that decouple at the threshold. The notation of $\eta$ and $\kappa$ is as in equation \eqref{eq:betas}. We will apply the former matching conditions at the threshold scale $\mu$ corresponding to the masses of the heavy gauge bosons, so that the contributions $\lambda_i^V$ can be ignored (up to subleading effects from possible lack of degeneracy of the massive gauge bosons from different groups, if the UV gauge group is not simple).

Next we  consider the case in which a $U^{IR}(1)$ group arises by combining two $U(1)$ subgroups in the UV, denoted as  $U_1(1)\in G^{\rm UV}_1$ and  $U_2(1)\in G^{\rm UV}_2$. The associated $U(1)$ generators
$T^{\rm IR}$, $T^{\rm UV}_1$, $T^{\rm UV}_2$ are all part of the Lie Algebra of the GUT group, and for GUT multiplets in representations $\rho$  of the GUT group with Dynkin index $S_{\rm GUT}(\rho)$, they satisfy
\begin{align}
{\rm Tr}_\rho (T^{\rm IR})^2=&\,\frac{1}{k_{IR}} \,S_{\rm GUT}(\rho), & {\rm Tr}_\rho (T^{\rm UV}_1)^2=&\,\frac{1}{k_1}\, S_{\rm GUT}(\rho),& {\rm Tr}_\rho (T^{\rm UV}_2)^2=&\,\frac{1}{k_2}\, S_{\rm GUT}(\rho).
\end{align}
The $k_i$ encode the normalisation of the U(1) generators when embedded into the GUT group, such that $\sqrt{k_{IR}}\,T^{\rm IR}$, $\sqrt{k_{1}}\,T^{\rm UV}_1$ and $\sqrt{k_{2}}\,T^{\rm UV}_2$ define GUT generators with the usual normalisation.
Assuming that $G^{\rm UV}_1$ and $G^{\rm UV}_2$ become  broken at the threshold to $G^{\rm IR}_1$ and $G^{\rm IR}_2$, respectively --so that part of the symmetry breaking is given by $G^{\rm UV}_1\otimes G^{\rm UV}_2\rightarrow G^{\rm IR}_1\otimes G^{\rm IR}_2\otimes U(1)^{\rm IR}$-- the matching of couplings goes as:
\begin{align}
\label{eq:threshold_2}
 \frac{1}{k_{IR}\,\alpha^{\rm IR}(\mu)}=\frac{1}{k_1\,\alpha^{\rm UV}_1(\mu)\sin^2\theta_{12}}-\frac{\tilde\lambda(\mu)}{12\pi}=\frac{1}{k_1\,\alpha^{\rm UV}_1(\mu)}+\frac{1}{k_2\,\alpha^{\rm UV}_2(\mu)}-\frac{\tilde\lambda(\mu)}{12\pi},
\end{align}
with
\begin{equation}\label{eq:lambda_threshold_2}
\begin{aligned}
           \tan^2\theta_{12}=&\,\frac{g_2^2k_2}{g_1^2 k_1},\\
           \tilde\lambda(\mu)=&\sum_{i=1}^2\,\left[\,\frac{C_2(G_i^{\rm UV})}{k_i}\right] -21 \sum_{j}Q^2_{IR}(V_j) \ln \frac{M_{V_j}}{\mu}\\ 
& \left.+\eta\sum_{k_{\rm phys}}Q^2_{IR}(S_{k_{\rm phys}})\ln \dfrac{M_{S_{k_{\rm phys}}}}{\mu}  +  8\kappa\sum_{l}Q^2_{IR}(F_l) \ln \dfrac{M_{F_l}}{\mu}\right].
\end{aligned}\end{equation}

In the above equation, $g_1$ and $g_2$ are the couplings of the groups $G^{\rm UV}_1$ and $G^{\rm UV}_2$, respectively, and $Q^2_{IR}$ represent the $U(1)$ charges under the generator $T^{\rm IR}$. The 
coupling $g_{IR}$ arising from the previous matching is in the GUT-compatible normalisation, as ensured by the factors of $k_i$.

Before moving on to the different models, we provide in table \ref{tab:thresholds} a summary of the decompositions of the different scalar multiplets under the gauge groups appearing in our breaking chains. The table also lists the scales at which the different representations decouple; decompositions are only provided for gauge groups which emerge at or above the decoupling scale, with the exception of fields decoupling at $M_{\rm PQ}$, since the latter may or may not break the gauge group. For fermion fields, the reader is referred to table \ref{tab:allfermions}.


\begin{table}[h]
\begin{align*}
\scriptsize
\begin{array}{|c|c|c|c|c|c|c|}
\hline
  SO(10)	& 4_C2_L2_R	&  4_C2_L1_R		&  3_C2_L1_R1_{B-L}	& 3_C2_L1_Y	&\text{Decoupling scale}           &\text{VEV}\\
  \hline
  \hline
   210_H        &(1,1,1)  	&			&			& 		&M_{\rm U}              &v_{\rm U}\\
                &(15,1,1)       &			&			&		&M_{\rm U}              &\\
                &(6,2,2)       &			&			&		&M_{\rm U}              &\\
                &(15,1,3)       &			&			&		&M_{\rm U}              &\\
                &(15,3,1)       &			&			&		&M_{\rm U}              &\\
                &(10,2,2)       &			&			&		&M_{\rm U}              &\\
                &(\overline{10},2,2)       &			&			&		&M_{\rm U}              &\\
  \hline
  10_H		&(6,1,1)	& 		&			&		&M_{\rm U}		&\\
  \cline{2-7}
  		&(1,2,2)	&(1,2,1/2)	& (1,2,1/2,0)	& (1,2,1/2)			&M_Z& v^{10}_u\\
		& 		& (1,2,-1/2)	&(1,2,-1/2,0)	&(1,2,-1/2) 	&M_Z & v^{10}_d\\
  \hline  
\overline{126}_H&(\overline{6},1,1)&		&			&			&M_{\rm U}		&\\
		&(\overline{10},3,1)&		&			&			&M_{\rm U}		&\\
\cline{2-7}
		&(10,1,3)	&(10,1,1)	&(1,1,1,-2)		&(1,1,0)		&M_{\rm BL}		&v_{\rm BL}\\
		 &		&		&(3,1,1,-2/3)		&(3,1,2/3)		&M_{\rm BL}		&\\
		 &		&		&(6,1,1,2/3)		&(6,1,4/3)		&M_{\rm BL}		&\\
\cline{3-7}
                 & 		&(10,1,0)	& (1,1,0,-2)		&(1,1,-1)		&{\rm Max}\{M_{\rm PQ},M_{\rm BL}\}		&\\
                 & 		&		& (3,1,0,-2/3)		&(3,1,-1/3)		&{\rm Max}\{M_{\rm PQ},M_{\rm BL}\}		&\\
                 & 		&		& (6,1,0,2/3)		&(6,1,1/3)		&{\rm Max}\{M_{\rm PQ},M_{\rm BL}\}		&\\
\cline{3-7}
                 & 		&(10,1,-1)	& (1,1,-1,-2)		&(1,1,-2)		&{\rm Max}\{M_{\rm PQ},M_{\rm BL}\}		&\\
                 & 		&		& (3,1,-1,-2/3)		&(3,1,-4/3)		&{\rm Max}\{M_{\rm PQ},M_{\rm BL}\}		&\\
                 & 		&		& (6,1,-1,2/3)		&(6,1,-2/3)		&{\rm Max}\{M_{\rm PQ},M_{\rm BL}\}		&\\
\cline{2-7}
		 &(15,2,2)	&(15,2,1/2)	&(1,2,1/2,0)		&(1,2,1/2)		& M_{\rm BL} 		&v^{126}_u\\
		 &		&			&(3,2,1/2,4/3)	&(3,2,7/6)		& M_{\rm BL} 		&\\
		&		&			&(\overline{3},2,1/2,-4/3)	&(\overline{3},2,-1/6)		& M_{\rm BL} 		&\\
		&		&			&(8,2,1/2,0)	&(8,2,1/2)		& M_{\rm BL} 		&\\
\cline{3-7}
		& 		& (15,2,-\frac{1}{2})	&(1,2,-1/2,0)	& (1,2,-1/2)   	&M_{\rm BL}			& v^{126}_d\\
		 &		&			&(3,2,-1/2,4/3)	&(3,2,1/6)		& M_{\rm BL} 		&\\
		&		&			&(\overline{3},2,-1/2,-4/3)	&(\overline{3},2,-7/6)		& M_{\rm BL} 		&\\
		&		&			&(8,2,-1/2,0)	&(8,2,-1/2)		& M_{\rm BL} 		&\\
  \hline  
  45_H		& (1,3,1)	&			&		&		&M_{\rm U}			&\\
		& (15,1,1)	&			&		&		&M_{\rm U}			&\\
		& (6,2,2)	&			&		&		&M_{\rm U}			&\\
\cline{2-7}
   		&(1,1,3) 	&(1,1,0)	& (1,1,0,0)		&(1,1,0)				&M_{\rm PQ}&v_{\rm PQ}\\
   		&	 	&(1,1,1)	& (1,1,1,0)		&(1,1,1)				&{\rm Max}\{M_{\rm PQ},M_{\rm BL}\}			&\\
   		&	 	&(1,1,-1)	& (1,1,-1,0)		&(1,1,-1)				&{\rm Max}\{M_{\rm PQ},M_{\rm BL}\}			&\\
\hline
 \end{array} 
 \end{align*}
 \caption{\label{tab:thresholds}Decomposition of the scalar multiplets according to the various subgroups in our breaking chains. We list the scales at which
 the different representations  decouple, and for a given representation we don't provide the decomposition under gauge groups that emerge below its decoupling scale, except for fields decoupling at $\MPQ$ (depending on the model, $\MPQ$ can lead to the breaking of the gauge group, or not).}
\end{table}

\subsection{\label{subsec:model1}Model 1}
In this minimal model the $45_H$ is not present. Between $M_{\rm BL}$ and $M_{\rm U}$, all particles mentioned in the second column of table \ref{tab:vevs} are included in the RG running. The resulting beta functions for the coupling constants of the gauge group $SU(4)_C\times SU(2)_L \times SU(2)_R$ are 
\begin{equation}
\label{runningminimalmodel}
a=\left(\begin{array}{r}
-\frac{7}{3}\\
2 \\ 
\frac{26}{3} 
\end{array} \right);\hspace{1cm} b=\left( \begin{array}{rcc}
\frac{2435}{6} & \frac{105}{2}& \frac{249}{2} \\
\frac{525}{2} & 73 & 48\\
\frac{1245}{2}& 48 & \frac{779}{3}
\end{array} \right).
\end{equation}
In this model, there are two high-scale thresholds associated with the breakings $SO(10)\rightarrow SU(4)_c\otimes SU(2)_L\otimes SU(2)_R\rightarrow SU(3)_C\otimes SU(2)_L\otimes U(1)_Y$.
The matching conditions of each gauge coupling at each threshold  are determined by the group structure and the particle content of the theory, following equations \eqref{eq:Threshold Def}, \eqref{eq:lambda_threshold_1}, \eqref{eq:threshold_2} and \eqref{eq:lambda_threshold_2}.
\subsubsection*{\label{subsec:MU_matching}Model 1 matching: $SO(10)\rightarrow SU(4)_c\otimes SU(2)_L\otimes SU(2)_R$}
This breaking is triggered at the  scale $M_U$ by the  $(1,1,1)$ VEV $v_U$ in the 210 representation, which, given its nonzero PQ charge (see \eqref{eq:PQGUT}), is taken as complex, as are the scalar representations $\overline{126}$ (complex to start with) and the $10$.
There are 24 broken generators, and correspondingly 24 Goldstone bosons inside the 210 representation, with the same quantum numbers as the broken generators. These Goldstones reside in the 
real part of the $(6,2,2)\subset 210$. According to the extended survival hypothesis, the scalar multiplets which don't get VEVs at lower scales should be integrated out. These are the multiplets  not included in table \ref{tab:vevs}, (see also table \ref{tab:thresholds}) and listed below:
\begin{equation}\begin{aligned}
 210\supset &\,\{(1,1,1), (15,1,1), {\rm Re }(6,2,2)\text{(G)}, {\rm Im}(6,2,2),(15,3,1),(15,1,3),(10,2,2),\\
 &(\overline{10},2,2)\},\\
 \overline{126}\supset&\,\{(\overline{6},1,1),(\overline{10},3,1)\},\\
 10\supset&\, (6,1,1),
\end{aligned}\end{equation}
where the $G$ indicates where the Goldstones reside. The relevant matching conditions are  \eqref{eq:Threshold Def}, \eqref{eq:lambda_threshold_1}, which give
\begin{equation}\label{eq:model1_U_lambdas}\begin{aligned}
&\left( \frac{1}{\alpha_{4C}(M_U)},\frac{1}{\alpha_{2R}(M_U)},\frac{1}{\alpha_{2L}(M_U)}\right)=(1,1,1)\,\frac{1}{\alpha_{G}(M_U)}-\frac{1}{12\pi}\,(\lambda^U_{4C},\lambda^U_{2R},\lambda^U_{2L}),\\
\, \\
&(\lambda^U_{4C},\lambda^U_{2R},\lambda^U_{2L})\,=\,(4,6,6)+(8,0,0)\,\log_U\, M_{(15,1,1)}+(4,6,6)\,\log_U\, M_{(6,2,2)}\\
&+(24,60,0)\,\log_U\, M_{(15,3,1)}+(24,0,60)\,\log_U\, M_{(15,1,3)}\\
&+(24,20,20)\,\log_U\,M_{(10,2,2)}M_{(\overline{10},2,2)}+(2,0,0)\,\log_U\, M_{(\overline{6},1,1)}M_{(6,1,1)}\\
&+(18,40,0)\,\log_U\, M_{(\overline{10},3,1)}.\\
\end{aligned}
\end{equation}
In the previous equations, we have defined 
\begin{align}
 \log_U A\cdot B\cdot\dots\equiv \log\left[\frac{A}{M_U}\frac{B}{M_U}\dots\right],
\end{align}
We have omitted threshold corrections depending on the masses of the heavy gauge bosons, as we assumed a choice of $\mu=M_U$ for which these contributions cancel; we will proceed similarly in the rest of the section.

\subsubsection*{\label{subsec:matching_BL}Model 1 matching: $SU(4)_c\otimes SU(2)_L\otimes SU(2)_R\rightarrow SU(3)_C\otimes SU(2)_L\otimes U(1)_Y$}

This breaking is triggered at the scale $\MBL$ by the VEV $v_R$ inside the $(1,1,0)_{SM}$ of the $\overline{126}$ (In the rest of this subsection, decompositions refer to the SM gauge group). There are 9 Goldstone bosons, contained in the real and imaginary parts of $\{(3,1,2/3),(1,1,-1)\}\supset \overline{126}$, and in the real part of $ (1,1,0)\subset \overline{126}$. All the $210$ fields were already integrated out at the previous threshold. Within the extended survival hypothesis, plus the assumption that $\Sigma_{u,d}$ decouple at $\MBL$, the scalar
fields to be integrated out at $\MBL$ are only inside the $\overline{126}$ --since the surviving ones from the $10$ include the fields $H_u,H_d$ that get VEVs at the electroweak scale-- and are given by (see table \ref{tab:thresholds}):
\begin{equation}\label{eq:matching_BL_fields}\begin{aligned}
 \overline{126}\subset&\,\{{\rm Re}(1,1,0),{\rm Im}(1,1,0)(G),(1,1,-1)(G),(1,1,-2),(3,1,2/3)(G)\\
 &(3,1,-1/3),(3,1,-4/3),(6,1,4/3),(6,1,1/3),(6,1,-2/3),(8,2,1/2)\\
 &(8,2,-1/2),(3,2,7/6),(3,2,1/6),(\overline{3},2,-1/6),(\overline{3},2,-7/6),(1,2,1/2)\\
 &(1,2,-1/2)\}.
\end{aligned}\end{equation}
The matching of the couplings of the groups $3_C$ and $2_L$ follows equations \eqref{eq:Threshold Def} and \eqref{eq:lambda_threshold_1}, which yield
\begin{equation}\label{eq:lambdas_32_BL}\begin{aligned}
&\left(\frac{1}{ \alpha_{3C}(\MBL)},\frac{1}{\alpha_{2L}(\MBL)}\right)=\left(\frac{1}{\alpha_{4C}(\MBL)},\frac{1}{\alpha_{2L}(\MBL)}\right)-\frac{1}{12\pi}\,(\lambda^{BL}_{3C},\lambda^{BL}_{2L}),\\
\,\\
&(\lambda^{BL}_{3C},\lambda^{BL}_{2L})\,=\,(1,0)+(1,0)\,\log_{BL}\, M_{(3,1,-1/3)} M_{(3,1,-4/3)}\\
&+(5,0)\,\log_{BL}\, M_{(6,1,4/3)} M_{(6,1,1/3)}M_{(6,1,-2/3)}+(12,8)\,\log_{BL}\, M_{(8,2,1/2)} M_{(8,2,-1/2)}\\
&+(2,3)\,\log_{BL}\, M_{(3,2,7/6)} M_{(3,2,1/6)}M_{(\overline{3},2,-7/6)}M_{(\overline{3},2,-1/6)}\\
&+(0,1)\,\log_{BL}\,M_{(1,2,1/2)}M_{(1,2,-1/2)},
\end{aligned}
\end{equation}
where now 
\begin{align}
\log_{BL} A^x\cdot B^y\cdot\dots\equiv \log\left[\left(\frac{A}{\MBL}\right)^x\left(\frac{B}{\MBL}\right)^y\dots\right].
\end{align}
The matching for the hypercharge coupling can be obtained by applying \eqref{eq:threshold_2} and \eqref{eq:lambda_threshold_2}. The relevant $U(1)$ generators in the UV are $T^{\rm UV}_1=(B-L)/2\subset SU(4)_C$ and
$T^{\rm UV}_2=T^3_{2R}\subset SU(2)_R$, with associated $k_1=3/2$, $k_2=1$. On the other hand, the GUT-normalised hypercharge coupling $g_Y$ has an associated $k_Y=3/5$. Then the matching goes as
\begin{equation}\label{eq:lambdaY}\begin{aligned}
\frac{1}{\alpha_{Y}(\MBL)}=&\,\frac{2}{5\alpha_{4C}(\MBL)\sin^2\theta_{BL}}-\frac{\lambda^{BL}_Y}{12\pi}=\frac{2}{5\alpha_{4C}(\MBL)}+\frac{3}{5\alpha_{2R}(\MBL)}-\frac{\lambda^{BL}_Y}{12\pi},\\
\tan^2\theta_{BL}=&\,\frac{2\alpha_{2R}}{3\alpha_{4C}},\\
\\
\lambda^{BL}_Y\,=&\,\frac{14}{5}+\log_{BL}M_{(1,1,-2)}^{24/5}M_{(3,1,-1/3)}^{2/5}M_{(6,1,1/3)}^{4/5}M_{(3,1,-4/3)}^{32/5}M_{(6,1,4/3)}^{64/5}\times\\
&\times M_{(6,1,-2/3)}^{16/5} M_{(8,2,1/2)}^{24/5}M_{(8,2,-1/2)}^{24/5}M_{(3,2,7/6)}^{49/5}M_{(3,2,1/6)}^{1/5}M_{(\overline{3},2,-7/6)}^{49/5}\times\\
&\times M_{(\overline{3},2,-1/6)}^{1/5}M_{(1,2,1/2)}^{3/5} M_{(1,2,-1/2)}^{3/5}.
\end{aligned}
\end{equation}
The above matching conditions, especially the matching of hypercharge to the higher gauge groups, are in agreement with existing literature \cite{Fonseca:2013bua,Bertolini:2013vta}.

\subsection{Model 2.1. Case A:  $\MPQ>\MBL$}

The $45_H$ breaks $4_C2_L2_R$ to $4_C2_L1_R$, so we need to consider the RG running for both groups.
Between $\MU$ and $\MPQ$, the beta functions for the coupling constants of $SU(4)_C\times SU(2)_L \times SU(2)_R$ are given by:
\begin{align}
 \label{rgeinthigh}
a=\left(\begin{array}{r}
-\frac{7}{3} \\
2 \\ 
\frac{28}{3} 
\end{array} \right);\hspace{1cm} b=\left( \begin{array}{rcc}
\frac{2435}{6} & \frac{105}{2}& \frac{249}{2} \\
\frac{525}{2} & 73 & 48\\
\frac{1245}{2}& 48 & \frac{835}{3}
\end{array} \right).
\end{align}
The differences between \eqref{runningminimalmodel} and \eqref{rgeinthigh} come from the inclusion of the $(1,1,3)$ multiplet of the complex $45_H$ (the rest of the fields in the $45$ are integrated out at $M_U$, to conform with the extended survival hypothesis --see table \ref{tab:thresholds}). Between $\MPQ$ and $\MBL$, the gauge group is $SU(4)_C\times SU(2)_L \times U(1)_R$ with beta functions
\begin{align}
 \label{rgeintlow}
a=\left(\begin{array}{r}
-\frac{13}{3} \\
2 \\ 
\frac{38}{3} 
\end{array} \right);\hspace{1cm} b=\left( \begin{array}{rcc}
\frac{1691}{6} & \frac{105}{2}& \frac{59}{2} \\
\frac{525}{2} & 73 & 16\\
\frac{885}{2}& 48 & 59
\end{array} \right).
\end{align}
The matching conditions are given next.

\subsubsection*{Model 2.1.A matching: $SO(10)\rightarrow SU(4)_c\otimes SU(2)_L\otimes SU(2)_R$}
Things go as in \ref{subsec:MU_matching}, but with the following differences:
in models such as the presently analyzed --and the ones that will follow-- the 210 is not charged under PQ and can be taken as real, which reduces the threshold corrections. Also, there are new fields in the $45$ (which is charged under PQ and thus complex) which have to be integrated out, as 
they don't get VEVs at lower scales. The scalar multiplets to be integrated out at the $M_U$ threshold are  then (see table \ref{tab:thresholds}):
\begin{equation}\begin{aligned}
 210\supset &\,\{(1,1,1), (15,1,1), (6,2,2)\text{(G)},(15,3,1),(15,1,3),(10,2,2),(\overline{10},2,2)\},\\
 \overline{126}\supset&\,\{(\overline{6},1,1),(\overline{10},3,1)\},\\
 10\supset&\, (6,1,1),\\
 45\supset&\, \{(6,2,2),(1,3,1),(15,1,1)\}.
\end{aligned}\end{equation}
The  matching conditions\eqref{eq:Threshold Def} and \eqref{eq:lambda_threshold_1} give now (using the same notation as before)
\begin{equation}\label{eq:matching_U_21}\begin{aligned}
&(\lambda^U_{4C},\lambda^U_{2R},\lambda^U_{2L})\,=\,(4,6,6)+(4,0,0)\,\log_U\, M_{(15,1,1)}+(12,30,0)\,\log_U\, M_{(15,3,1)}\\
&+(12,0,30)\,\log_U\, M_{(15,1,3)}+(12,10,10)\,\log_U\,M_{(10,2,2)}M_{(\overline{10},2,2)}\\
&+(2,0,0)\,\log_U\, M_{(\overline{6},1,1)}M_{(6,1,1)}+(18,40,0)\,\log_U\, M_{(\overline{10},3,1)}\\
&+(8,0,0)\,\log_U\, M'_{(15,1,1)}+(0,4,0)\,\log_U \,M_{(1,3,1)}+(8,12,12)\,\log_U \,M_{(6,2,2)}.\\
\end{aligned}
\end{equation}

\subsubsection*{\label{subsec:21A_MPQ}Model 2.1.A matching: $ SU(4)_c\otimes SU(2)_L\otimes SU(2)_R\rightarrow SU(4)_C\otimes SU(2)_L \otimes U(1)_R$}
This breaking is triggered at a scale $M_{\rm PQ}$ by the VEV $v_{\rm PQ}$ within the $(1,1,0)_{4_C 2_L 1_R}\subset 45$ (We consider decompositions along $4_C 2_L 1_R$ in the rest of this subsection). There are two broken generators of $SU(2)_R$, whose Goldstones are in real part of the $(1,1,1)$ and $(1,1,-1)$  components of the 45. The scalar fields to be integrated out are (see table \ref{tab:thresholds})
\begin{equation}\label{eq:21A_MPQ_fields}\begin{aligned}
 45\supset&\, \{{\rm Re}(1,1,1)(G),{\rm Im}(1,1,1),{\rm Re}(1,1,-1)(G),{\rm Im}(1,1,-1),(1,1,0)\},\\
 \overline{126}\supset&\,\{(10,1,0),(10,1,-1)\}.
\end{aligned}\end{equation}
This gives threshold corrections (with notation that should be clear from the above cases)
\begin{equation}\begin{aligned}
\label{eq:21A_MPQ_lambda}
(\lambda^{PQ}_{4C}, \lambda^{PQ}_{2L},\lambda^{PQ}_{1R})=&\,(0,0,2)+(0,0,1)\log_{PQ} M_{1,1,1} M_{1,1,-1}+(6,0,0)\log_{PQ}M_{10,1,0}\\
&+(6,0,20)\log_{PQ}M_{10,1,-1}.
\end{aligned}\end{equation}

\subsubsection*{Model 2.1.A matching: $SU(4)_c\otimes SU(2)_L\otimes U(1)_R\rightarrow SU(3)_C\otimes SU(2)_L\otimes U(1)_Y$}

This case is similar to that in section \ref{subsec:matching_BL}, with the following differences. First, the Goldstones from the breaking of $SU(2)_R$ are now shared between the $45$ (whose Goldstones were 
integrated out at the
previous thresholds) and the $\overline{126}$. Additionally, now one must exclude from the loops the heavy gauge bosons that were decoupled at the scale $M_{\rm PQ}$. 
All the $45$ fields were integrated out at the latter scale, so that the fields that acquire a mass at the scale $\MBL$ are (see table \ref{tab:thresholds}):
\begin{equation}\begin{aligned}
 \overline{126}\supset&\,\{{\rm Re}(1,1,0),{\rm Im}(1,1,0)(G),(3,1,2/3)(G),(6,1,4/3),(8,2,1/2),(8,2,-1/2),\\
 &(3,2,7/6),(3,2,1/6),(\overline{3},2,-1/6),(\overline{3},2,-7/6),(1,2,1/2),(1,2,-1/2)\}.
\end{aligned}\end{equation}
Note the difference in Goldstone mode counting with respect to \eqref{eq:matching_BL_fields}, and the absence of the descendants of the $(10,1,0)_{4_C 2_L 1_R}, (10,1,-1)_{4_C 2_L 1_R}$. The values of $\lambda^{BL}_{3C},\lambda^{BL}_{2L}$ are, as follows follows from equations \eqref{eq:Threshold Def} and \eqref{eq:lambda_threshold_1}, 
\begin{equation}\label{eq:lambdas_32_BL_21A}\begin{aligned}
(\lambda^{BL}_{3C},\lambda^{BL}_{2L})\,=&\,(1,0)+(5,0)\,\log_{BL}\, M_{(6,1,4/3)} +(12,8)\,\log_{BL}\, M_{(8,2,1/2)} M_{(8,2,-1/2)}\\
&+(2,3)\,\log_{BL}\, M_{(3,2,7/6)} M_{(3,2,1/6)}M_{(\overline{3},2,-7/6)}M_{(\overline{3},2,-1/6)}\\
&+(0,1)\,\log_{BL}\,M_{(1,2,1/2)}M_{(1,2,-1/2)},
\end{aligned}
\end{equation}
 while for 
$\lambda_Y$ we now have 
\begin{equation}\label{eq:lambdaY2}\begin{aligned}
\lambda_Y\,=&\,\frac{8}{5}+\log_{BL}M_{(6,1,4/3)}^{64/5}M_{(8,2,1/2)}^{24/5}M_{(8,2,-1/2)}^{24/5}M_{(3,2,7/6)}^{49/5}M_{(3,2,1/6)}^{1/5}\times\\
&\times\!M_{(\overline{3},2,-7/6)}^{49/5}M_{(\overline{3},2,-1/6)}^{1/5}M_{(1,2,1/2)}^{3/5}M_{(1,2,-1/2)}^{3/5}.
\end{aligned}
\end{equation}


\subsection{Model 2.1. Case B:  $\MBL>\MPQ$}

 Since the $45_H$ in this case acquires its VEV only after the $126_H$, there is only one intermediate gauge symmetry group to consider, $SU(4)_C\times SU(2)_L \times SU(2)_R$. The beta functions  between $M_U$ and $\MBL$ are given by \eqref{rgeinthigh}, while for scales below $\MBL$ they are given by \eqref{eq:beta coefficients 2HDM} (the only field in the $45$ surviving below the $\MBL$ threshold is a SM singlet, and thus does not contribute to the beta functions).
Since the symmetry breaking chain is the same is in model \ref{gutaxionmodel}, the matching conditions are similar. However we have to take into account additional heavy particles from the $45$ multiplet.

\subsubsection*{Model 2.1.B matching: $SO(10)\rightarrow SU(4)_c\otimes SU(2)_L\otimes SU(2)_R$}
The matching goes in this case as in equation \eqref{eq:matching_U_21}.

\subsubsection*{Model 2.1.B matching: $SU(4)_c\otimes SU(2)_L\otimes SU(2)_R\rightarrow SU(3)_C\otimes SU(2)_L\otimes U(1)_Y$}

The matching is similar to that in section \ref{subsec:matching_BL}, but with the difference that now the components of the $45$ which do not get a VEV below the $\MBL$ threshold have to be integrated out --in addition to the fields in \eqref{eq:matching_BL_fields}-- so as to comply with the extended survival hypothesis. As the new components are singlets under $SU(3)_C$ and $SU(2)_L$, the matching of $\alpha_{3C}$
and $\alpha_{2L}$ is as in \eqref{eq:lambdas_32_BL}. On the other hand, the threshold correction for $\alpha_Y$ receives extra contributions:
\begin{equation}\label{eq:lambdaY3}\begin{aligned}
\lambda^{BL}_Y\,=&\,\frac{14}{5}+\log_{BL}M_{(1,1,-2)}^{24/5}M_{(3,1,-1/3)}^{2/5}M_{(6,1,1/3)}^{4/5}M_{(3,1,-4/3)}^{32/5}M_{(6,1,4/3)}^{64/5}\times\\
&\times M_{(6,1,-2/3)}^{16/5} M_{(8,2,1/2)}^{24/5}M_{(8,2,-1/2)}^{24/5}M_{(3,2,7/6)}^{49/5}M_{(3,2,1/6)}^{1/5}M_{(\overline{3},2,-7/6)}^{49/5}\times\\
&\times M_{(\overline{3},2,-1/6)}^{1/5}M_{(1,2,1/2)}^{3/5} M_{(1,2,-1/2)}^{3/5}M_{(1,1,1)}^{6/5}M_{(1,1,-1)}^{6/5}.
\end{aligned}
\end{equation}

\subsubsection*{Model 2.1.B matching across the PQ threshold (no group breaking)}
At the $\MPQ$ threshold there is only one field component getting a VEV, $(1,1,0)_{SM}\subset 45$. This is a singlet under all SM groups, and thus it contributes to no finite threshold corrections. The matching is then trivial.

\subsection{Model 2.2. Case A:  $\MPQ>\MBL$}

In contrast to  Model 2.1.A one has to consider the additional fermions in the $10$ representation, which contribute to the running between $M_U$ and $\MPQ$. The beta functions of $SU(4)_C\times SU(2)_L \times SU(2)_R$ are changed accordingly:
\begin{align}
 \label{rgehlshigh}
a=\left(\begin{array}{r}
-1 \\
\frac{10}{3} \\ 
\frac{32}{3} 
\end{array} \right);\hspace{1cm} b=\left( \begin{array}{rcc}
\frac{885}{2} & \frac{105}{2}& \frac{249}{2} \\
\frac{525}{2} & \frac{268}{3} & 51\\
\frac{1245}{2}& 51& \frac{884}{3}
\end{array} \right).
\end{align}
Between $\MPQ$ and $\MBL$, the heavy fermions have been integrated out and do not contribute to the running anymore. The beta functions are given by \eqref{rgeintlow}. Below $\MBL$ the running is that in equation \eqref{eq:beta coefficients 2HDM}. The matching conditions are discussed next.

\subsubsection*{Model 2.2.A matching: $SO(10)\rightarrow SU(4)_c\otimes SU(2)_L\otimes SU(2)_R$}
As in \eqref{eq:matching_U_21}.

\subsubsection*{Model 2.2.A matching: $ SU(4)_c\otimes SU(2)_L\otimes SU(2)_R\rightarrow SU(4)_C\otimes SU(2)_L \otimes U(1)_R$}
The difference with the matching in 2.1.A (section \ref{subsec:21A_MPQ}) comes from the heavy fermions in the 10 of $SO(10)$, which acquire  masses due to the VEV $v_{\rm PQ}$. The Weyl fermions from the two multiplets in the $10$ can be grouped into massive Dirac fermions. Then in addition to the fields in \eqref{eq:21A_MPQ_fields}, 
one has to integrate out the  heavy Dirac fermions from the $10_F$ in the following representations of $SU(4)_C\otimes SU(2)_L \otimes U(1)_R$ (see table \ref{tab:allfermions}):
\begin{align}
\label{eq:fermions10_PS}
 \{10_F,10_F\}\supset \{(6,1,0),(1,2,1/2),(1,2,-1/2)\}.
\end{align}
As a consequence of the extra fields above, \eqref{eq:21A_MPQ_lambda} must be modified to
\begin{equation}\begin{aligned}
\label{eq:21B_MPQ_lambda}
(\lambda^{PQ}_{4C}, \lambda^{PQ}_{2L},\lambda^{PQ}_{1R})=&\,(0,0,2)+(0,0,1)\log_{PQ} M_{1,1,1} M_{1,1,-1}+(6,0,0)\log_{PQ}M_{10,1,0}\\
&+(6,0,20)\log_{PQ}M_{10,1,-1}+(8,0,0)\log_{PQ}M_{(6,1,0)}\\
&+(0,4,4)\log_{PQ}M_{(1,2,1/2)}M_{(1,2,-1/2)}.
\end{aligned}\end{equation}


\subsubsection*{Model 2.2.A matching: $SU(4)_c\otimes SU(2)_L\otimes U(1)_R\rightarrow SU(3)_C\otimes SU(2)_L\otimes U(1)_Y$}

With the extra fermions already integrated out, the matching goes as in \eqref{eq:lambdas_32_BL_21A} and \eqref{eq:lambdaY2}.


\subsection{Model 2.2. Case B:  $\MBL>\MPQ$}

The fermions contribute the the RG running down to the scale at which they acquire their masses - $\MPQ$.
Between $M_U$ and $\MBL$, the relevant gauge group is $SU(4)_C\times SU(2)_L\times SU(2)_R$ and the beta functions the same as \eqref{rgehlshigh}. At lower scales between $\MBL$ and $\MPQ$ however, the additional fermions are still active and contribute to the coupling evolution. The corresponding beta functions for the gauge group $SU(3)_C\times SU(2)_L\times U(1)_Y$ are 

\begin{align}
 \label{rgehlslow}
a=\left(\begin{array}{r}
-\frac{17}{3} \\
-\frac{5}{3} \\ 
\frac{83}{15} 
\end{array} \right);\hspace{1cm} b=\left( \begin{array}{rcc}
-\frac{2}{3} & \frac{9}{2}& \frac{41}{30} \\
12 & \frac{73}{3} & \frac{9}{5}\\
\frac{164}{15}& \frac{27}{5}& \frac{347}{75}
\end{array} \right).
\end{align}

\subsubsection*{Model 2.2.B matching: $SO(10)\rightarrow SU(4)_c\otimes SU(2)_L\otimes SU(2)_R$}
As in equation \eqref{eq:matching_U_21}.

\subsubsection*{Model 2.2.B matching: $SU(4)_c\otimes SU(2)_L\otimes SU(2)_R\rightarrow SU(3)_C\otimes SU(2)_L\otimes U(1)_Y$}

As in \eqref{eq:lambdaY3}.

\subsubsection*{Model 2.2.B matching across the PQ threshold (no group breaking)}
From the 45, only the SM singlet was left below the $\MBL$ threshold. This field  cannot contribute to one-loop finite corrections of the gauge couplings.
Still, one has to integrate out the fermions in the $10$ of $SO(10)$, whose decomposition under the SM gauge group is  (see table \ref{tab:allfermions})
\begin{align}
 \{10_F,10_F\}\supset \{(3,1,1/3),(\bar{3},1,-1/3),(1,2,1/2),(1,2,-1/2)\}.
\end{align}
The matching is then
\begin{equation}
\label{eq:lambdas_SM_22B}
\begin{aligned}
 (\lambda^{PQ}_{3C}, \lambda^{PQ}_{2L},\lambda^{PQ}_{Y})=&\,\left(4,0,\frac{8}{5}\right)\!\log_{PQ} M_{(3,1,1/3)}M_{(\bar{3},1,-1/3)}\\
 &+\left(0,4,\frac{12}{5}\right)\!\log_{PQ} M_{(1,2,1/2)}M_{(1,2,-1/2)}.
\end{aligned}\end{equation}
\subsection{Model 3.1}
The model differs from Model 1 by the addition of a singlet under all gauge symmetries. Then one can take the 
beta functions and matching conditions  as in section \ref{subsec:model1} (if the 210 is again assumed to be complex). 
\subsection{Model 3.2. Case A:  $\MPQ>\MBL$}

With the heavy quarks acquiring their masses before the B-L scale is broken, we obtain the following beta function coefficients for scales between $\MU$ and $\MPQ$:
\begin{align}
 \label{rge32ahigh}
a=\left(\begin{array}{r}
-1 \\
\frac{10}{3} \\ 
10 
\end{array} \right);\hspace{1cm} b=\left( \begin{array}{rcc}
\frac{885}{2} & \frac{105}{2}& \frac{249}{2} \\
\frac{525}{2} & \frac{268}{3} & 51\\
\frac{1245}{2}& 51& 276
\end{array} \right).
\end{align}
At scales below $\MPQ$ and above $\MBL$, the beta coefficients are given by \eqref{runningminimalmodel}.
At the lowest scales above $M_Z$, we have two Higgs doublet running given by \eqref{eq:beta coefficients 2HDM} as usual. The matching will only
differ from that of Model 1 due to the effects of the fermions, as summarised next.

\subsubsection*{Model 3.2.A matching: $SO(10)\rightarrow SU(4)_c\otimes SU(2)_L\otimes SU(2)_R$}
As in \eqref{eq:model1_U_lambdas}.

\subsubsection*{Model 3.2.A matching at $\MPQ$ (no group breaking)}

The only effect comes from the fermions in the $10$, whose representations under the Pati-Salam group are
\begin{align}
 \{10_F,10_F\}\supset \{(6,1,1),(1,2,2)\}.
\end{align}
Their effect
on the matching follows from \eqref{eq:lambda_threshold_1}:
\begin{equation}
\label{eq:PQ_matching_31}
 \begin{aligned}
  (\lambda^U_{4C},\lambda^U_{2R},\lambda^U_{2L})\,=\,(8,0,0)\log_{PQ}M_{(6,1,1)}+(0,8,8)\log_{PQ}M_{(1,2,2)}.
 \end{aligned}
\end{equation}


\subsubsection*{Model 3.2.A matching: $SU(4)_c\otimes SU(2)_L\otimes SU(2)_R\rightarrow SU(3)_C\otimes SU(2)_L\otimes U(1)_Y$}

With the extra fermions already integrated out, the matching is as in \eqref{eq:lambdas_32_BL} and \eqref{eq:lambdaY}.

\subsection{Model 3.2. Case B:  $\MBL>\MPQ$}

Between $\MU$ and $\MBL$, the beta coefficients are given by \eqref{rge32ahigh}. At scales between $\MBL$ and $\MPQ$,  we obtain \eqref{rgehlslow} (in model 2.2.B, the only component from the $45$ left below the $\MBL$ threshold is the SM singlet, which does not contribute to the beta functions). Below $\MPQ$ one has the 2HDM running of \eqref{eq:beta coefficients 2HDM}.

\subsubsection*{Model 3.2.B matching: $SO(10)\rightarrow SU(4)_c\otimes SU(2)_L\otimes SU(2)_R$}
As in equation \eqref{eq:model1_U_lambdas}.


\subsubsection*{Model 3.2.B matching: $SU(4)_c\otimes SU(2)_L\otimes SU(2)_R\rightarrow SU(3)_C\otimes SU(2)_L\otimes U(1)_Y$}

As in in \eqref{eq:lambdas_32_BL} and \eqref{eq:lambdaY}.

\subsubsection*{Model 3.2.B matching at $\MPQ$ (no group breaking)}

As in \eqref{eq:lambdas_SM_22B}.

%% file: app_discrete_symmetry.tex
\section{Higher dimensional PQ-violating operators}
\label{discretesymmetry}
In models where the Peccei-Quinn symmetry is a low-energy remnant of a discrete global symmetry --which can protect the axion sector from gravitational corrections \cite{Babu:1992cu,Barr:1992qq,Kamionkowski:1992mf}-- one can derive constraints on charges of the scalar fields under such discrete symmetries, see e.g. \cite{Ringwald:2015dsf}. (For other works using discrete symmetries to protect the axion's interactions in models with extended gauge groups, see for example \cite{Dias:2002gg,Babu:2002ic,Dias:2003iq}, and the recent \cite{Bjorkeroth:2017tsz,DiLuzio:2017tjx}).

For the derivation we need to know how the higher-order Peccei-Quinn violating operators that are allowed by the discrete symmetry enter in the axion effective potential. In order to keep in line with the observations of the electric dipole moment of the neutron, one has to ensure that the contributions of these higher order operators are small enough. In the models described in \cite{Ringwald:2015dsf}, the VEV that breaks the accidental Peccei-Quinn symmetry is the VEV of the additional scalar $\sigma$ whose phase eventually becomes the axion. The dominant contribution to the axion potential then comes from the PQ violating operator $\frac{\sigma^N}{M_p^{N-4}}$. In Models 2.1 and 3.1 of the present paper, it is not quite so obvious which operator is dominant, as additional fields that acquire large VEVs are present. In particular, the $\overline{126}_H$ aquires a VEV $v_{\rm BL}$ that can even be larger than $v_{\rm PQ}$.  In the following we will analyze the dominant contributions derived from the symmetry defined in table \ref{tab:Z10} for Model 2.1.

The discrete symmetry allows for Peccei-Quinn violating operators in the Lagrangian of the form
\begin{equation}
 \frac{{\overline{126}_H}^{n_{\rm BL}}\, 45_H^{n_{\rm PQ}}}{M_p^{D-4}} + h.c. \supset \frac{v_{\rm PQ}^{n_{\rm PQ}} v_{\rm BL}^{n_{\rm BL}}}{M_p^{D-4}}+ h.c.,
\end{equation}
with 
\begin{equation}
 D=n_{\rm PQ}+n_{\rm BL}.
\end{equation}
Not all of these operators are allowed by the gauge symmetries of the models.
From table \ref{tab:Z10} we read that
\begin{equation}
 \frac{1}{10} n_{\rm PQ}+\frac{1}{20} n_{\rm BL}=Z \hspace{20pt} \leftrightarrow \hspace{20pt} 2 n_{\rm PQ}+ n_{\rm BL}=20 Z
\end{equation}
must be satisfied for some positive integer $Z$. (Negative $Z$ just correspond to complex conjugates of these operators, $Z=0$ describes Peccei-Quinn conserving operators). The lowest dimensional such operator is $\mathcal{O}_{10}:=\frac{45_H^{10}}{M_p^6}\supset \frac{v_{\rm PQ}^{10}}{M_p^6}$. At dimension 10, this is the only PQ violating operator. We now impose that all higher order operators should be suppressed with respect to $\mathcal{O}_{10}$:
\begin{equation}
\label{comp1}
 \frac{v_{\rm PQ}^{10}}{M_p^6}>\frac{v_{\rm PQ}^{n_{\rm PQ}} v_{\rm BL}^{n_{\rm BL}}}{M_p^{D-4}} \hspace{10 pt} \leftrightarrow \hspace{10pt} v_{\rm BL}^{20 Z-2n_{\rm PQ}} M_p^{10 -n_{\rm PQ}-n_{\rm BL}}<v_{\rm PQ}^{10-n_{\rm PQ}}.
\end{equation}

Let us first consider the case $Z=1$. In this case, $n_{\rm PQ} \leq 10$ and \eqref{comp1} becomes 
\begin{equation}
 v_{\rm BL}^2 M_p^{-1}<v_{\rm PQ}.
\end{equation}
Using the upper bound  $v_{\rm BL}<10^{13} \GeV$ derived in sections \ref{gaugecouplingunification} and \ref{discussionandconclusions}, we obtain a lower bound on $v_{\rm PQ}$:
\begin{equation}
(10^{13} \GeV)^2 (10^{18} \GeV)^{-1} = 10^8 \GeV <v_{\rm PQ}.
\end{equation}
This lower bound is fulfilled if the axion is the dominant component of dark matter (compare figure \ref{fig:summaryshort}). The case $Z=1$ covers all operators up to dimension 19. Operators of even higher dimensions are suppressed by higher orders of $M_p$ and can therefore be neglected. We can conclude that $\mathcal{O}_{10}$ is the dominating PQ violating operator in this model. 

Our model is a special case of the DFSZ axion model and the calculation of the relic abundance goes through as in reference \cite{Ringwald:2015dsf} (however with $N=10$ and $N_{\rm DW}=3$). The same argument can be applied to Model 3.1, replacing $45_H\rightarrow S$.

%% file: axion_mass_so10.bbl
\providecommand{\href}[2]{#2}\begingroup\raggedright\begin{thebibliography}{100}

\bibitem{Ballesteros:2016euj}
G.~Ballesteros, J.~Redondo, A.~Ringwald and C.~Tamarit, \emph{{Unifying
  inflation with the axion, dark matter, baryogenesis and the seesaw
  mechanism}},
  \href{https://doi.org/10.1103/PhysRevLett.118.071802}{\emph{Phys. Rev. Lett.}
  {\bfseries 118} (2017) 071802},
  [\href{https://arxiv.org/abs/1608.05414}{{\ttfamily 1608.05414}}].

\bibitem{Ballesteros:2016xej}
G.~Ballesteros, J.~Redondo, A.~Ringwald and C.~Tamarit, \emph{{Standard
  Model-Axion-Seesaw-Higgs Portal Inflation. Five problems of particle physics
  and cosmology solved in one stroke}},
  \href{https://doi.org/10.1088/1475-7516/2017/08/001}{\emph{JCAP} {\bfseries
  1708} (2017) 001}, [\href{https://arxiv.org/abs/1610.01639}{{\ttfamily
  1610.01639}}].

\bibitem{Minkowski:1977sc}
P.~Minkowski, \emph{{$\mu \to e\gamma$ at a Rate of One Out of $10^{9}$ Muon
  Decays?}}, \href{https://doi.org/10.1016/0370-2693(77)90435-X}{\emph{Phys.
  Lett.} {\bfseries 67B} (1977) 421--428}.

\bibitem{GellMann:1980vs}
M.~Gell-Mann, P.~Ramond and R.~Slansky, \emph{{Complex Spinors and Unified
  Theories}}, {\emph{Conf. Proc.} {\bfseries C790927} (1979) 315--321},
  [\href{https://arxiv.org/abs/1306.4669}{{\ttfamily 1306.4669}}].

\bibitem{Yanagida:1979as}
T.~Yanagida, \emph{{HORIZONTAL SYMMETRY AND MASSES OF NEUTRINOS}}, {\emph{Conf.
  Proc.} {\bfseries C7902131} (1979) 95--99}.

\bibitem{Mohapatra:1979ia}
R.~N. Mohapatra and G.~Senjanovic, \emph{{Neutrino Mass and Spontaneous Parity
  Violation}}, \href{https://doi.org/10.1103/PhysRevLett.44.912}{\emph{Phys.
  Rev. Lett.} {\bfseries 44} (1980) 912}.

\bibitem{Peccei:1977hh}
R.~D. Peccei and H.~R. Quinn, \emph{{CP Conservation in the Presence of
  Instantons}}, \href{https://doi.org/10.1103/PhysRevLett.38.1440}{\emph{Phys.
  Rev. Lett.} {\bfseries 38} (1977) 1440--1443}.

\bibitem{Weinberg:1977ma}
S.~Weinberg, \emph{{A New Light Boson?}},
  \href{https://doi.org/10.1103/PhysRevLett.40.223}{\emph{Phys. Rev. Lett.}
  {\bfseries 40} (1978) 223--226}.

\bibitem{Wilczek:1977pj}
F.~Wilczek, \emph{{Problem of Strong p and t Invariance in the Presence of
  Instantons}}, \href{https://doi.org/10.1103/PhysRevLett.40.279}{\emph{Phys.
  Rev. Lett.} {\bfseries 40} (1978) 279--282}.

\bibitem{Fukugita:1986hr}
M.~Fukugita and T.~Yanagida, \emph{{Baryogenesis Without Grand Unification}},
  \href{https://doi.org/10.1016/0370-2693(86)91126-3}{\emph{Phys. Lett.}
  {\bfseries B174} (1986) 45--47}.

\bibitem{Preskill:1982cy}
J.~Preskill, M.~B. Wise and F.~Wilczek, \emph{{Cosmology of the Invisible
  Axion}}, \href{https://doi.org/10.1016/0370-2693(83)90637-8}{\emph{Phys.
  Lett.} {\bfseries 120B} (1983) 127--132}.

\bibitem{Abbott:1982af}
L.~F. Abbott and P.~Sikivie, \emph{{A Cosmological Bound on the Invisible
  Axion}}, \href{https://doi.org/10.1016/0370-2693(83)90638-X}{\emph{Phys.
  Lett.} {\bfseries 120B} (1983) 133--136}.

\bibitem{Dine:1982ah}
M.~Dine and W.~Fischler, \emph{{The Not So Harmless Axion}},
  \href{https://doi.org/10.1016/0370-2693(83)90639-1}{\emph{Phys. Lett.}
  {\bfseries 120B} (1983) 137--141}.

\bibitem{Borsanyi:2016ksw}
S.~Borsanyi et~al., \emph{{Calculation of the axion mass based on
  high-temperature lattice quantum chromodynamics}},
  \href{https://doi.org/10.1038/nature20115}{\emph{Nature} {\bfseries 539}
  (2016) 69--71}, [\href{https://arxiv.org/abs/1606.07494}{{\ttfamily
  1606.07494}}].

\bibitem{Klaer:2017ond}
V.~B. Klaer and G.~D. Moore, \emph{{The dark-matter axion mass}},
  \href{https://doi.org/10.1088/1475-7516/2017/11/049}{\emph{JCAP} {\bfseries
  1711} (2017) 049}, [\href{https://arxiv.org/abs/1708.07521}{{\ttfamily
  1708.07521}}].

\bibitem{Boucenna:2017fna}
S.~M. Boucenna and Q.~Shafi, \emph{{Axion Inflation, Proton Decay and
  Leptogenesis in $SU(5)\times U(1)_{PQ}$}},
  \href{https://arxiv.org/abs/1712.06526}{{\ttfamily 1712.06526}}.

\bibitem{Georgi:1974my}
H.~Georgi, \emph{{The State of the Art - Gauge Theories}},
  \href{https://doi.org/10.1063/1.2947450}{\emph{AIP Conf. Proc.} {\bfseries
  23} (1975) 575--582}.

\bibitem{Fritzsch:1974nn}
H.~Fritzsch and P.~Minkowski, \emph{{Unified Interactions of Leptons and
  Hadrons}}, \href{https://doi.org/10.1016/0003-4916(75)90211-0}{\emph{Annals
  Phys.} {\bfseries 93} (1975) 193--266}.

\bibitem{Reiss:1981nd}
D.~B. Reiss, \emph{{INVISIBLE AXION AT AN INTERMEDIATE SYMMETRY BREAKING
  SCALE}}, \href{https://doi.org/10.1016/0370-2693(82)91091-7}{\emph{Phys.
  Lett.} {\bfseries 109B} (1982) 365--368}.

\bibitem{Mohapatra:1982tc}
R.~N. Mohapatra and G.~Senjanovic, \emph{{The Superlight Axion and Neutrino
  Masses}}, \href{https://doi.org/10.1007/BF01577819}{\emph{Z. Phys.}
  {\bfseries C17} (1983) 53--56}.

\bibitem{Holman:1982tb}
R.~Holman, G.~Lazarides and Q.~Shafi, \emph{{Axions and the Dark Matter of the
  Universe}}, \href{https://doi.org/10.1103/PhysRevD.27.995}{\emph{Phys. Rev.}
  {\bfseries D27} (1983) 995}.

\bibitem{Bajc:2005zf}
B.~Bajc, A.~Melfo, G.~Senjanovic and F.~Vissani, \emph{{Yukawa sector in
  non-supersymmetric renormalizable SO(10)}},
  \href{https://doi.org/10.1103/PhysRevD.73.055001}{\emph{Phys. Rev.}
  {\bfseries D73} (2006) 055001},
  [\href{https://arxiv.org/abs/hep-ph/0510139}{{\ttfamily hep-ph/0510139}}].

\bibitem{Altarelli:2013aqa}
G.~Altarelli and D.~Meloni, \emph{{A non supersymmetric SO(10) grand unified
  model for all the physics below $M_{GUT}$}},
  \href{https://doi.org/10.1007/JHEP08(2013)021}{\emph{JHEP} {\bfseries 08}
  (2013) 021}, [\href{https://arxiv.org/abs/1305.1001}{{\ttfamily 1305.1001}}].

\bibitem{Babu:2015bna}
K.~S. Babu and S.~Khan, \emph{{Minimal nonsupersymmetric $SO(10)$ model: Gauge
  coupling unification, proton decay, and fermion masses}},
  \href{https://doi.org/10.1103/PhysRevD.92.075018}{\emph{Phys. Rev.}
  {\bfseries D92} (2015) 075018},
  [\href{https://arxiv.org/abs/1507.06712}{{\ttfamily 1507.06712}}].

\bibitem{Fong:2014gea}
C.~S. Fong, D.~Meloni, A.~Meroni and E.~Nardi, \emph{{Leptogenesis in SO(10)}},
  \href{https://doi.org/10.1007/JHEP01(2015)111}{\emph{JHEP} {\bfseries 01}
  (2015) 111}, [\href{https://arxiv.org/abs/1412.4776}{{\ttfamily 1412.4776}}].

\bibitem{Deshpande:1992au}
N.~G. Deshpande, E.~Keith and P.~B. Pal, \emph{{Implications of LEP results for
  SO(10) grand unification}},
  \href{https://doi.org/10.1103/PhysRevD.46.2261}{\emph{Phys. Rev.} {\bfseries
  D46} (1993) 2261--2264}.

\bibitem{Deshpande:1992em}
N.~G. Deshpande, E.~Keith and P.~B. Pal, \emph{{Implications of LEP results for
  SO(10) grand unification with two intermediate stages}},
  \href{https://doi.org/10.1103/PhysRevD.47.2892}{\emph{Phys. Rev.} {\bfseries
  D47} (1993) 2892--2896},
  [\href{https://arxiv.org/abs/hep-ph/9211232}{{\ttfamily hep-ph/9211232}}].

\bibitem{Bertolini:2009qj}
S.~Bertolini, L.~Di~Luzio and M.~Malinsky, \emph{{Intermediate mass scales in
  the non-supersymmetric SO(10) grand unification: A Reappraisal}},
  \href{https://doi.org/10.1103/PhysRevD.80.015013}{\emph{Phys. Rev.}
  {\bfseries D80} (2009) 015013},
  [\href{https://arxiv.org/abs/0903.4049}{{\ttfamily 0903.4049}}].

\bibitem{Leontaris:2016jty}
G.~K. Leontaris, N.~Okada and Q.~Shafi, \emph{{Non-minimal quartic inflation in
  supersymmetric $SO(10)$}},
  \href{https://doi.org/10.1016/j.physletb.2016.12.038}{\emph{Phys. Lett.}
  {\bfseries B765} (2017) 256--259},
  [\href{https://arxiv.org/abs/1611.10196}{{\ttfamily 1611.10196}}].

\bibitem{Pati:1974yy}
J.~C. Pati and A.~Salam, \emph{{Lepton Number as the Fourth Color}},
  \href{https://doi.org/10.1103/PhysRevD.10.275,
  10.1103/PhysRevD.11.703.2}{\emph{Phys. Rev.} {\bfseries D10} (1974)
  275--289}.

\bibitem{Senjanovic:2006nc}
G.~Senjanovic, \emph{{SO(10): A Theory of fermion masses and mixings}},  in
  \emph{{CP Violation and the Flavour Puzzle: Symposium in Honour of Gustavo C.
  Branco. GustavoFest 2005, Lisbon, Portugal, July 2005}}, 2006,
  \href{https://arxiv.org/abs/hep-ph/0612312}{{\ttfamily hep-ph/0612312}},
  \href{http://inspirehep.net/record/735521/files/arXiv:hep-ph\_0612312.pdf}{http://inspirehep.net/record/735521/files/arXiv:hep-ph\_0612312.pdf}.

\bibitem{DiLuzio:2011my}
L.~Di~Luzio, \emph{{Aspects of symmetry breaking in Grand Unified Theories}},
  Ph.D. thesis, SISSA, Trieste, 2011.
\newblock \href{https://arxiv.org/abs/1110.3210}{{\ttfamily 1110.3210}}.

\bibitem{Babu:1992ia}
K.~S. Babu and R.~N. Mohapatra, \emph{{Predictive neutrino spectrum in minimal
  SO(10) grand unification}},
  \href{https://doi.org/10.1103/PhysRevLett.70.2845}{\emph{Phys. Rev. Lett.}
  {\bfseries 70} (1993) 2845--2848},
  [\href{https://arxiv.org/abs/hep-ph/9209215}{{\ttfamily hep-ph/9209215}}].

\bibitem{Joshipura:2011nn}
A.~S. Joshipura and K.~M. Patel, \emph{{Fermion Masses in SO(10) Models}},
  \href{https://doi.org/10.1103/PhysRevD.83.095002}{\emph{Phys. Rev.}
  {\bfseries D83} (2011) 095002},
  [\href{https://arxiv.org/abs/1102.5148}{{\ttfamily 1102.5148}}].

\bibitem{Dueck:2013gca}
A.~Dueck and W.~Rodejohann, \emph{{Fits to SO(10) Grand Unified Models}},
  \href{https://doi.org/10.1007/JHEP09(2013)024}{\emph{JHEP} {\bfseries 09}
  (2013) 024}, [\href{https://arxiv.org/abs/1306.4468}{{\ttfamily 1306.4468}}].

\bibitem{Chakrabortty:2017mgi}
J.~Chakrabortty, R.~Maji, S.~Mohanty, S.~K. Patra and T.~Srivastava,
  \emph{{Roadmap of left-right models rooted in GUT}},
  \href{https://arxiv.org/abs/1711.11391}{{\ttfamily 1711.11391}}.

\bibitem{Adler:1969gk}
S.~L. Adler, \emph{{Axial vector vertex in spinor electrodynamics}},
  \href{https://doi.org/10.1103/PhysRev.177.2426}{\emph{Phys. Rev.} {\bfseries
  177} (1969) 2426--2438}.

\bibitem{Bell:1969ts}
J.~S. Bell and R.~Jackiw, \emph{{A PCAC puzzle: pi0 --> gamma gamma in the
  sigma model}}, \href{https://doi.org/10.1007/BF02823296}{\emph{Nuovo Cim.}
  {\bfseries A60} (1969) 47--61}.

\bibitem{Bardeen:1969md}
W.~A. Bardeen, \emph{{Anomalous Ward identities in spinor field theories}},
  \href{https://doi.org/10.1103/PhysRev.184.1848}{\emph{Phys. Rev.} {\bfseries
  184} (1969) 1848--1857}.

\bibitem{Fujikawa:1979ay}
K.~Fujikawa, \emph{{Path Integral Measure for Gauge Invariant Fermion
  Theories}}, \href{https://doi.org/10.1103/PhysRevLett.42.1195}{\emph{Phys.
  Rev. Lett.} {\bfseries 42} (1979) 1195--1198}.

\bibitem{Breitenlohner:1977hr}
P.~Breitenlohner and D.~Maison, \emph{{Dimensional Renormalization and the
  Action Principle}}, \href{https://doi.org/10.1007/BF01609069}{\emph{Commun.
  Math. Phys.} {\bfseries 52} (1977) 11--38}.

\bibitem{Perez:2014fja}
P.~Fileviez~Perez and H.~H. Patel, \emph{{The Electroweak Vacuum Angle}},
  \href{https://doi.org/10.1016/j.physletb.2014.03.064}{\emph{Phys. Lett.}
  {\bfseries B732} (2014) 241--243},
  [\href{https://arxiv.org/abs/1402.6340}{{\ttfamily 1402.6340}}].

\bibitem{Baker:2006ts}
C.~A. Baker et~al., \emph{{An Improved experimental limit on the electric
  dipole moment of the neutron}},
  \href{https://doi.org/10.1103/PhysRevLett.97.131801}{\emph{Phys. Rev. Lett.}
  {\bfseries 97} (2006) 131801},
  [\href{https://arxiv.org/abs/hep-ex/0602020}{{\ttfamily hep-ex/0602020}}].

\bibitem{Vafa:1984xg}
C.~Vafa and E.~Witten, \emph{{Parity Conservation in QCD}},
  \href{https://doi.org/10.1103/PhysRevLett.53.535}{\emph{Phys. Rev. Lett.}
  {\bfseries 53} (1984) 535}.

\bibitem{diCortona:2015ldu}
G.~Grilli~di Cortona, E.~Hardy, J.~Pardo~Vega and G.~Villadoro, \emph{{The QCD
  axion, precisely}},
  \href{https://doi.org/10.1007/JHEP01(2016)034}{\emph{JHEP} {\bfseries 01}
  (2016) 034}, [\href{https://arxiv.org/abs/1511.02867}{{\ttfamily
  1511.02867}}].

\bibitem{Weinberg:1996kr}
S.~Weinberg, \emph{{The quantum theory of fields. Vol. 2: Modern
  applications}}.
\newblock Cambridge University Press, 2013.

\bibitem{Srednicki:1985xd}
M.~Srednicki, \emph{{Axion Couplings to Matter. 1. CP Conserving Parts}},
  \href{https://doi.org/10.1016/0550-3213(85)90054-9}{\emph{Nucl. Phys.}
  {\bfseries B260} (1985) 689--700}.

\bibitem{Dias:2014osa}
A.~G. Dias, A.~C.~B. Machado, C.~C. Nishi, A.~Ringwald and P.~Vaudrevange,
  \emph{{The Quest for an Intermediate-Scale Accidental Axion and Further
  ALPs}}, \href{https://doi.org/10.1007/JHEP06(2014)037}{\emph{JHEP} {\bfseries
  06} (2014) 037}, [\href{https://arxiv.org/abs/1403.5760}{{\ttfamily
  1403.5760}}].

\bibitem{Donnelly:1978ty}
T.~W. Donnelly, S.~J. Freedman, R.~S. Lytel, R.~D. Peccei and M.~Schwartz,
  \emph{{Do Axions Exist?}},
  \href{https://doi.org/10.1103/PhysRevD.18.1607}{\emph{Phys. Rev.} {\bfseries
  D18} (1978) 1607}.

\bibitem{Kim:1981cr}
J.~E. Kim, \emph{{NATURAL EMBEDDING OF PECCEI-QUINN SYMMETRY IN FLAVOR GRAND
  UNIFICATION}}, \href{https://doi.org/10.1103/PhysRevD.26.3221}{\emph{Phys.
  Rev.} {\bfseries D26} (1982) 3221}.

\bibitem{Zeldovich:1974uw}
{\relax Ya}.~B. Zeldovich, I.~{\relax Yu}. Kobzarev and L.~B. Okun,
  \emph{{Cosmological Consequences of the Spontaneous Breakdown of Discrete
  Symmetry}}, {\emph{Zh. Eksp. Teor. Fiz.} {\bfseries 67} (1974) 3--11}.

\bibitem{Vilenkin:1982ks}
A.~Vilenkin and A.~E. Everett, \emph{{Cosmic Strings and Domain Walls in Models
  with Goldstone and PseudoGoldstone Bosons}},
  \href{https://doi.org/10.1103/PhysRevLett.48.1867}{\emph{Phys. Rev. Lett.}
  {\bfseries 48} (1982) 1867--1870}.

\bibitem{Patrignani:2016xqp}
{\scshape Particle Data Group} collaboration, C.~Patrignani et~al.,
  \emph{{Review of Particle Physics}},
  \href{https://doi.org/10.1088/1674-1137/40/10/100001}{\emph{Chin. Phys.}
  {\bfseries C40} (2016) 100001}.

\bibitem{Zhitnitsky:1980tq}
A.~R. Zhitnitsky, \emph{{On Possible Suppression of the Axion Hadron
  Interactions. (In Russian)}}, {\emph{Sov. J. Nucl. Phys.} {\bfseries 31}
  (1980) 260}.

\bibitem{Dine:1981rt}
M.~Dine, W.~Fischler and M.~Srednicki, \emph{{A Simple Solution to the Strong
  CP Problem with a Harmless Axion}},
  \href{https://doi.org/10.1016/0370-2693(81)90590-6}{\emph{Phys. Lett.}
  {\bfseries 104B} (1981) 199--202}.

\bibitem{Geng:1990dv}
C.~Q. Geng and J.~N. Ng, \emph{{THE DOMAIN WALL NUMBER IN VARIOUS INVISIBLE
  AXION MODELS}}, \href{https://doi.org/10.1103/PhysRevD.41.3848}{\emph{Phys.
  Rev.} {\bfseries D41} (1990) 3848--3850}.

\bibitem{Lazarides:1982tw}
G.~Lazarides and Q.~Shafi, \emph{{Axion Models with No Domain Wall Problem}},
  \href{https://doi.org/10.1016/0370-2693(82)90506-8}{\emph{Phys. Lett.}
  {\bfseries B115} (1982) 21--25}.

\bibitem{Mathematica}
W.~R. Inc., ``Mathematica, {V}ersion 11.2.''

\bibitem{Feger:2012bs}
R.~Feger and T.~W. Kephart, \emph{{LieART - A Mathematica application for Lie
  algebras and representation theory}},
  \href{https://doi.org/10.1016/j.cpc.2014.12.023}{\emph{Comput. Phys. Commun.}
  {\bfseries 192} (2015) 166--195},
  [\href{https://arxiv.org/abs/1206.6379}{{\ttfamily 1206.6379}}].

\bibitem{delAguila:1980qag}
F.~del Aguila and L.~E. Ibanez, \emph{{Higgs Bosons in SO(10) and Partial
  Unification}},
  \href{https://doi.org/10.1016/0550-3213(81)90266-2}{\emph{Nucl. Phys.}
  {\bfseries B177} (1981) 60--86}.

\bibitem{Miura:2016krn}
{\scshape Super-Kamiokande} collaboration, K.~Abe et~al., \emph{{Search for
  proton decay via $p \to e^+\pi^0$ and $p \to \mu^+\pi^0$ in
  0.31  megaton·years exposure of the Super-Kamiokande water Cherenkov
  detector}}, \href{https://doi.org/10.1103/PhysRevD.95.012004}{\emph{Phys.
  Rev.} {\bfseries D95} (2017) 012004},
  [\href{https://arxiv.org/abs/1610.03597}{{\ttfamily 1610.03597}}].

\bibitem{Arvanitaki:2009fg}
A.~Arvanitaki, S.~Dimopoulos, S.~Dubovsky, N.~Kaloper and J.~March-Russell,
  \emph{{String Axiverse}},
  \href{https://doi.org/10.1103/PhysRevD.81.123530}{\emph{Phys. Rev.}
  {\bfseries D81} (2010) 123530},
  [\href{https://arxiv.org/abs/0905.4720}{{\ttfamily 0905.4720}}].

\bibitem{Arvanitaki:2010sy}
A.~Arvanitaki and S.~Dubovsky, \emph{{Exploring the String Axiverse with
  Precision Black Hole Physics}},
  \href{https://doi.org/10.1103/PhysRevD.83.044026}{\emph{Phys. Rev.}
  {\bfseries D83} (2011) 044026},
  [\href{https://arxiv.org/abs/1004.3558}{{\ttfamily 1004.3558}}].

\bibitem{Arvanitaki:2014wva}
A.~Arvanitaki, M.~Baryakhtar and X.~Huang, \emph{{Discovering the QCD Axion
  with Black Holes and Gravitational Waves}},
  \href{https://doi.org/10.1103/PhysRevD.91.084011}{\emph{Phys. Rev.}
  {\bfseries D91} (2015) 084011},
  [\href{https://arxiv.org/abs/1411.2263}{{\ttfamily 1411.2263}}].

\bibitem{Ayala:2014pea}
A.~Ayala, I.~Domínguez, M.~Giannotti, A.~Mirizzi and O.~Straniero,
  \emph{{Revisiting the bound on axion-photon coupling from Globular
  Clusters}}, \href{https://doi.org/10.1103/PhysRevLett.113.191302}{\emph{Phys.
  Rev. Lett.} {\bfseries 113} (2014) 191302},
  [\href{https://arxiv.org/abs/1406.6053}{{\ttfamily 1406.6053}}].

\bibitem{Dixit:1989ff}
V.~V. Dixit and M.~Sher, \emph{{The Futility of High Precision SO(10)
  Calculations}}, \href{https://doi.org/10.1103/PhysRevD.40.3765}{\emph{Phys.
  Rev.} {\bfseries D40} (1989) 3765}.

\bibitem{Abe:2014oxa}
{\scshape Hyper-Kamiokande Working Group} collaboration, K.~Abe et~al.,
  \emph{{A Long Baseline Neutrino Oscillation Experiment Using J-PARC Neutrino
  Beam and Hyper-Kamiokande}},  2014,
  \href{https://arxiv.org/abs/1412.4673}{{\ttfamily 1412.4673}},
  \href{https://inspirehep.net/record/1334360/files/arXiv:1412.4673.pdf}{https://inspirehep.net/record/1334360/files/arXiv:1412.4673.pdf}.

\bibitem{Saikawa:2017lzn}
K.~Saikawa, \emph{{Axion as a non-WIMP dark matter candidate}},  in \emph{{2017
  European Physical Society Conference on High Energy Physics (EPS-HEP 2017)
  Venice, Italy, July 5-12, 2017}}, 2017,
  \href{https://arxiv.org/abs/1709.07091}{{\ttfamily 1709.07091}},
  \href{http://inspirehep.net/record/1624646/files/arXiv:1709.07091.pdf}{http://inspirehep.net/record/1624646/files/arXiv:1709.07091.pdf}.

\bibitem{Ringwald:2015dsf}
A.~Ringwald and K.~Saikawa, \emph{{Axion dark matter in the post-inflationary
  Peccei-Quinn symmetry breaking scenario}},
  \href{https://doi.org/10.1103/PhysRevD.93.085031,
  10.1103/PhysRevD.94.049908}{\emph{Phys. Rev.} {\bfseries D93} (2016) 085031},
  [\href{https://arxiv.org/abs/1512.06436}{{\ttfamily 1512.06436}}].

\bibitem{Budker:2013hfa}
D.~Budker, P.~W. Graham, M.~Ledbetter, S.~Rajendran and A.~Sushkov,
  \emph{{Proposal for a Cosmic Axion Spin Precession Experiment (CASPEr)}},
  \href{https://doi.org/10.1103/PhysRevX.4.021030}{\emph{Phys. Rev.} {\bfseries
  X4} (2014) 021030}, [\href{https://arxiv.org/abs/1306.6089}{{\ttfamily
  1306.6089}}].

\bibitem{Stern:2016bbw}
I.~Stern, \emph{{ADMX Status}}, {\emph{PoS} {\bfseries ICHEP2016} (2016) 198},
  [\href{https://arxiv.org/abs/1612.08296}{{\ttfamily 1612.08296}}].

\bibitem{Chung:2016ysi}
W.~Chung, \emph{{CULTASK, The Coldest Axion Experiment at CAPP/IBS in Korea}},
  {\emph{PoS} {\bfseries CORFU2015} (2016) 047}.

\bibitem{Kahn:2016aff}
Y.~Kahn, B.~R. Safdi and J.~Thaler, \emph{{Broadband and Resonant Approaches to
  Axion Dark Matter Detection}},
  \href{https://doi.org/10.1103/PhysRevLett.117.141801}{\emph{Phys. Rev. Lett.}
  {\bfseries 117} (2016) 141801},
  [\href{https://arxiv.org/abs/1602.01086}{{\ttfamily 1602.01086}}].

\bibitem{TheMADMAXWorkingGroup:2016hpc}
{\scshape MADMAX Working Group} collaboration, A.~Caldwell, G.~Dvali,
  B.~Majorovits, A.~Millar, G.~Raffelt, J.~Redondo et~al., \emph{{Dielectric
  Haloscopes: A New Way to Detect Axion Dark Matter}},
  \href{https://doi.org/10.1103/PhysRevLett.118.091801}{\emph{Phys. Rev. Lett.}
  {\bfseries 118} (2017) 091801},
  [\href{https://arxiv.org/abs/1611.05865}{{\ttfamily 1611.05865}}].

\bibitem{Arvanitaki:2014dfa}
A.~Arvanitaki and A.~A. Geraci, \emph{{Resonantly Detecting Axion-Mediated
  Forces with Nuclear Magnetic Resonance}},
  \href{https://doi.org/10.1103/PhysRevLett.113.161801}{\emph{Phys. Rev. Lett.}
  {\bfseries 113} (2014) 161801},
  [\href{https://arxiv.org/abs/1403.1290}{{\ttfamily 1403.1290}}].

\bibitem{Armengaud:2014gea}
E.~Armengaud et~al., \emph{{Conceptual Design of the International Axion
  Observatory (IAXO)}},
  \href{https://doi.org/10.1088/1748-0221/9/05/T05002}{\emph{JINST} {\bfseries
  9} (2014) T05002}, [\href{https://arxiv.org/abs/1401.3233}{{\ttfamily
  1401.3233}}].

\bibitem{Giannotti:2017hny}
M.~Giannotti, I.~G. Irastorza, J.~Redondo, A.~Ringwald and K.~Saikawa,
  \emph{{Stellar Recipes for Axion Hunters}},
  \href{https://doi.org/10.1088/1475-7516/2017/10/010}{\emph{JCAP} {\bfseries
  1710} (2017) 010}, [\href{https://arxiv.org/abs/1708.02111}{{\ttfamily
  1708.02111}}].

\bibitem{Turner:1990uz}
M.~S. Turner and F.~Wilczek, \emph{{Inflationary axion cosmology}},
  \href{https://doi.org/10.1103/PhysRevLett.66.5}{\emph{Phys. Rev. Lett.}
  {\bfseries 66} (1991) 5--8}.

\bibitem{Beltran:2006sq}
M.~Beltran, J.~Garcia-Bellido and J.~Lesgourgues, \emph{{Isocurvature bounds on
  axions revisited}},
  \href{https://doi.org/10.1103/PhysRevD.75.103507}{\emph{Phys. Rev.}
  {\bfseries D75} (2007) 103507},
  [\href{https://arxiv.org/abs/hep-ph/0606107}{{\ttfamily hep-ph/0606107}}].

\bibitem{Hertzberg:2008wr}
M.~P. Hertzberg, M.~Tegmark and F.~Wilczek, \emph{{Axion Cosmology and the
  Energy Scale of Inflation}},
  \href{https://doi.org/10.1103/PhysRevD.78.083507}{\emph{Phys. Rev.}
  {\bfseries D78} (2008) 083507},
  [\href{https://arxiv.org/abs/0807.1726}{{\ttfamily 0807.1726}}].

\bibitem{Hamann:2009yf}
J.~Hamann, S.~Hannestad, G.~G. Raffelt and Y.~Y.~Y. Wong, \emph{{Isocurvature
  forecast in the anthropic axion window}},
  \href{https://doi.org/10.1088/1475-7516/2009/06/022}{\emph{JCAP} {\bfseries
  0906} (2009) 022}, [\href{https://arxiv.org/abs/0904.0647}{{\ttfamily
  0904.0647}}].

\bibitem{Abe:2011ts}
K.~Abe et~al., \emph{{Letter of Intent: The Hyper-Kamiokande Experiment ---
  Detector Design and Physics Potential ---}},
  \href{https://arxiv.org/abs/1109.3262}{{\ttfamily 1109.3262}}.

\bibitem{Kemp:2017kbm}
E.~Kemp, \emph{{The Deep Underground Neutrino Experiment -- DUNE: the precision
  era of neutrino physics}},  in \emph{{4th Caribbean Symposium on Cosmology,
  Gravitation, Nuclear and Astroparticle Physics (STARS2017) Havana, Cuba, May
  7-13, 2017}}, 2017, \href{https://arxiv.org/abs/1709.09385}{{\ttfamily
  1709.09385}},
  \href{http://inspirehep.net/record/1626104/files/arXiv:1709.09385.pdf}{http://inspirehep.net/record/1626104/files/arXiv:1709.09385.pdf}.

\bibitem{Ghigna:1992iv}
S.~Ghigna, M.~Lusignoli and M.~Roncadelli, \emph{{Instability of the invisible
  axion}}, \href{https://doi.org/10.1016/0370-2693(92)90019-Z}{\emph{Phys.
  Lett.} {\bfseries B283} (1992) 278--281}.

\bibitem{Barr:1992qq}
S.~M. Barr and D.~Seckel, \emph{{Planck scale corrections to axion models}},
  \href{https://doi.org/10.1103/PhysRevD.46.539}{\emph{Phys. Rev.} {\bfseries
  D46} (1992) 539--549}.

\bibitem{Kamionkowski:1992mf}
M.~Kamionkowski and J.~March-Russell, \emph{{Planck scale physics and the
  Peccei-Quinn mechanism}},
  \href{https://doi.org/10.1016/0370-2693(92)90492-M}{\emph{Phys. Lett.}
  {\bfseries B282} (1992) 137--141},
  [\href{https://arxiv.org/abs/hep-th/9202003}{{\ttfamily hep-th/9202003}}].

\bibitem{Holman:1992us}
R.~Holman, S.~D.~H. Hsu, T.~W. Kephart, E.~W. Kolb, R.~Watkins and L.~M.
  Widrow, \emph{{Solutions to the strong CP problem in a world with gravity}},
  \href{https://doi.org/10.1016/0370-2693(92)90491-L}{\emph{Phys. Lett.}
  {\bfseries B282} (1992) 132--136},
  [\href{https://arxiv.org/abs/hep-ph/9203206}{{\ttfamily hep-ph/9203206}}].

\bibitem{Brubaker:2016ktl}
B.~M. Brubaker et~al., \emph{{First results from a microwave cavity axion
  search at 24 $\mu$eV}},
  \href{https://doi.org/10.1103/PhysRevLett.118.061302}{\emph{Phys. Rev. Lett.}
  {\bfseries 118} (2017) 061302},
  [\href{https://arxiv.org/abs/1610.02580}{{\ttfamily 1610.02580}}].

\bibitem{McAllister:2017lkb}
B.~T. McAllister, G.~Flower, E.~N. Ivanov, M.~Goryachev, J.~Bourhill and M.~E.
  Tobar, \emph{{The ORGAN Experiment: An axion haloscope above 15 GHz}},
  \href{https://doi.org/10.1016/j.dark.2017.09.010}{\emph{Phys. Dark Univ.}
  {\bfseries 18} (2017) 67--72},
  [\href{https://arxiv.org/abs/1706.00209}{{\ttfamily 1706.00209}}].

\bibitem{Barbieri:2016vwg}
R.~Barbieri, C.~Braggio, G.~Carugno, C.~S. Gallo, A.~Lombardi, A.~Ortolan
  et~al., \emph{{Searching for galactic axions through magnetized media: the
  QUAX proposal}},
  \href{https://doi.org/10.1016/j.dark.2017.01.003}{\emph{Phys. Dark Univ.}
  {\bfseries 15} (2017) 135--141},
  [\href{https://arxiv.org/abs/1606.02201}{{\ttfamily 1606.02201}}].

\bibitem{Alesini:2017ifp}
D.~Alesini, D.~Babusci, D.~Di~Gioacchino, C.~Gatti, G.~Lamanna and C.~Ligi,
  \emph{{The KLASH Proposal}},
  \href{https://arxiv.org/abs/1707.06010}{{\ttfamily 1707.06010}}.

\bibitem{Rybka:2014cya}
G.~Rybka, A.~Wagner, A.~Brill, K.~Ramos, R.~Percival and K.~Patel,
  \emph{{Search for dark matter axions with the Orpheus experiment}},
  \href{https://doi.org/10.1103/PhysRevD.91.011701}{\emph{Phys. Rev.}
  {\bfseries D91} (2015) 011701},
  [\href{https://arxiv.org/abs/1403.3121}{{\ttfamily 1403.3121}}].

\bibitem{Mambrini:2015vna}
Y.~Mambrini, N.~Nagata, K.~A. Olive, J.~Quevillon and J.~Zheng, \emph{{Dark
  matter and gauge coupling unification in nonsupersymmetric SO(10) grand
  unified models}},
  \href{https://doi.org/10.1103/PhysRevD.91.095010}{\emph{Phys. Rev.}
  {\bfseries D91} (2015) 095010},
  [\href{https://arxiv.org/abs/1502.06929}{{\ttfamily 1502.06929}}].

\bibitem{Nagata:2015dma}
N.~Nagata, K.~A. Olive and J.~Zheng, \emph{{Weakly-Interacting Massive
  Particles in Non-supersymmetric SO(10) Grand Unified Models}},
  \href{https://doi.org/10.1007/JHEP10(2015)193}{\emph{JHEP} {\bfseries 10}
  (2015) 193}, [\href{https://arxiv.org/abs/1509.00809}{{\ttfamily
  1509.00809}}].

\bibitem{Arbelaez:2015ila}
C.~Arbelaez, R.~Longas, D.~Restrepo and O.~Zapata, \emph{{Fermion dark matter
  from SO(10) GUTs}},
  \href{https://doi.org/10.1103/PhysRevD.93.013012}{\emph{Phys. Rev.}
  {\bfseries D93} (2016) 013012},
  [\href{https://arxiv.org/abs/1509.06313}{{\ttfamily 1509.06313}}].

\bibitem{Boucenna:2015sdg}
S.~M. Boucenna, M.~B. Krauss and E.~Nardi, \emph{{Dark matter from the vector
  of SO (10)}},
  \href{https://doi.org/10.1016/j.physletb.2016.02.008}{\emph{Phys. Lett.}
  {\bfseries B755} (2016) 168--176},
  [\href{https://arxiv.org/abs/1511.02524}{{\ttfamily 1511.02524}}].

\bibitem{Slansky:1981yr}
R.~Slansky, \emph{{Group Theory for Unified Model Building}},
  \href{https://doi.org/10.1016/0370-1573(81)90092-2}{\emph{Phys. Rept.}
  {\bfseries 79} (1981) 1--128}.

\bibitem{Machacek:1983tz}
M.~E. Machacek and M.~T. Vaughn, \emph{{Two Loop Renormalization Group
  Equations in a General Quantum Field Theory. 1. Wave Function
  Renormalization}},
  \href{https://doi.org/10.1016/0550-3213(83)90610-7}{\emph{Nucl. Phys.}
  {\bfseries B222} (1983) 83--103}.

\bibitem{Hall:1980kf}
L.~J. Hall, \emph{{Grand Unification of Effective Gauge Theories}},
  \href{https://doi.org/10.1016/0550-3213(81)90498-3}{\emph{Nucl. Phys.}
  {\bfseries B178} (1981) 75--124}.

\bibitem{Weinberg:1980wa}
S.~Weinberg, \emph{{Effective Gauge Theories}},
  \href{https://doi.org/10.1016/0370-2693(80)90660-7}{\emph{Phys. Lett.}
  {\bfseries 91B} (1980) 51--55}.

\bibitem{Fonseca:2013bua}
R.~M. Fonseca, M.~Malinský and F.~Staub, \emph{{Renormalization group
  equations and matching in a general quantum field theory with kinetic
  mixing}}, \href{https://doi.org/10.1016/j.physletb.2013.09.042}{\emph{Phys.
  Lett.} {\bfseries B726} (2013) 882--886},
  [\href{https://arxiv.org/abs/1308.1674}{{\ttfamily 1308.1674}}].

\bibitem{Bertolini:2013vta}
S.~Bertolini, L.~Di~Luzio and M.~Malinsky, \emph{{Light color octet scalars in
  the minimal SO(10) grand unification}},
  \href{https://doi.org/10.1103/PhysRevD.87.085020}{\emph{Phys. Rev.}
  {\bfseries D87} (2013) 085020},
  [\href{https://arxiv.org/abs/1302.3401}{{\ttfamily 1302.3401}}].

\bibitem{Babu:1992cu}
K.~S. Babu and S.~M. Barr, \emph{{Family symmetry, gravity, and the strong CP
  problem}}, \href{https://doi.org/10.1016/0370-2693(93)91347-P}{\emph{Phys.
  Lett.} {\bfseries B300} (1993) 367--372},
  [\href{https://arxiv.org/abs/hep-ph/9212219}{{\ttfamily hep-ph/9212219}}].

\bibitem{Dias:2002gg}
A.~G. Dias, V.~Pleitez and M.~D. Tonasse, \emph{{Naturally light invisible
  axion in models with large local discrete symmetries}},
  \href{https://doi.org/10.1103/PhysRevD.67.095008}{\emph{Phys. Rev.}
  {\bfseries D67} (2003) 095008},
  [\href{https://arxiv.org/abs/hep-ph/0211107}{{\ttfamily hep-ph/0211107}}].

\bibitem{Babu:2002ic}
K.~S. Babu, I.~Gogoladze and K.~Wang, \emph{{Stabilizing the axion by discrete
  gauge symmetries}},
  \href{https://doi.org/10.1016/S0370-2693(03)00411-8}{\emph{Phys. Lett.}
  {\bfseries B560} (2003) 214--222},
  [\href{https://arxiv.org/abs/hep-ph/0212339}{{\ttfamily hep-ph/0212339}}].

\bibitem{Dias:2003iq}
A.~G. Dias, C.~A. de~S.~Pires and P.~S. Rodrigues~da Silva, \emph{{Discrete
  symmetries, invisible axion and lepton number symmetry in an economic 3 3 1
  model}}, \href{https://doi.org/10.1103/PhysRevD.68.115009}{\emph{Phys. Rev.}
  {\bfseries D68} (2003) 115009},
  [\href{https://arxiv.org/abs/hep-ph/0309058}{{\ttfamily hep-ph/0309058}}].

\bibitem{Bjorkeroth:2017tsz}
F.~Björkeroth, E.~J. Chun and S.~F. King, \emph{{Accidental Peccei-Quinn
  symmetry from discrete flavour symmetry and Pati-Salam}},
  \href{https://doi.org/10.1016/j.physletb.2017.12.058}{\emph{Phys. Lett.}
  {\bfseries B777} (2018) 428--434},
  [\href{https://arxiv.org/abs/1711.05741}{{\ttfamily 1711.05741}}].

\bibitem{DiLuzio:2017tjx}
L.~Di~Luzio, E.~Nardi and L.~Ubaldi, \emph{{Accidental Peccei-Quinn symmetry
  protected to arbitrary order}},
  \href{https://doi.org/10.1103/PhysRevLett.119.011801}{\emph{Phys. Rev. Lett.}
  {\bfseries 119} (2017) 011801},
  [\href{https://arxiv.org/abs/1704.01122}{{\ttfamily 1704.01122}}].

\end{thebibliography}\endgroup
